\documentclass[10pt]{article}
\usepackage{epsfig}
\input{Ijmp_d.tex}

\def\Journal#1#2#3#4{{#1}{\bf #2} #4 (#3)}
\def\JP{{\it J. Phys.\ }}

\def\EPJ{{\it Eur.\ Phys.\ J.\ }}
\def\NIM{{\it Nucl.\ Instr.\ Meth.\ }}
\def\NP{{\it Nucl.\ Phys.\ }}
\def\PL{{\it Phys.\ Lett.\ }}
\def\PRL{{\it Phys.\ Rev.\ Lett.\ }}
\def\PR{{\it Phys.\ Rev.\ }}
\def\ZP{{\it Z. Phys.\ }}
\def\JETP{{\it JETP\ }}
\def\JETPLett{{\it JETP Lett.\ }}
\def\JHEP{{\it J.\ High En.\ Phys.\ }}

\def\SJNP{{\it Sov.\ J.\ Nucl. Phys.\ }}
\def\SPJ{{\it Sov.\ Phys.\ JETP\ }}
\def\IJMP{{\it Int.\ J.\ Mod.\ Phys.\ } A}

\def\PLB{{\it Phys. Lett.\ }B}

\def\CPC{{\it Comp.\ Phys.\ Comm.\ }}

\newlength{\gap}
\newlength{\numwid}
\newcommand{\QQ}{\mbox{$Q^2$}}
\newcommand{\qh}{\mbox{$q_h$}}
\newcommand{\qhbar}{\mbox{$\bar{q}_h$}}
\newcommand{\pT}{\mbox{$p_T$}}

\newcommand{\ET}{\mbox{$E_T$}}

\newcommand{\gs}{\mbox{$\gamma^*$}} 
\newcommand{\xg}{\mbox{$x_\gamma$}}
\newcommand{\xgO}{\mbox{$x_\gamma^{\mbox{\tiny OBS}}$}} 
\newcommand{\LQCD}{\mbox{$\Lambda_\mathit{QCD}$}} 
\newcommand{\Jpsi}{\mbox{$J/\psi$}} 
\newcommand{\HAd}{H1 Collab., C. Adloff et al., }
\newcommand{\ZEUSMXD}{ZEUS Collab., M. Derrick et al., }
\newcommand{\ZEUSBre}{ZEUS Collab., J. Breitweg et al., }
\newcommand{\ZEUSCh} {ZEUS Collab., S. Chekanov et al., }
\newcommand{\ZBud}[1]{ZEUS Collab., contributed paper  #1, 
Int.\ Europhys.\  Conf.\  on HEP, Buda\-pest, 2001.}
\newcommand{\ZBuds}[1]{ZEUS Collab., contributed papers  #1, 
Int.\ Europhys.\  Conf.\  on HEP, Buda\-pest, 2001 }
\def\HBud#1{H1 Collab., contributed paper  #1, 
Int.\ Europhys.\  Conf.\  on HEP, Buda\-pest, 2001.}
\def\ZOsa#1{ZEUS Collab., contributed paper  #1, 
XXX Int.\ Conf.\  on HEP, Osaka, 2000.}
\def\HOsa#1{H1 Collab., contributed paper  #1, 
XXX Int.\ Conf.\  on HEP, Osaka, 2000}
\def\HTam#1{H1 Collab., contributed paper  #1, 
Int.\ Europhys.\  Conf.\  on HEP, Tampere, 1999.}
\newcommand{\Bologna} {proc.\ DIS 2001, Bologna 
(World Scientific, to be published)}

\newcommand{\avval}[1]{\mbox{$<\!#1\!>$}}

\newcommand{\sleq}{\raisebox{-.4ex}{$\;\stackrel{<}{\scriptstyle \sim}\;$}}

\begin{document}
\alph{footnote}
\runninghead{Heavy Flavour Physics at HERA$\ldots$} 
{Heavy Flavour Physics at HERA$\ldots$}

\normalsize\textlineskip
\thispagestyle{empty}
\setcounter{page}{1}

\copyrightheading{}			%{Vol. 0, No. 0 (1993) 000--000}

\vspace*{0.88truein}

\fpage{1}
\centerline{\bf HEAVY FLAVOUR PHYSICS AT HERA -- A SURVEY}
\vspace*{0.37truein}
\centerline{\footnotesize P. J. BUSSEY}
\vspace*{0.015truein}
\centerline{\footnotesize\it Department of Physics and Astronomy, 
University of Glasgow,}
\baselineskip=10pt
\centerline{\footnotesize\it Glasgow G12 8QQ, UK}
\vspace*{0.225truein}
%%%%%\centerline{{\footnotesize Received 25 Sep.\ 2001}}\vspace*{5mm}

\vspace*{0.21truein}
 
%%%%%%%%%%%%%%%%%%%%% next line special for d version %%%%%%%%
\centerline{{\footnotesize\bf Abstract}}\vspace*{1mm}
\abstracts{    
At the HERA collider at DESY, high
energy electron and positron beams interact with proton beams.  A
review is presented of the variety of ways in which these collisions
produce final states containing  charm and beauty quarks.\\ }{}{}

%\textlineskip			%) USE THIS MEASUREMENT WHEN THERE IS
%\vspace*{12pt}			%) NO SECTION HEADING

\vspace*{1pt}\textlineskip	%) USE THIS MEASUREMENT WHEN THERE IS
%\section{General Appearance}	%) A SECTION HEADING

\section{Introduction}
\subsection{The HERA collider}
\begin{figure} 
\centerline{
\epsfig{file=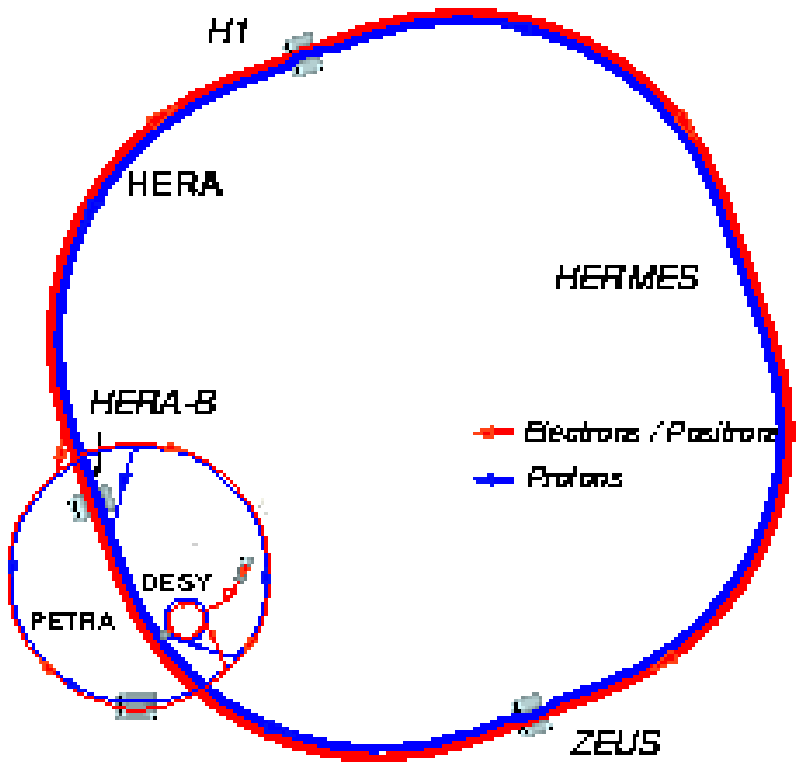%
,width=7cm,bbllx=60pt,bblly=460pt,bburx=300pt,bbury=690pt,clip=}
\hspace*{-5mm}\raisebox{20mm}{\epsfig{file=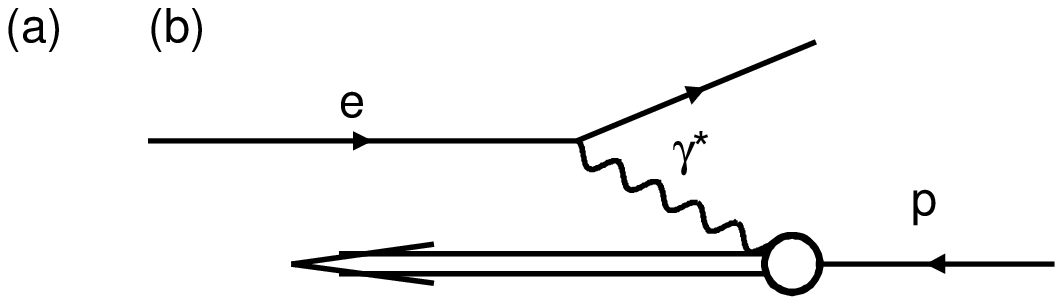,%
width=6cm,bbllx=160pt,bblly=310pt,bburx=470pt,bbury=560pt,clip=}}
}\vspace*{3mm}
\fcaption{(a) The HERA collider complex at DESY, showing the $ep$ ring and
the lower energy accelerator PETRA.  The H1 and ZEUS experiments are
in the North and South halls respectively.  
The proton beam energy was raised
from 820 to 920 GeV at the end of 1998. (b) $ep$ interaction
mechanism through the radiation of a virtual photon. The
broad arrow denotes the final-state products. }
\label{heraring}\end{figure}

\noindent
The HERA collider is the only experimental facility in the world where
high-energy electron-proton and positron-proton interactions can be
studied.  Electrons (or positrons) and protons are accelerated in the
DESY ring complex to energies of 27.5 GeV and 920 GeV respectively
(fig.~\ref{heraring}a).  The circulating beams are then brought to a
focus at the two collision points where the H1 and ZEUS experiments
are located.  In most of the $e^\pm p$ interactions, the incoming
$e^\pm $ interacts with the proton by first radiating a virtual photon
(fig.~\ref{heraring}b).  The virtuality \QQ\ of this exchanged photon
(for definition see section 1.5) lies within a wide range; its minimum
value is limited by the kinematics of the radiation process to be of
the order of the electron mass squared, while its maximum value is
limited to $10^5$ GeV$^2$ by the energies of the incoming beams.

 During the period 1992-2000, the H1 and ZEUS experiments each
accumulated approximately 135 pb$^{-1}$ of integrated luminosity, which
has enabled a broad range  of physics to be investigated.  At
low \QQ, the exchanged photon resembles a real photon, and we shall
usually refer to such collisions simply as $\gamma p$ or
``photoproduction''.  Real photons, as will be discussed below, can
display hadronic as well as electromagnetic properties.  The hadronic
behaviour falls away over a transition region where \QQ\ is of the
order of 1 GeV$^2$, above which we are in the regime of deep inelastic
scattering (DIS).  Here the photon is primarily an electromagnetic
object which couples to quark-antiquark ($q\bar q$) pairs.  At very
high \QQ\ the exchange of $Z$ and $W$ bosons also becomes
important.  The $\gamma^*p$ luminosity is approximately uniform in
$\log \QQ$, giving a prolific production of low-\QQ\ photons.

Differences between $e^+p$ and $e^-p$ collisions are of significance
in some new particle searches and at the very high \QQ\ values where
electroweak effects occur.  In the physics reviewed here, however, the
sign of the incoming $e$ beam is nearly always unimportant.  For
convenience, we shall normally refer to the incoming $e^\pm$ as a
positron, including electrons by implication, since most of the
integrated luminosity was taken with positrons.  The symbol $\gamma^*$
will denote any virtual photon, while $\gamma$ will refer, where
appropriate, to the particular case of low-\QQ, quasi-real photons.

\subsection{QCD physics with heavy quarks at HERA}
\noindent
In the present survey, we consider those processes at HERA in which
heavy flavours are produced, namely the production of particles
containing charm $(c)$ and beauty $(b)$ quarks.  HERA is {\it par
excellence\/} a laboratory for the study of quantum chromodynamics
(QCD), our present theory of the strong interaction.  While $e^+e^-$
colliders such as LEP have enabled the production of $q\bar q$ pairs
to be studied in depth, along with the associated radiation of
gluons, the presence of the proton in the initial state at HERA allows
a wide variety of further QCD processes to be investigated.  The
incoming $\gamma^*$, with its range of virtualities, probes powerfully
the partonic properties of the proton and the reactions
initiated through the quarks and gluons of which the proton is a
source.

Heavy quark production can also, of course, be studed in hadron-hadron
colliders such as the Tevatron.  However a virtual photon is a
cleaner probe than a second high-energy hadron, and it is
experimentally possible to trigger on photon-induced processes that
occur down to lower transverse momenta, down to a few GeV.

The behaviour of QCD processes is governed by the running coupling
constant $\alpha_s$, whose value is $\approx$0.12 at momentum
transfers corresponding to the $Z$ mass, though significantly higher
at a few GeV.  Consequently, the lowest-order (LO) perturbative
calculations, which suffice to a first approximation, require
frequently to be replaced by next-to-leading order (NLO) calculations
or higher.  This can present formidable technical difficulties, which
can limit our experimental understanding of QCD.  With light quark
systems, a certain impasse may by now have been reached, but the
consideration of $b$ or $c$ quarks presents a number of new factors.
These arise because the production of a heavy object requires a
corresponding momentum transfer which may be significantly higher than
the QCD scale parameter $\Lambda_{QCD}\approx 300$ MeV.  Such
processes offer a different perspective on the operation of QCD.  The
measured final-state hadron containing a heavy quark relates directly
to the properties of the QCD-governed hard scatter, and although the
presence of a second momentum scale can complicate the theory, it
provides a means of calculating cross sections using perturbative
approaches that are unavailable, or less reliable, when dealing with
exclusively light quark systems.

There remain a number of difficulties, however.  
The effective mass of the $c$ quark is uncertain within
the range 1.3 to 1.5 GeV, a value which is itself not overwhelmingly
higher than $\Lambda_{QCD}$. These facts translate into
significant uncertainties in the perturbative calculations, comparable
to the effects of higher-order QCD diagrams.  As will be seen, this
makes the interpretation of many experiments less than
straight\-forward.  The higher mass ($\approx 4.75$ GeV) of the $b$
quark should reduce these problems; nevertheless it will be seen
that even the production of $b$ hadrons currently presents challenges
to theory.

A major aspect of heavy quark physics concerns the weak decays of the
hadrons that contain these quarks.  Such decays may be studied at any
collider that is able to produce the relevant hadrons in sufficiently large
numbers.  Many of the most interesting decay channels have low
branching ratios, however, and HERA has not yet produced data in
enough quantity to be competitive here; CP violating effects,
especially, lie in the domain of the $B$ factories and the Tevatron.
The present survey therefore concentrates on the production mechanisms
of heavy flavour systems, an area where HERA is able to play a
prominent role.  The author's aim is to provide an account which is
accessible to non-specialist readers and gives  an up-to-date
picture of the state of this area of HERA physics.  Experimental
results are preliminary where indicated.

\subsection{The H1 and ZEUS detectors}
\noindent
The H1 and ZEUS detectors\cite{h1,z} are situated respectively at the
north and south collision regions of HERA.  Each detector comprises of
a cylindrical central tracking chamber, surrounded by calorimeters.
Both central tracking chambers are of drift chamber design.  In the
case of H1, the calorimeters consist of lead and steel converters in
liquid argon, and are mainly located inside the solenoidal magnet
which provides a magnetic field of 1.15 Tesla.  In ZEUS, the
calorimeter is a uranium-scintillator sandwich. It lies outside the
solenoid, within which there is a field of 1.4 Tesla.  The return
yokes are part of the muon detection system, with muon detection
arrays outside them.  H1 has a silicon vertex detector array; ZEUS is
currently commissioning its own silicon system.

Both detectors have a design which is approximately cylindrically
symmetric around the beam line, but with a forward-backward asymmetry.
In the forward (proton beam) direction there is a need for good hadron
calorimetry in order to detect energetic particles associated with the
proton, its remnants, and particles that are in general boosted in
this direction.  In the rear direction, the best possible
electromagnetic calorimetry is required to measure accurately the deep
inelastic scattered positrons.

In both experiments the calorimetry is designed to be as hermetic as
possible, with holes of a minimum size for the beam pipes.

\subsection{Types of photon-hadron interaction}
\begin{figure}[t!]
\begin{center}
\epsfig{file=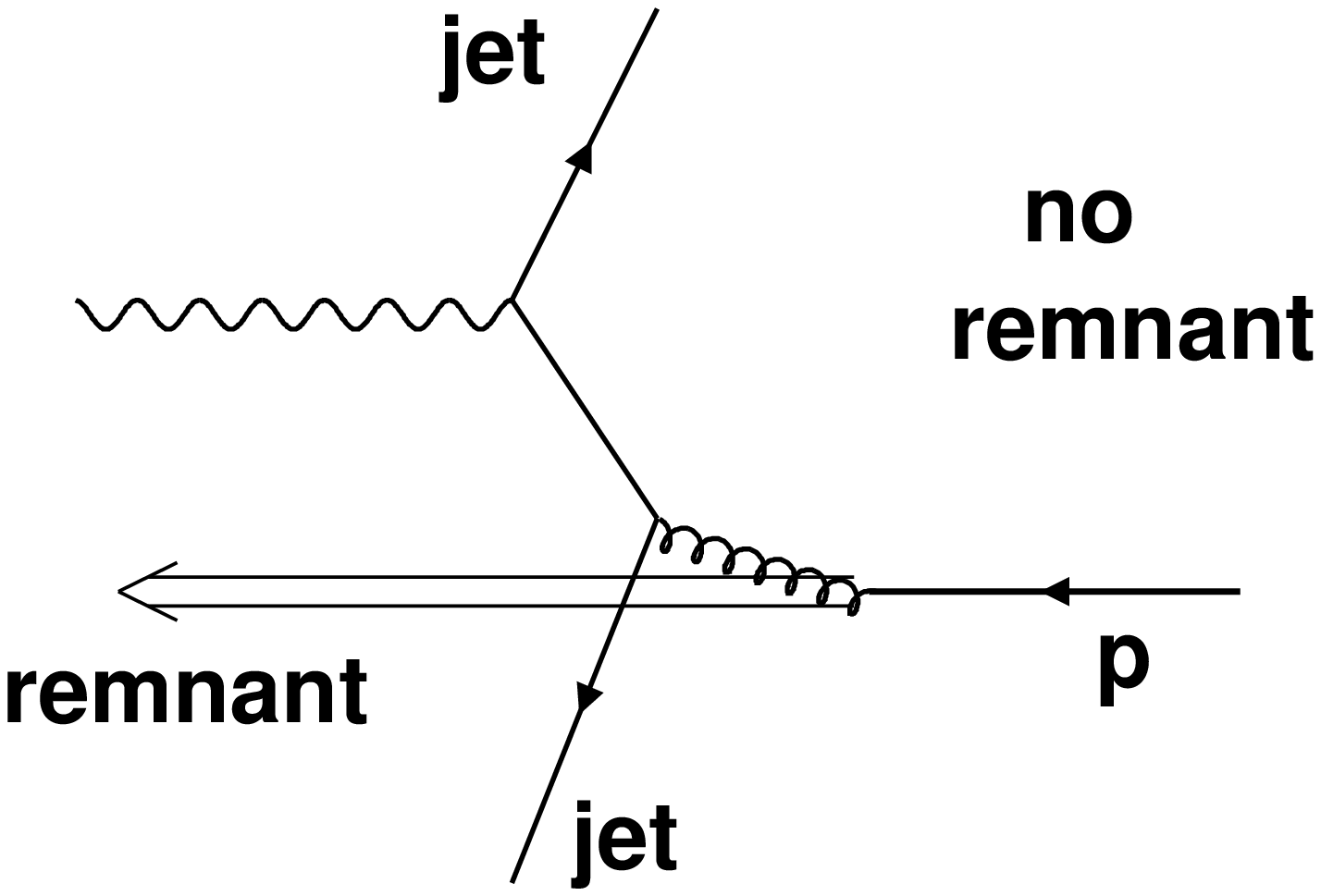,width=3.6cm%
,bbllx=100pt,bblly=90pt,bburx=540pt,bbury=425pt,clip=}
\hspace*{6mm} 
\epsfig{file=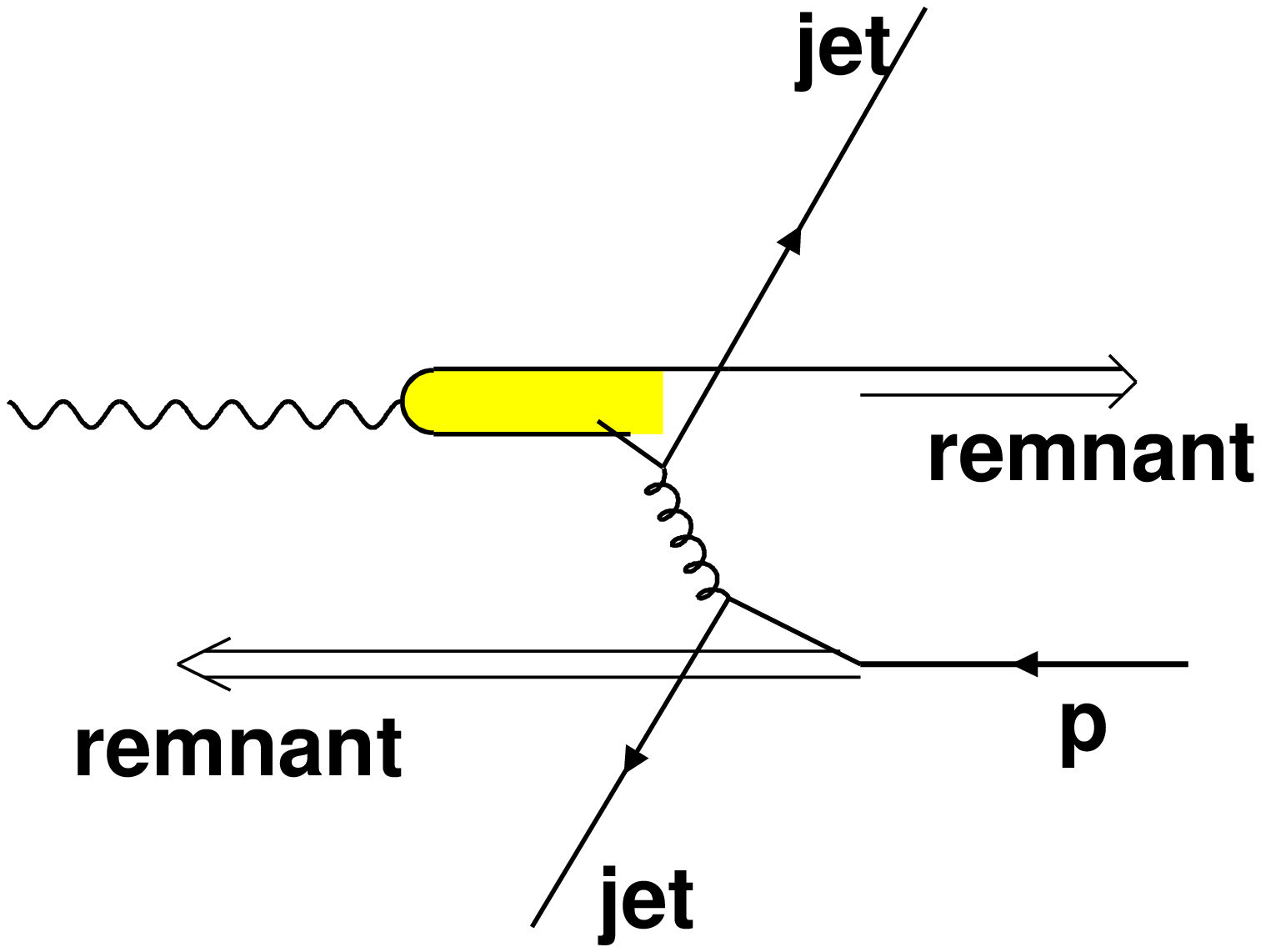,width=3.6cm%
,bbllx=100pt,bblly=90pt,bburx=540pt,bbury=425pt,clip=} 
\hspace*{6mm} 
\epsfig{file=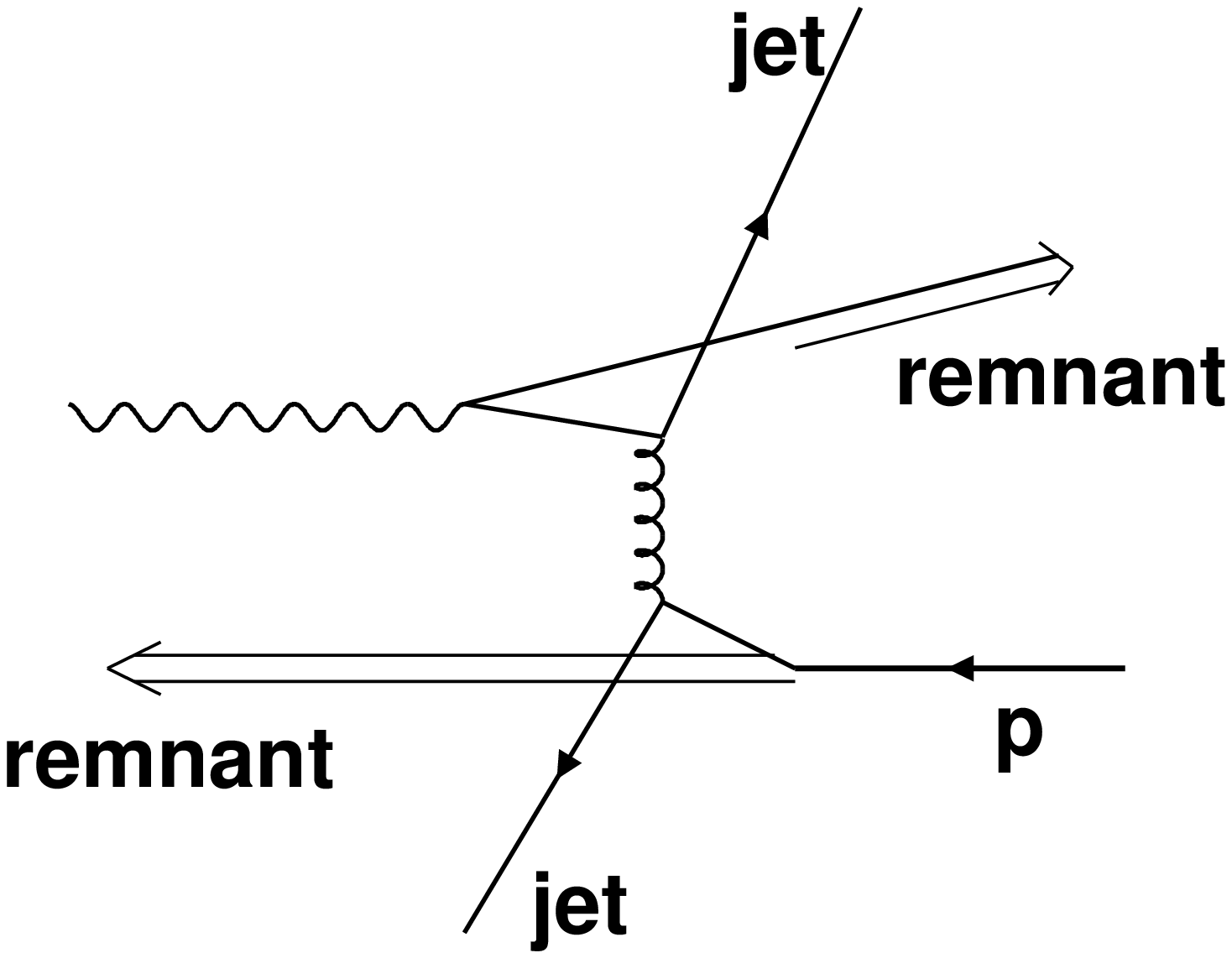,width=3.6cm% 
,bbllx=100pt,bblly=90pt,bburx=540pt,bbury=425pt,clip=} 
\mbox{\sf \hspace*{25mm} 
(a)\hspace*{38mm}(b)\hspace*{42mm}(c)\hspace*{5mm}}\\[-3mm]
\epsfig{file=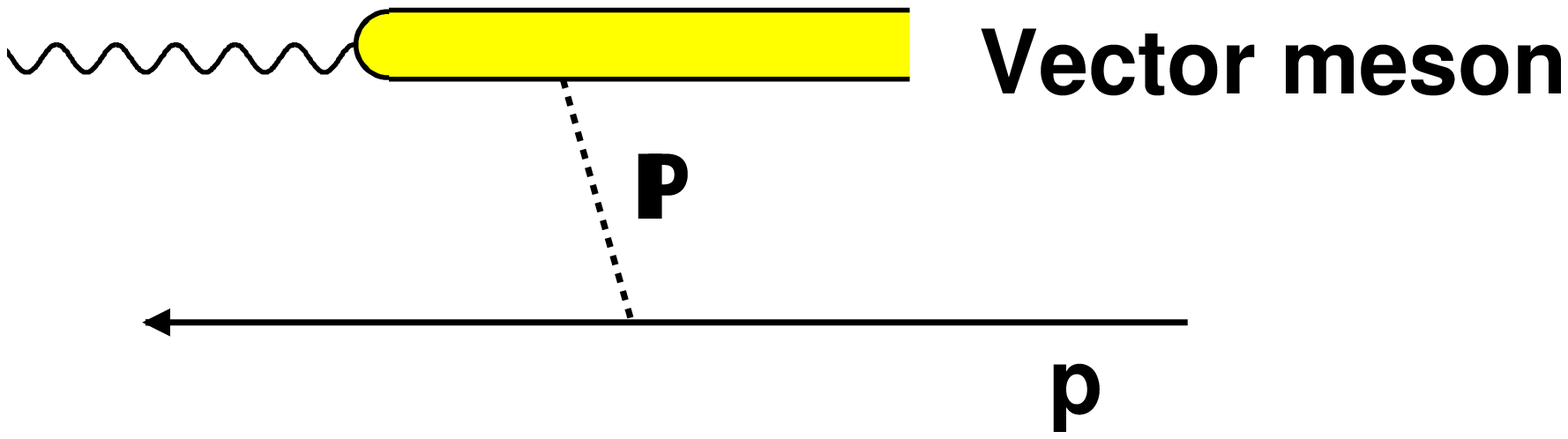,width=4.4cm%
,bbllx=50pt,bblly=150pt,bburx=570pt,bbury=425pt,clip=} 
\mbox{\sf (d)\hspace*{5mm}}\\[5mm]
\end{center}
\fcaption{Examples of different types of photon-proton interaction: 
(a) direct (b) resolved (c) anomalous (d) elastic diffractive.}\label{resdir}
\end{figure}

\noindent
Since the $ep$ processes of concern here are mediated almost entirely
by photons, some basic classes of photon-proton interactions must be
considered.  The most important of these are illustrated in
fig.~\ref{resdir}, where the first three diagrams depict different
hard QCD scattering processes.  {\it Direct\/} processes (a) are those
in which the entire photon couples to a quark-antiquark line at high
transverse momentum \pT.  
%===================================================================
%=================================================================
Alternatively (b), the proton may interact
through an intermediate hadronic state which is a source of
partons that can scatter off those in the proton.  These are known as
{\it resolved\/} photon processes, and here the photon possesses a
parton structure like any other hadron. The concept of a resolved
photon can be extended to situations where the photon couples to a
relatively low-\pT\ quark pair (c). These perturbatively calculable
processes are sometimes referred to as {\it anomalous} photon
processes.\cite{anom}

Ultimately, the outgoing partons hadronise as jets or beam remnants.
The signature for a resolved or direct photon process at LO is the
presence or absence, respectively, of a photon remnant in the event.
The direct and resolved classes of photon interaction are fully
distinct only in LO processes such as those illustrated.  At NLO the
two classes can still be defined for calculational purposes, but a
continuum of diagrams exists connecting them.  In photoproduction, the
resolved processes occur at higher rates than the direct process,
but as the \QQ\ of the photon increases, the resolved cross sections
decrease and the reactions become predominantly direct in nature.

The fraction \xg\ of the incoming photon energy that enters a
given QCD subprocess is an important parton-level quantity.  In direct
processes it is by definition unity, while in resolved processes it
can take any value in the range (0,1).  The parameter \xg\ is clearly
defined in LO diagrams but at higher order there can be ambiguities.
Experimentally the hadronic final state is measured, and a suitable 
final-state quantity is required which correlates with \xg\ in a
chosen theoretical description of the process.  For dijet final
states, two such experimental estimators\cite{zobs,zmeas} are
$$ \xgO = 
\frac{\Sigma_\mathit{jets}E_T^\mathit{\;jet} 
 e^{-\eta^\mathit{jet}}}{2 E_\gamma}
\mbox{\hspace*{7mm}and\hspace*{7mm}}  
x_\gamma^{\;\mathit{meas}} = \frac{\Sigma_\mathit{jets}(E -p_z)}
{\Sigma_\mathit{event}(E -p_z)}.$$ 
Here as elsewhere \ET\ denotes transverse energy, namely $E
\sin\theta$ for a single particle, summed over the particles in a jet.   
The laboratory pseudorapidity $\eta$ is given by
$$ \eta = - \ln \tan \frac{\theta}{2},$$ 
where $\theta$ is the polar angle and the proton beam defines the
forward ($+z$) direction.  The proton has $E-p_z = 0$ to an excellent
approximation, and a quasi-real incoming photon has $E - p_z =
2E_\gamma =\Sigma_\mathit{event}(E -p_z)$.  Snowmass --- i.e.\
\ET-averaged --- quantities\cite{sno} are used in \xgO, while in
$x_\gamma^\mathit{\; meas}$ the sums are over all particles in the
jets and in the event as a whole.  One notes that for a massless
object, $E_T e^{-\eta}\equiv(E - p_z)$; the two estimators differ in
that \xgO\ neglects jet-mass effects.  The estimator \xgO\ has been
mostly used in HERA analyses.

A further class of processes in which $b\bar b$ and $c\bar c$ systems
are produced comprises {\it diffractive\/} processes.  An example is
illustrated in fig.~\ref{resdir}d, where the photon forms an
intermediate vector meson state $V$ which scatters off the proton by
the exchange of a pomeron.  Both quasi-real and highly virtual photons
may take part in diffractive processes at HERA; these will be the
subject of Section 6 of this paper.

\subsection{DIS kinematics}
\noindent
Let the incoming and the scattered positron have four-momenta $k$ and
$k'$ respectively.  Then the exchanged photon virtuality is 
\begin{equation}
\QQ = -q^2 = -(k-k')^2 = 4 EE'\cos^2 (\theta_e/2),\end{equation}
where $\theta_e$ is the polar angle of the scattered positron relative
to the proton beam direction.  If the overall $ep$ centre-of-mass
energy is $s = (k+P)^2$, where $P$ is the proton four-momentum, then
Bjorken $x$ is defined as $\QQ/(2P\cdot q)$.  The quantity
\begin{equation}
y=\frac{\QQ}{sx} = 1 - \frac{E'}{E}\frac{1 - \cos\theta_e}{2},
\end{equation}
represents the virtual photon energy, as a fraction of the positron
energy, in the proton rest frame. 

The cross sections for unpolarised DIS can be expressed in terms of
structure functions $F_2$ and $F_1$.  It is usually valid to neglect
the contribution from $F_1$; one then obtains:
\begin{equation}
\frac{d^2\sigma}{dx\,dQ^2} = \frac {2\pi\alpha^2}{Q^4x}\left(1 + 
(1-y)^2\right) F_2(x,Q^2)
\end{equation}
in terms of the electromagnetic coupling constant $\alpha$.
The overall structure function will be denoted by $F_2$ and the
structure function for charm-containing events by $F_2^c$

The $\gamma^*p$ centre-of-mass
energy or, equivalently, that of the final-state hadronic system,
is termed  $W$.

\section{QCD Aspects of Heavy Flavour Production}
\noindent
Perturbative methods in QCD calculations require the occurrence of a
momentum transfer that is much greater than the QCD scale parameter
\LQCD.  We shall first outline an approach that is widely used in
calculating heavy quark cross sections for processes involving a
high-\pT\ scatter.  Let a hadron $H$, containing a heavy quark of
flavour $h$, be produced from two initial-state partons $i$, $j$.  The
cross section is then expressed as the product of two basic terms:
\begin{equation}  \sigma_{QCD}(ij\to hx)\;f(h\to H).  \label{eqqcd}
\end{equation}   
where $\sigma_{QCD}$ is the perturbatively calculated QCD cross
section, i.e.\ at the parton level.  Here the symbol $x$ denotes further
parton-level products, and the fragmentation factor $f(h\to H)$ is the
probability for $h$ to fragment into $H$ in an outgoing
jet. The inclusive fragmentation probability for $x$ is taken as
unity; if we are interested in a particular final-state product $X$
then a further factor $f(x\to X)$ is needed.  All these quantities are
kinematics-dependent.

In HERA physics, the initial-state partons $i$, $j$ can
be taken as quarks, gluons or the incoming photon itself.  A parton
distribution function (pdf), namely $f(p\to j)$, is required for the
proton.  In resolved photon processes, a pdf $f(\gamma \to i)$ must be
supplied similarly to denote the probability that the photon gives
rise to the quark or gluon $i$ that interacts.

Finally, expressions of type (\ref{eqqcd}) are summed over all the
relevant parton states.  Thus for direct inclusive photoproduction of
a given $D$ meson one can write 
\begin{equation} \sigma(\gamma p
\to Dx) = \sum_{j,x}f(p\to j)\;\sigma_{QCD}(\gamma j\to cx)\;f(c\to D).
 \label{dircharm} \end{equation}  

The following points need to be noted: 
\begin{romanlist} 
\item The QCD term $\sigma_{QCD}$ may be calculated at LO or to
higher orders; 
\item The assumption that a process may be split into
separate terms in this way is known as {\it factorisation.}  A given
fragmentation factor or pdf is normally assumed to be independent of
the perturbative QCD process $ij\to hx$, i.e.\ it is {\it universal.}
A mention of factorisation usually implies universality.

\item The pdf's and fragmentation factors incorporate a variety of
initial-state and final-state processes.  For example gluon radiation
may occur and, if sufficiently hard, might legitimately be regarded as
constituting a higher-order QCD process.  By imposing a {\it
factorisation scale,} one defines up to what momentum transfers such a
gluon or its products will still be called part of the proton
structure or part of the jet.

\item In general, these are non-perturbative quantities and must
be obtained either from models or from experimental measurement.  The
assumption of universality implies that a fragmentation factor
measured in one experiment can be used correspondingly in another
experimental context.  The pdf's are usually obtained from fits to
large collections of experimental data.

\item As the order of the QCD calculation rises, 
the validity of the above assumptions becomes more questionable.
\end{romanlist}

Behind the kind of scheme outlined above there lies the physical
insight that hard processes take place over a shorter time-scale than
soft processes.  Thus the production of the heavy quark may be
regarded as prior to, and physically distinguishable from, the
subsequent slower hadronisation phase.

\subsection{Methods and models}
\noindent
At this point we present some theoretical models and procedures
which are in common  use in the analysis of heavy quark production.
Some further topics will be discussed later in connection with the
specific areas where they are met.
\begin{itemlist}
\item{\it PYTHIA, HERWIG.}
These well-known Monte Carlo models can calculate a large variety of
QCD processes at LO. Initial and final state leading-logarithm parton
showers are incorporated to simulate certain kinds of higher order
effect.  Hadronisation is performed in two different ways.  In
PYTHIA,\cite{PYTHIA} colour strings are constructed 
between the final-state partons, and then hadronised according to a
set of phenomenological prescriptions.  In HERWIG,\cite{HERWIG} a
series of parton clusters are produced and allowed to decay.  Both
these models are extensively used, and form the basis of further
models.  The PYTHIA colour-string fragmentation is also known by the
name of JETSET.\\[-2mm]
\item{\it Peterson fragmentation model.}  
A standard representation of the fragmentation factor $f(h\to H)$,
based on phenomenological considerations, is the Peterson
formula:\cite{peterson}
\begin{equation} 
f(h\to H) \;=\; PD(z) \;=\; P\frac{A}{z\left[1 -1/z-\epsilon/(1-z)\right]^2},
\end{equation}
in which  the hadron of interest, $H$, is produced at a given Feynman $z$
relative to the momentum of the heavy quark $h$, with $z = (E +
p_{||})_H/(E + p_{||})_h$ where $ p_{||}$ denotes the momentum
component along the $h$ direction.  $A$ is a normalisation constant,
and $P$ is the total probability for $h$ to fragment to $H$.  The
so-called Peterson parameter $\epsilon$ is determined from
experiment:\cite{petpar} a small value means that the fragmentation is
peaked near $z = 1$. For $c$ and $b$ quarks, respectively, typical
values of $\epsilon$ are 0.035 and 0.006. There is a question of
whether the hadron is produced, on average, along the same direction
as the quark, as taken in Peterson fragmentation, or whether the
colour string from the proton remnant might pull the hadron to a
higher rapidity (the ``beam-drag effect''\cite{beamdrag}).  H1 have
concluded that the effect is unimportant in DIS charm
production.\cite{p791} On the other hand ZEUS\cite{p493a} have
obtained an improved description of charm rapidity distributions when
the hadronisation procedure from JETSET or HERWIG is used in place of
the Peterson method.  These issues have been discussed further by
Bodwin and Harris.\cite{bodhar} An analytic fragmentation function
with possibly better relativistic properties is that of
Bowler.\cite{bowler} Others have been given by
Kartvelishvili et al.\cite{kartv} and Collins et al.;\cite{pcollins} see also the
review by Frixione et al.\cite{frixetal} 
\\[-2mm]

\item{\it Fixed flavour-number schemes.} 
In heavy quark production up to NLO, two main schemes have been
proposed.  In the {\it fixed flavour-number scheme,} often referred to
in the charm case as the {\it massive charm\/} scheme, the incoming
photon and proton are given hadronic structures which contain only
three quark flavours ($u,$ $d,$ $s$).  QCD interactions are then
generated which produce heavy quark pairs (\qh\qhbar), whose dynamics
is calculated using a realistic quark mass assignment (e.g.\ $m_c =
1.5$ GeV).  The heavy quark then fragments --- e.g.\ using the Peterson
formula --- into an observable hadron.  Since heavy quark excitation
processes (see Section 3) are not treated, and may be important at
high energies, these schemes are expected to work best at $\pT =
O(m_h)$.\\[-2mm]

\item {\it Variable flavour-number schemes.}
To enable quark excitation to take place, charm and beauty are treated
as active flavours in the proton or photon.  In the more common {\it
massless\/} (``zero-mass'') versions, the mass $m_h$ is treated as
zero up to the final hadronisation which produces a massive $H$
hadron.  QCD processes taking place at $\pT\approx m_h$ may therefore not
be accurately described, but this approach should work well at high
\pT.\\[-2mm]

\item{\it DGLAP evolution.}
Given a set of pdf's or a structure function which apply at low \QQ,
one may wish to know the corresponding quantities at higher \QQ\ for a
given Bjorken $x$ value.  The DGLAP equations\cite{dglap} evaluate the
necessary gluon radiations and splittings that occur in this
transition, using perturbative QCD to a given order.
\\[-2mm]

\item{\it CCFM evolution.}
As an alternative to DGLAP evolution, the CCFM scheme\cite{ccfm} uses
a different form of factorisation in which parton densities are
explicitly treated as a function of transverse momentum (which is
integrated over in the DGLAP treatment) and an angular ordering
parameter is introduced.  This is a development of the earlier BFKL
evolution scheme,\cite{bfkl} which evolves amplitudes over $x$ instead
of \QQ\ and claims advantages at small $x$.  With CCFM there is a
matched transition to the DGLAP regime; it is implemented in
the CASCADE Monte Carlo among others.\cite{cascade,lonnblad}\\[-2mm]

\item{\it Parton distribution functions.}
A variety of pdf's are on the market to describe the quark and gluon 
densities in the proton and  the photon.  They must be provided in
versions suited to the order of the QCD calculation to which they are
to be applied.  In the case of the proton, the pdf's are usually
obtained from fits to large collections of experimental data, and are
updated when new experimental results are announced.  The constraints
of DGLAP evolution are normally applied.  Two groups active in this
area are CTEQ and MRST, whose publications may be consulted for
further details.\cite{cteq,mrst} In addition the H1 and ZEUS
collaborations have calculated their own proton pdf
fits on the basis of the DGLAP equations.\cite{h1pdf,zeuspdf}\\ 
Photon pdf's are normally obtained based on the Vector Meson
Dominance model (see Section 6), so that the partonic structure of the
photon at low virtualities uses that taken for virtual mesons.
An additional direct coupling of the photon to quark pairs (``box
diagram'') is included, and allowance is made for free parameters,
such as the gluon density in the photon at low virtuality.  Evolution
in \QQ\ from a base value $Q_0^2$ is performed.  The models cited in
the analyses quoted here, namely GRV,\cite{grv} GS,\cite{gs}
AFG,\cite{afg} and SaS,\cite{sas} differ in a number of technical
aspects, such as the value of $Q_0$, the factorisation scheme,
treatment of heavy quark thresholds, and the number of free parameters
that have been fitted to published data.  The SaS model extends the
pdf's to virtual photons.  A review of some of the issues has been
given by Vogt.\cite{vogtlund}

\end{itemlist}

\section{Open Charm Production at HERA}
\noindent
The production of heavy quarks ($q_h$) in $ep$ collisions takes place
through three main mechanisms, which
%are illustrated in fig.~\ref{bdiags}, and 
are here briefly outlined in turn:

In {\it photon-gluon fusion\/} (fig.\ \ref{resdir}a), the incoming
virtual photon interacts with a gluon from the proton, so as to form an
outgoing $\qhbar \qh$ pair through the reaction $\gamma^*g\to
\qh\qhbar$. The LO diagram often offers a good description when the quark
pair emerges at high \pT.  The fusion process can give a direct
measurement of the gluon content of the proton; this complements the
determinations through QCD fits to the proton structure function $F_2$
as a function of \QQ.  We shall refer to ``photon-gluon fusion'' in
the context of photoproduction, and ``boson-gluon fusion'' in DIS,
where $\gamma^*$ and $W/Z$ exchanges may both be present.

In {\it excitation\/} processes, the incoming \gs\ interacts with a
heavy quark that is an effective part of the proton structure and
scatters it out of the proton.  This may occur through resolved photon
processes and through the direct ``QCD Compton'' process,
$\gamma^*\qh\to g\qh$, whose cross section is somewhat smaller than that
of the fusion process.  The proton has only light valence quarks, and
so any $c$ or $b$ content in its pdf is generated by DGLAP evolution --- 
specifically, through a gluon fluctuating into a $\qh\qhbar$ pair.
Similar considerations apply to the photon.

In production by {\it fragmentation,} the heavy quark is found in a
jet that originated from a light parton: specifically, an outgoing
gluon splits into a $c\bar c$ or $b\bar b$ pair.  Fragmentation may be
treated entirely phenomenologically, obtaining the distributions from
fits to data, or it may be calculated in perturbative QCD, e.g.\ in
the process $\gamma q \to gq$ with $g\to\qh\qhbar$, namely ``gluon
splitting''.  This is a higher order QCD process than the LO fusion
term $\gamma g \to\qh\qhbar,$.

The distribution of an observed hadron $H$ in a jet requires in any
case a phenomenological treatment of the hadronisation process.

\subsection{Charm meson states}
\begin{figure}\centerline{
\epsfig{file=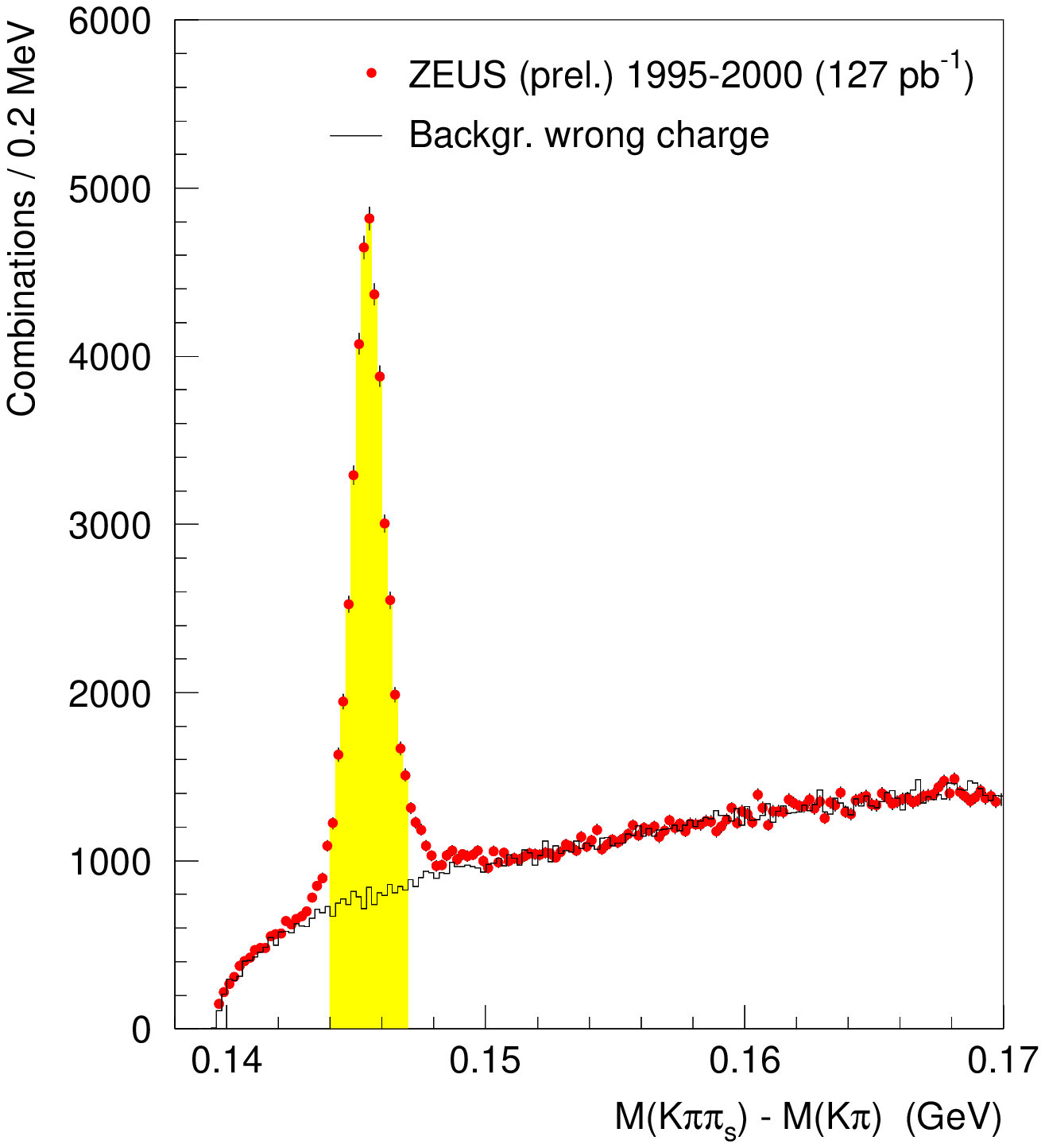,width=7cm%
%,bbllx=80pt,bblly=220pt,bburx=540pt,bbury=666pt,clip=%
}}
\fcaption{$D^*-D$ meson mass difference, $\Delta M$, illustrated from ZEUS 
photoproduction data (preliminary),\protect\cite{p497} using $K\pi$ pairs selected to have invariant masses in the range 1.83-1.90 GeV.}
\label{Dmass}\end{figure}

\noindent
The lowest charm states are the $cd$ and $cu$ $D^0$ and $D^\pm$
mesons, which are produced along with excited states such as the
$D^*$.  The $D^*$ mesons are easier to identify because of their
decays to the $D$ states through emission of a pion that is almost at
rest relative to the $D^*$.  The decay chains $D^{*\pm}(2010) \to
D^0(1864) + \pi^{\pm}$, with $D^0 \to K^\mp\pi^\pm$ or $D^0 \to
K^\mp\pi^\pm\pi^+\pi^-$ (+ c.c.)  are commonly selected since all the
final state particles are charged, and hence accurately measurable.  A
typical mass-difference plot is shown in fig.~\ref{Dmass}.  No peak is
seen in the distribution that uses the wrong-sign combinations: this 
quantifies the background to the signal.

\begin{figure}\vspace*{0.5mm}\centerline{\hspace*{6mm}
\epsfig{file=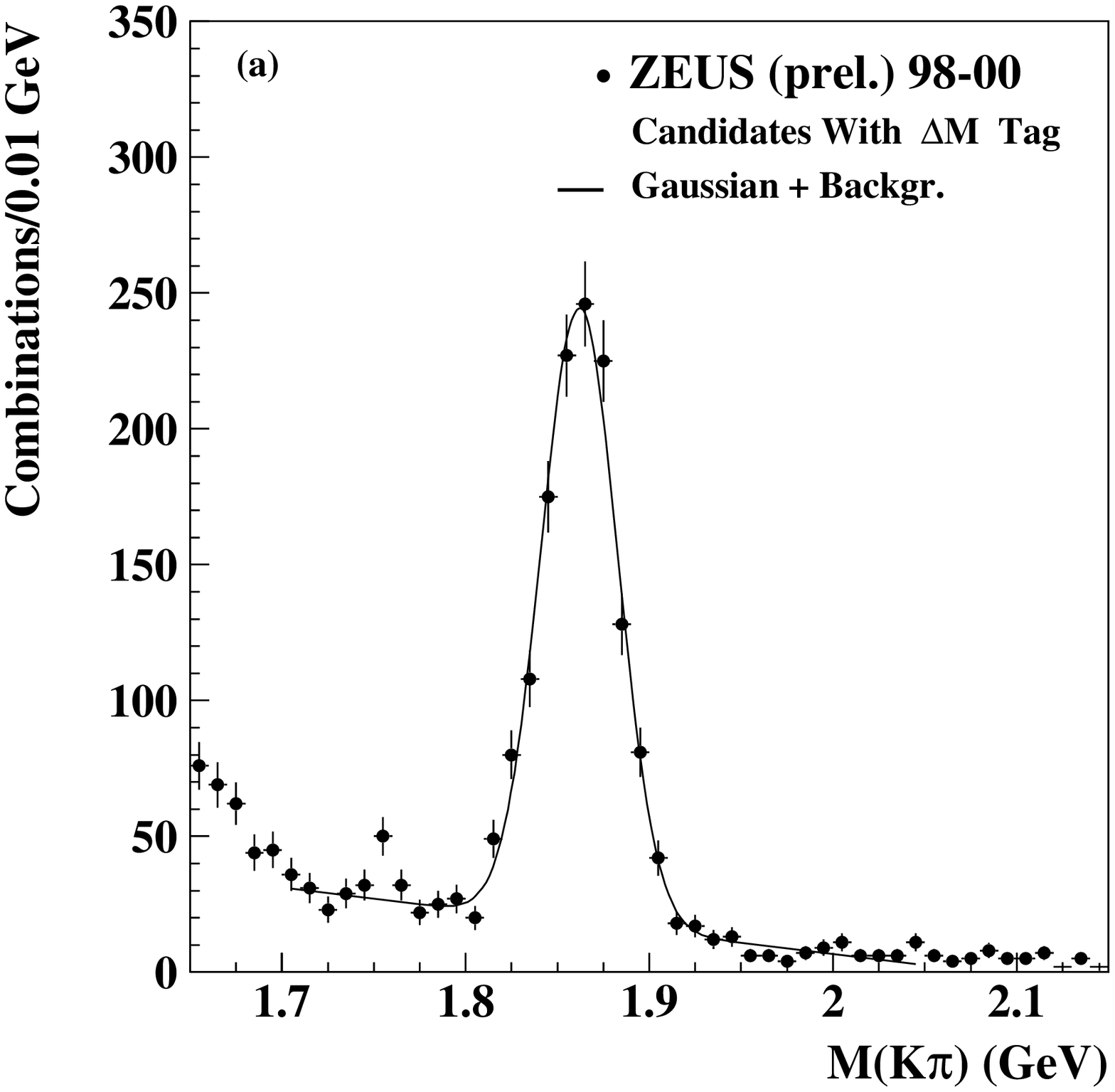,width=7cm%
%,bbllx=80pt,bblly=220pt,bburx=540pt,bbury=666pt,clip=%
}\hspace*{-6mm}
\epsfig{file=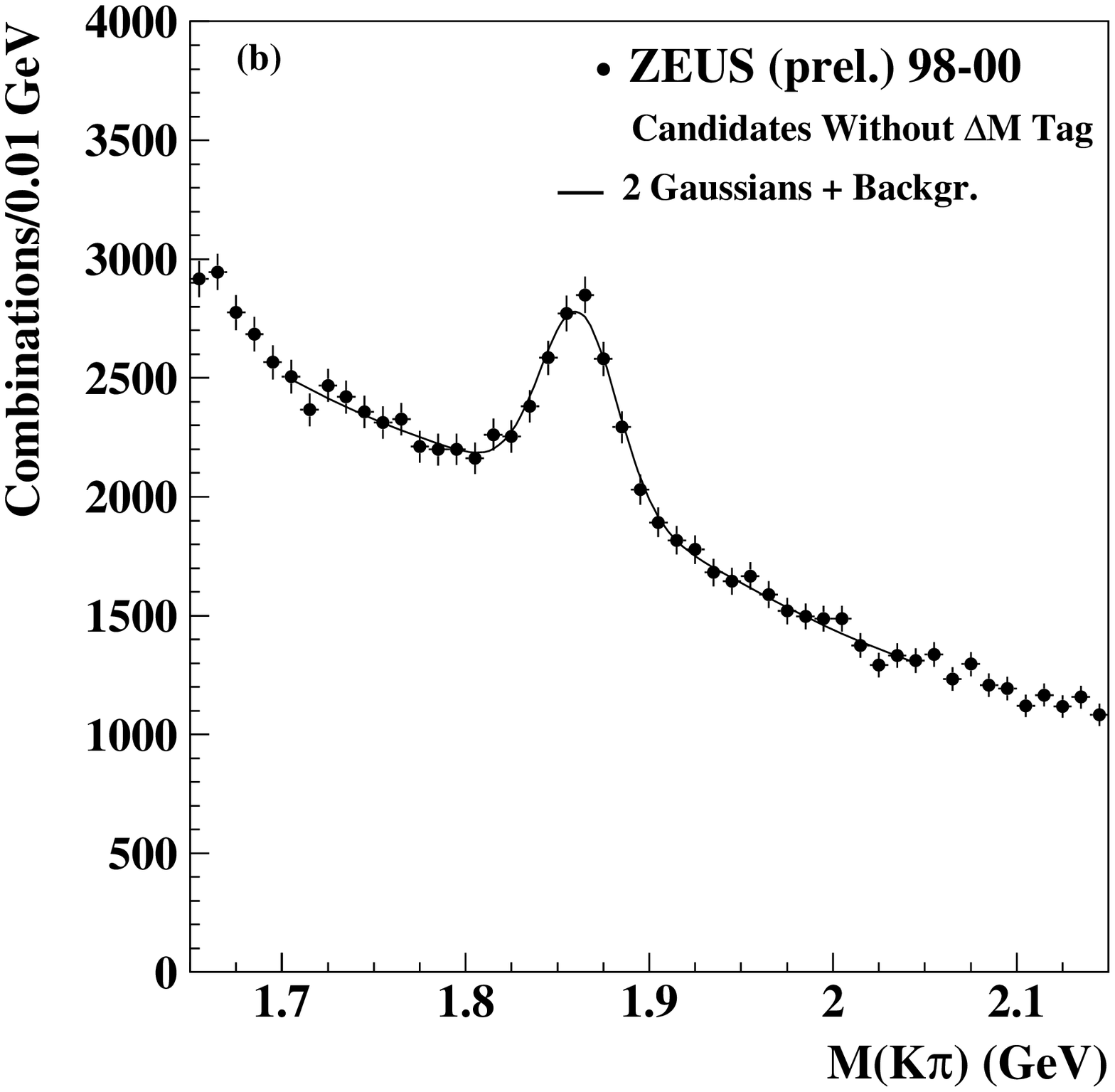,width=7cm%
%,bbllx=80pt,bblly=220pt,bburx=540pt,bbury=666pt,clip=%
}} 
\fcaption{ZEUS data (preliminary) on $D$ meson production from $K^\mp\pi^\pm$
pairs (a) accompanied by an extra pion giving a mass difference
$\Delta M$ consistent  with $D^*$ decay, (b) without an extra pion of
this type.}
\label{p501-2}\end{figure}

In recent ZEUS photoproduction data, the $D^0$ has been measured both
as the product of $D^*$ decay and in direct production.\cite{p501}
Figure \ref{p501-2} shows the $K\pi$ mass distributions for these two
cases.  After selection on the $\Delta M$ peak, a dramatic $D^0$
signal is seen, while the directly-produced $D^0$ signal lies on a
much higher background, although the number of events in the peak is
much larger.  In the latter case the wrongly-assigned $K\pi$
combinations remain in the plotted background.  From these plots, ZEUS
have evaluated the ratio $P_v = V/(V+PS)$ representing the production
of the vector $D^*$ (spin-1) relative to the pseudoscalar $D$.  This
is predicted to be 0.75 from simple spin statistics, but the inclusion
of decays from heavier charm states and the effects of hadronisation
can reduce this value.  ZEUS obtains the value $P_v =
0.546\pm0.045\pm0.028$ (preliminary),\cite{pv} in agreement with
measurements at LEP.\footnote{Here and elsewhere, unless otherwise
stated, the first of two quoted uncertainties is statistical and the
second is systematic.}

\begin{figure}
\centerline{\hspace*{6mm}
\raisebox{2mm}{\epsfig{file=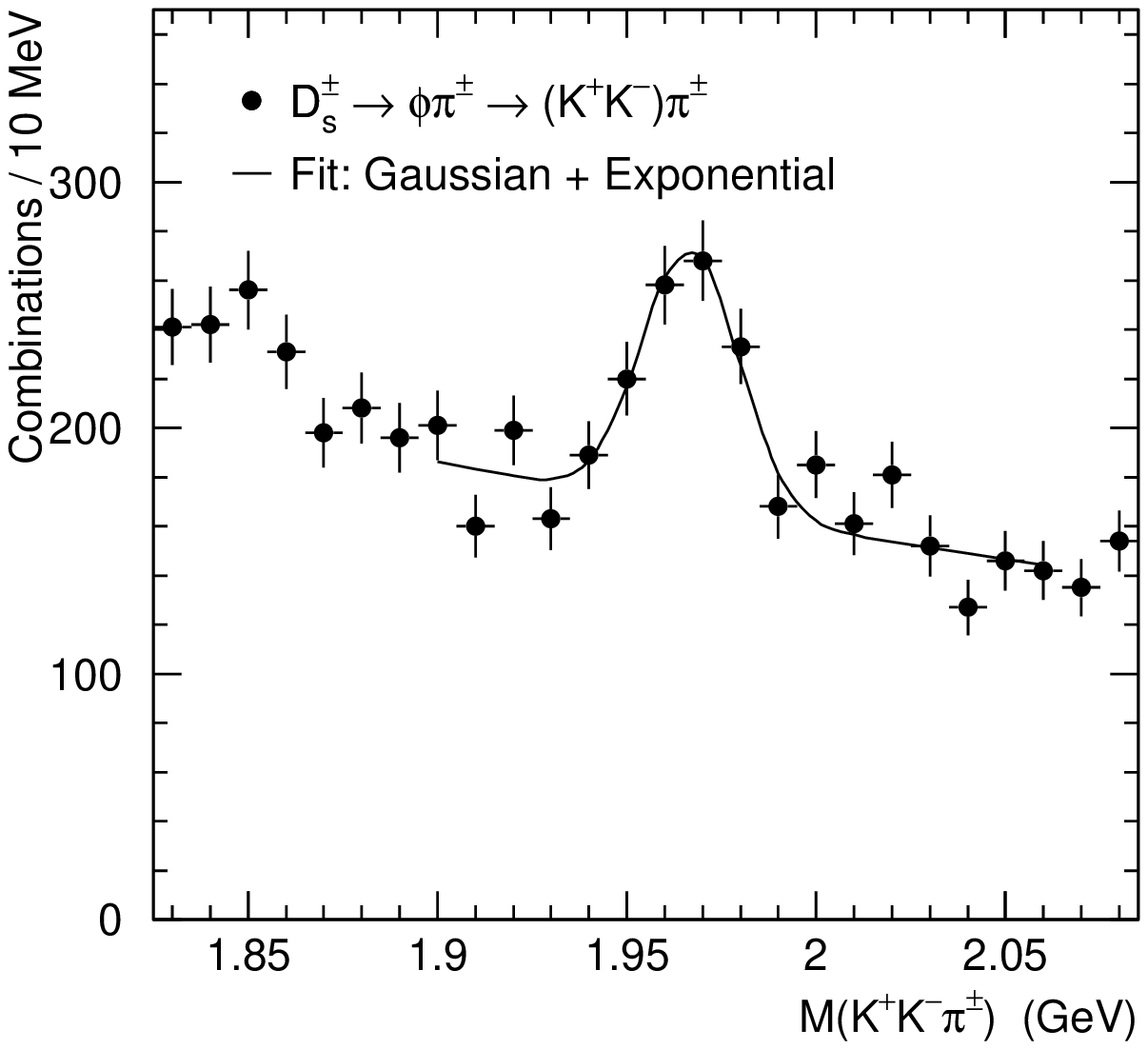,width=6.1cm%
%,bbllx=80pt,bblly=220pt,bburx=540pt,bbury=666pt,clip=%
}}
\epsfig{file=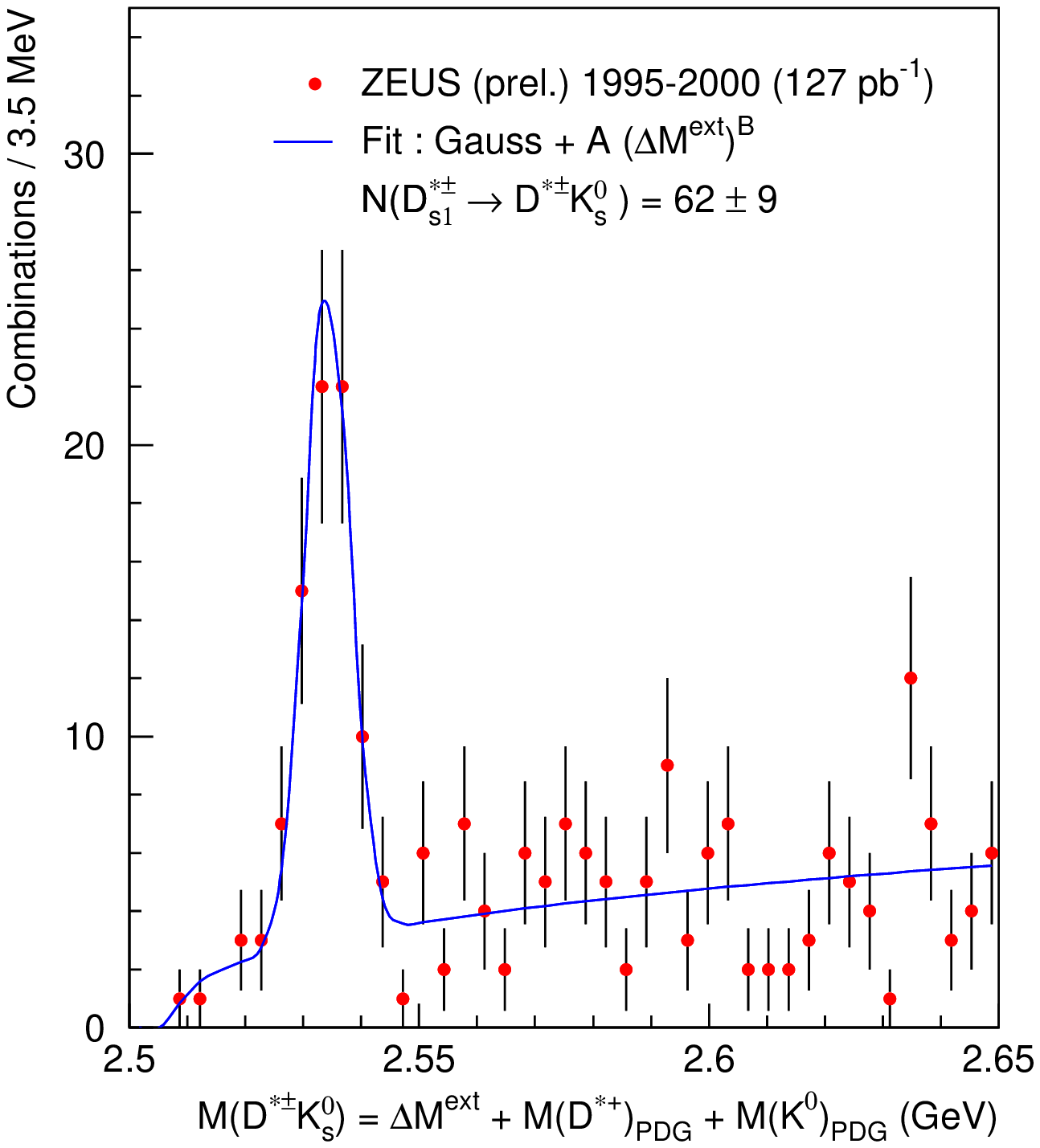,width=6.5cm%
%,bbllx=80pt,bblly=220pt,bburx=540pt,bbury=666pt,clip=%
}} 
\vspace*{1mm}
\fcaption{ZEUS data on $cs$ meson photoproduction: 
(a) $D_s^\pm(1969)$ signal seen in the decay $D_s^\pm\to \phi\pi^\pm
\to K^+K^-\pi^\pm$ (b) $D_{s1}^\pm(2536)$ signal seen in the decay
$D_{s1}^\pm\to D^{*\pm}K^0_S$ with $D^{*\pm}\to D^0\pi^\pm$ and
$D^0\to K^\mp\pi^\pm$ + c.c. (preliminary). The $\Delta M$ between
the states before and after the $K_S$ emission is added to the
standard mass values of the decay products.}  \label{ds}\end{figure}

Other mesons with a single $c$ quark are also observed at HERA.  The
$cs$ system has as its lowest state the $D_s^\pm(1969)$ which has been
detected by ZEUS in its $\phi\pi^\pm$ decay mode
(fig.~\ref{ds}a).\cite{p498} Details of the production characteristics
will be discussed below.  The ratio of the cross sections for
$D_s^\pm$ and $D^{\pm*}$ production can give a measure of the
strangeness-suppression ratio, a phenomenological parameter of the
Lund string fragmentation scheme.  The cross section ratio is found by
ZEUS to be $0.41\pm0.07^{+0.03}_{-0.05}\pm0.10$(b.r.), where the last
error is the uncertainty on the $D_s\to\phi \pi$ branching ratio.
This value is in close agreement with the corresponding result from
$e^+e^-$ data. It corresponds to a strangeness-suppression ratio of
$0.27\pm0.04^{+0.02}_{-0.03}\pm0.07$(b.r.), in agreement with most
other experiments and with the standard default value of 0.3 in
JETSET.  Both these and the above results support the concepts of
factorisation and universality in charm fragmentation.

A further charm state recently reported by ZEUS is the
\mbox{$D^\pm_{s1}(2536)$}, which decays by emission of a $K^0$ meson
to a $D^{*\pm}$.\cite{p497} This is one of a cluster of $L=1$ states
of the $cs$ system.  A clear signal is seen in photoproduction,
(fig.~\ref{ds}b).  The production rate and helicity characteristics
are consistent with other measurements.\cite{dscleo}
%but appear high compared with the rates for similar non-strange $c$ states.
%
Evidence for the detection of other $P$-wave charm meson states has
also been presented by ZEUS.\cite{o854}  These are neutral states in
the mass range 2.4 to 2.5 GeV.  Further statistics are required to
clarify the observations.

\subsection{Inclusive photoproduction of open charm}
\noindent
%=========================================================================
\begin{figure}
\vspace*{-3mm}\centerline{
\raisebox{14mm}{\epsfig{file=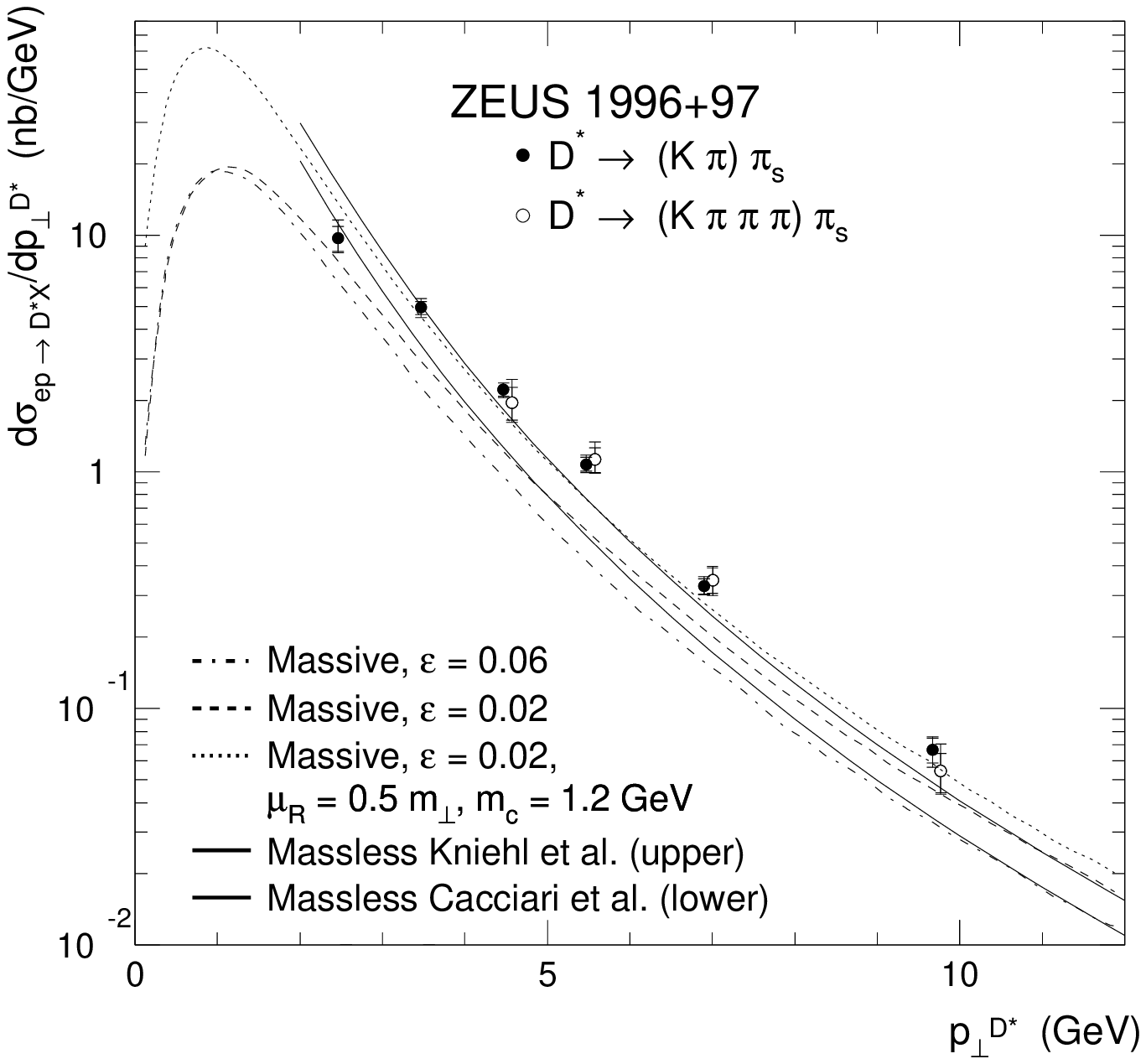,width=7.0cm%
%,bbllx=80pt,bblly=220pt,bburx=540pt,bbury=666pt,clip=%
}}}\vspace*{-10mm}
\fcaption{$D^*$ photoproduction: ZEUS distributions in \pT\ 
compared with models (see text).
Outer error bars are statistical + systematic combined in quadrature;
an overall normalisation uncertainty of $\pm4\%$ is not shown. Note that the data here and in the next figure are quoted as $ep$ cross sections. }
\label{p499apt}\end{figure}
%=========================================================================
\begin{figure}
\centerline{\hspace*{6mm}
\raisebox{0mm}{\epsfig{file=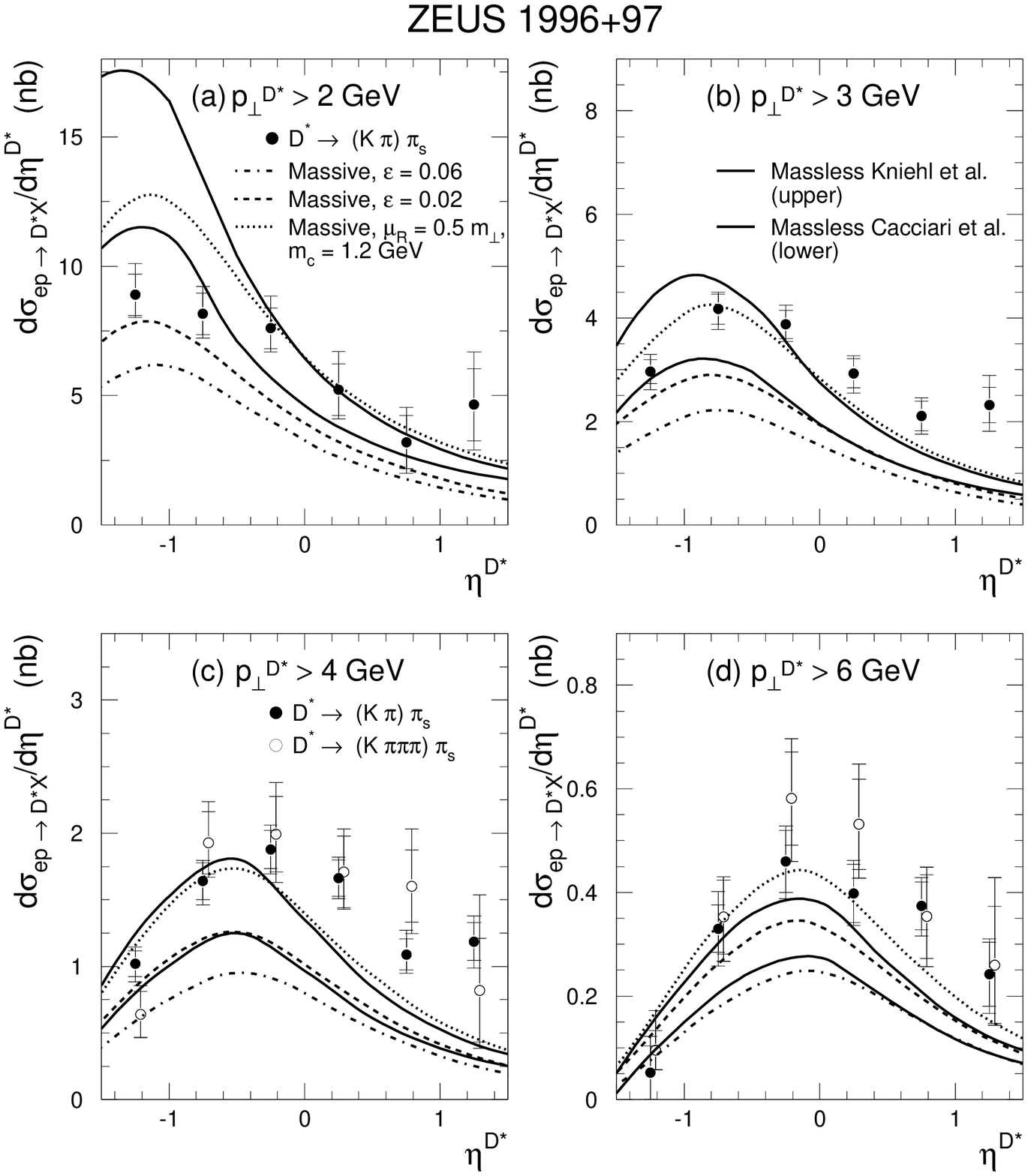,width=8.0cm%
%,bbllx=80pt,bblly=220pt,bburx=540pt,bbury=666pt,clip=%
}}}
\centerline{\hspace*{6mm}
\raisebox{0mm}{\epsfig{file=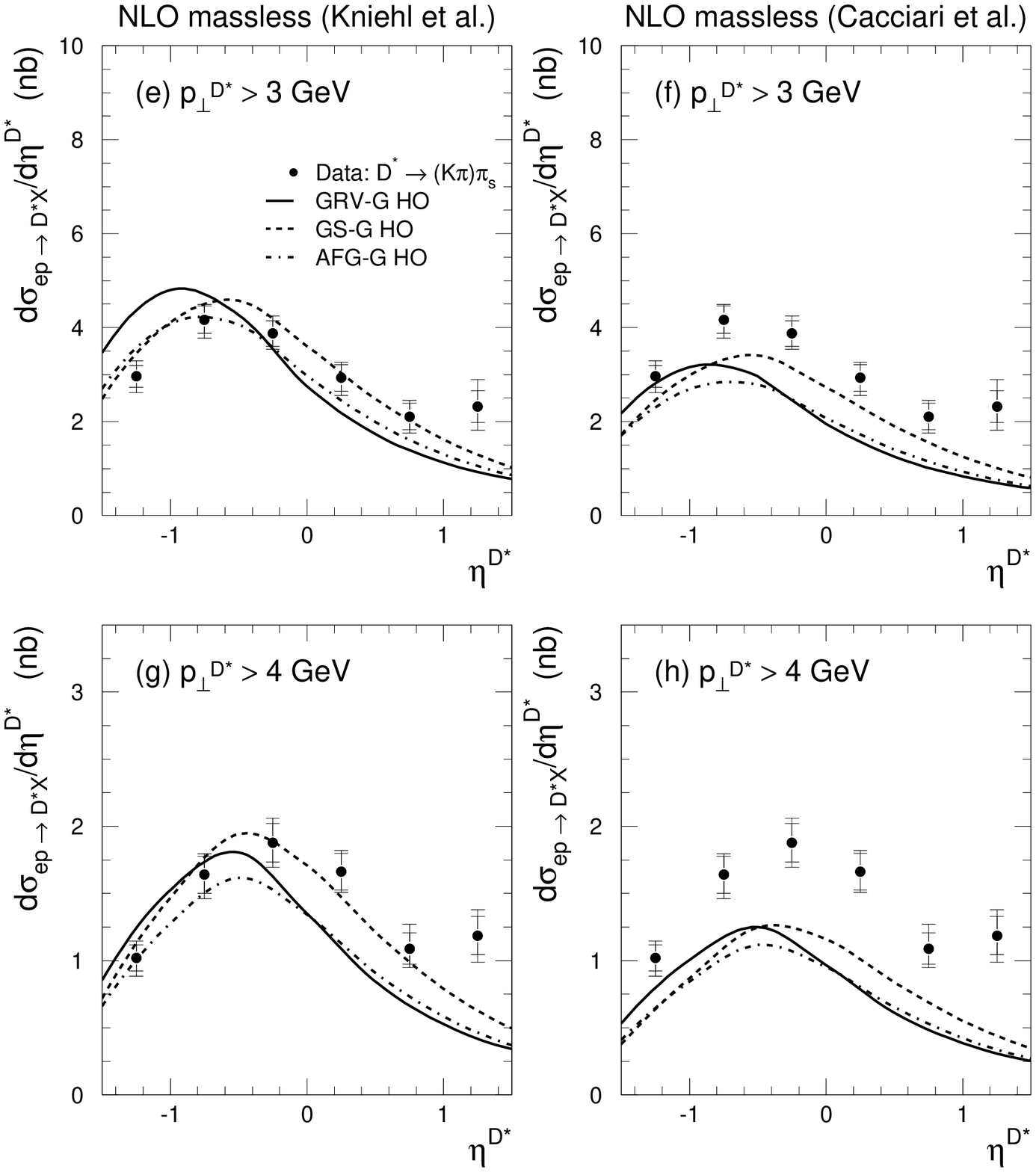,width=8.0cm%
%,bbllx=80pt,bblly=220pt,bburx=540pt,bbury=666pt,clip=%
}}}
\fcaption{$D^*$ photoproduction: ZEUS pseudorapidity distributions compared 
with models (see text). Details as fig.~\ref{p499apt}; $ep$ cross
sections are given.  GRV-HO photon pdf's are used in (a)-(d) and in the previous figure. }
\label{p499a}\end{figure}
%=========================================================================
ZEUS have published a study in which inclusive differential cross
sections for $D^{*\pm}$ are compared with several ``massive'' and
``massless'' charm schemes in NLO QCD.\cite{p499a} The $K\pi\pi\pi$
decay channel of the $D$ provides a check on the higher-quality measurements
using $K\pi$.  The massive charm, fixed flavour-number scheme of
Frixione et al.\ (FMNR)\cite{fmnr} is taken with several values of the
Peterson fragmentation parameter $\epsilon$, along with the massless,
variable flavour-number calculations of Kniehl et
al.\cite{kniehl,kniehl2} and of Cacciari et al.\cite{cacc,cacc2} The
two massless calculations differ in the factorisation schemes applied
to the fragmentation.

The shape, although not the normalisation, of the \pT\ distribution of
the $D^{*\pm}$ is fairly well described by all the models, although
the description cannot be said to be perfect (fig.~\ref{p499apt}).
The issues have been recently discussed by some of the theoretical
authors.\cite{fcn} Pronounced differences, however, appear in the $\eta$
distributions; these are shown in fig.~\ref{p499a}a
\ref{p499a}d integrated over
\pT\ from different thresholds.  In the highest \pT\ range (d), the
scheme of Kniehl et al.\cite{kniehl2} is more successful than that of
Cacciari et al.,\cite{cacc2} whose predictions are too low.  Although
the latter scheme does appear to work in the lowest \pT\ range
measured (a), this is probably of little significance since neither
massless calculation is appropriate for $\pT\approx m_c$.  In fact even the
massive scheme can be brought to reasonable agreement only by the
artificial recourse of setting $m_c$ as low as 1.2 GeV, while the
massless schemes are unconvincing for $\pT < 6$ GeV.

\begin{figure}
\vspace*{-1mm}\centerline{\hspace*{0mm}
\raisebox{0mm}{\epsfig{file=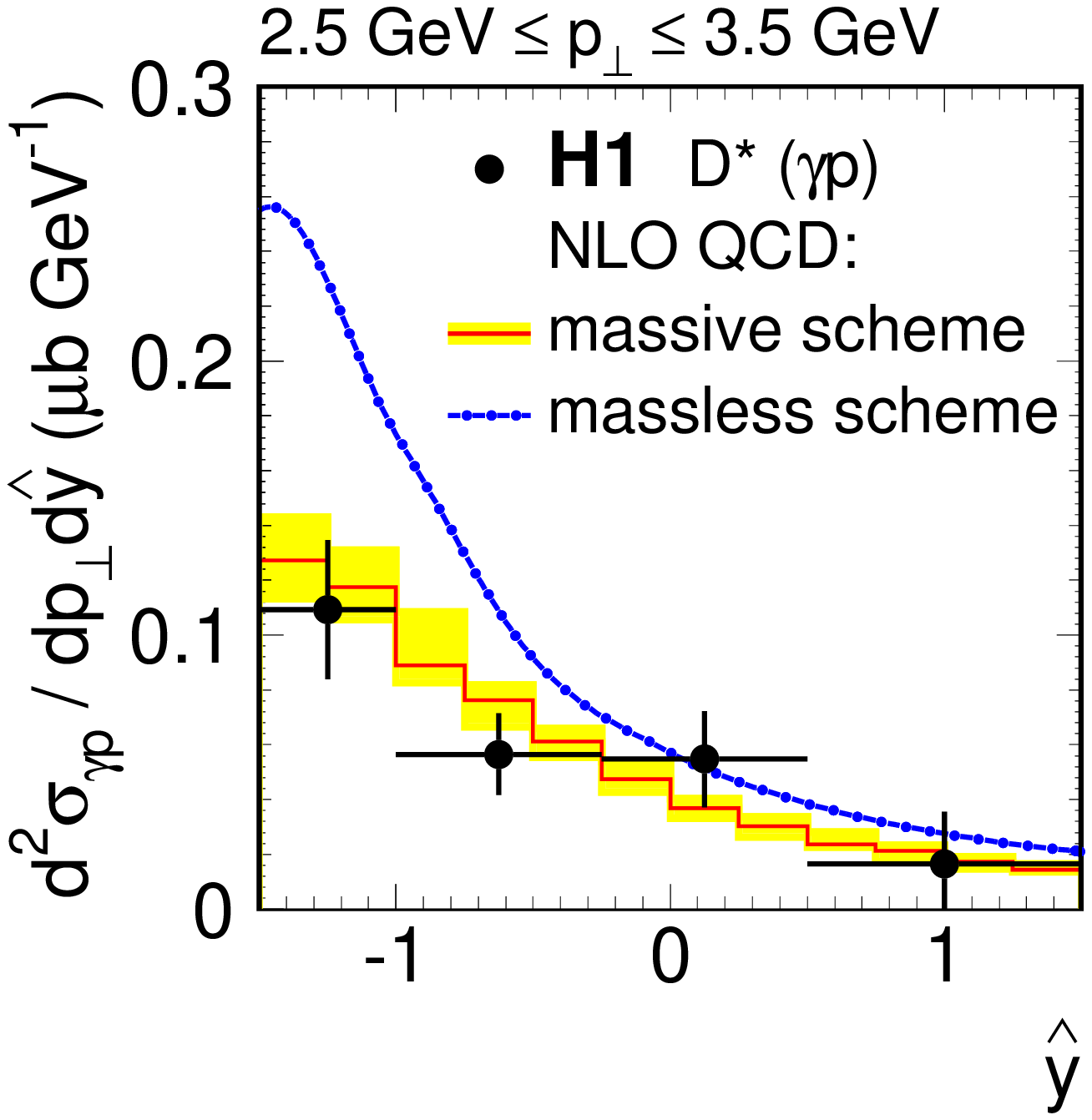,width=4.4cm%
%,bbllx=80pt,bblly=220pt,bburx=540pt,bbury=666pt,clip=%
}}
\raisebox{0mm}{\epsfig{file=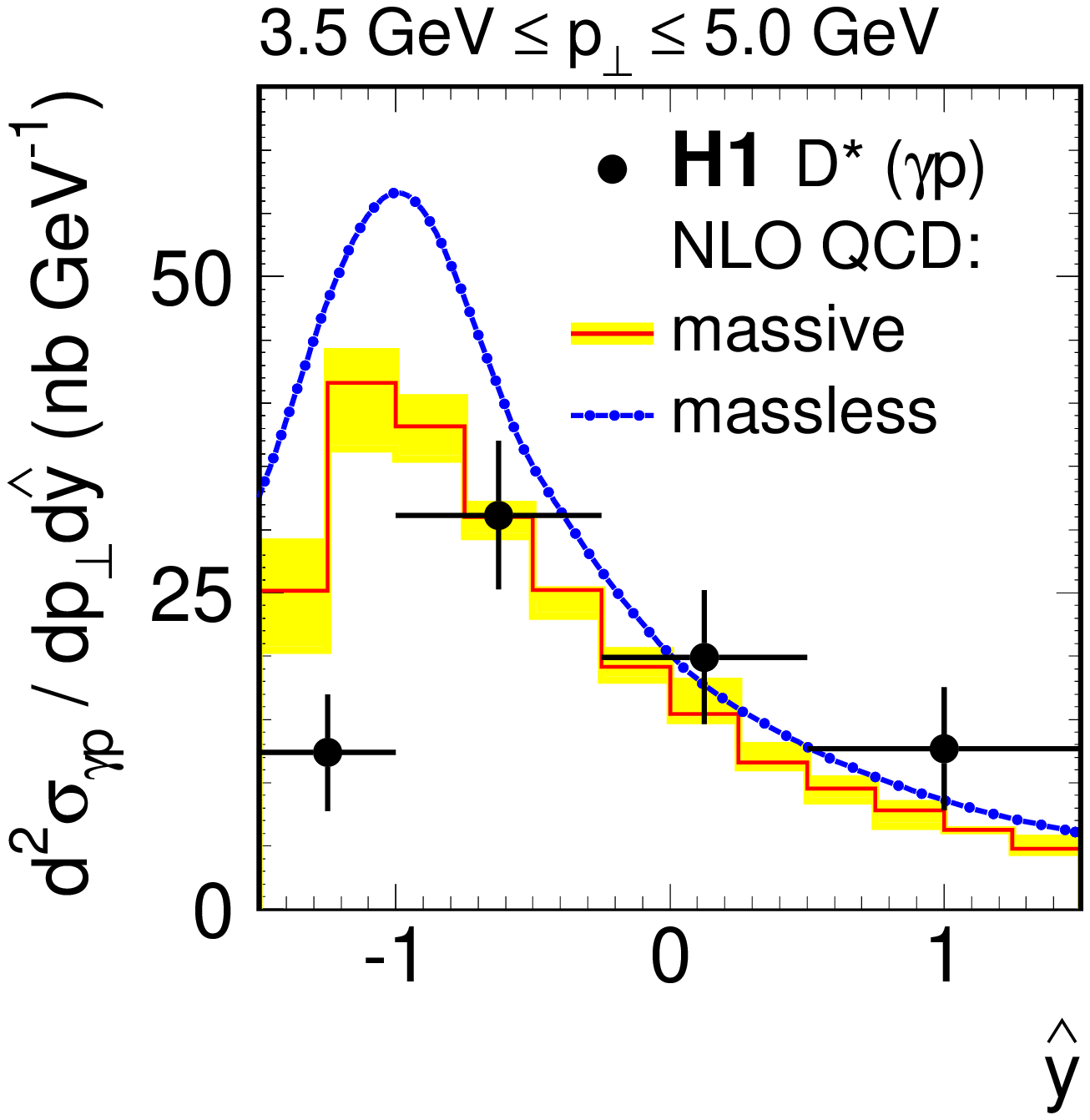,width=4.4cm%
%,bbllx=80pt,bblly=220pt,bburx=540pt,bbury=666pt,clip=%
}}
\raisebox{0mm}{\epsfig{file=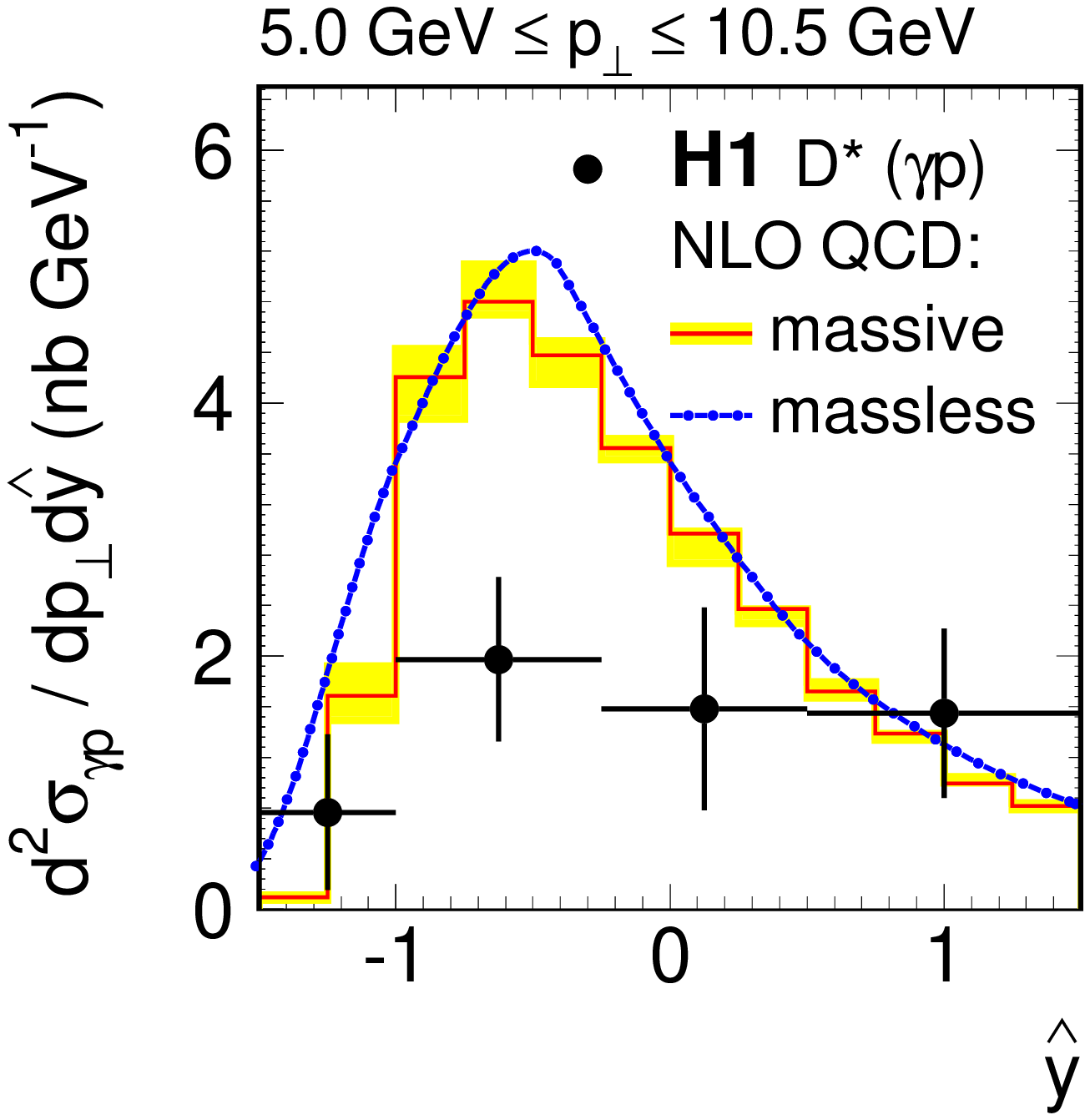,width=4.4cm%
%,bbllx=80pt,bblly=220pt,bburx=540pt,bbury=666pt,clip=%
}}}
\fcaption{Inclusive $\gamma p\to D^*$ cross sections from H1 vs.\ laboratory 
rapidity, compared with ``massive'' and
``massless'' models (see text).  The shaded bands indicate
variation of $m_c$ between 1.3 and 1.7 GeV.}
\label{p502_4}\end{figure}

Differences at the 20-25\% level can be made by varying the photon
pdf, as illustrated in fig.~\ref{p499a}e - \ref{p499a}h, where (e), (g) use
the calculation of Kniehl et al. and (f), (h) use that
of Cacciari et al.  For $ \pT\ < 6$ GeV, the GS-G photon structure
gives better results for the Kniehl et al.\ calculation due to its
high $c$ quark density, similar to that of the $u$ quark.  Since the
quasi-real photon is believed to couple mainly to light-quark hadronic
states, this structure may lack plausibility.  It
cannot, of course, be used in the massive-charm, fixed
three-flavour-number scheme.  

Overall, the message from these results is that the massive scheme has
trouble between 3 and 6 GeV, where its chances of working are best,
while the massless scheme of Kniehl et al.\ does seem to become more
successful as \pT\ increases: higher \pT\ measurements are required to
confirm this trend.  Results from H1\cite{p502} have also been
compared with the massive and massless charm calculations of
FMNR\cite{fmnr} and Kniehl et al.\cite{kniehl2}   Cross
sections as a function of laboratory rapidity are shown in
fig.~\ref{p502_4}.  At low \pT\ the results are qualitative similar to
those of ZEUS, but at high \pT\ the data lie significantly above the
calculations, in strong contrast to those of ZEUS.  There are apparent
differences between the two experiments here which need explanation.

\begin{figure}[t!]
\centerline{\hspace*{0mm}
\raisebox{0mm}{\epsfig{file=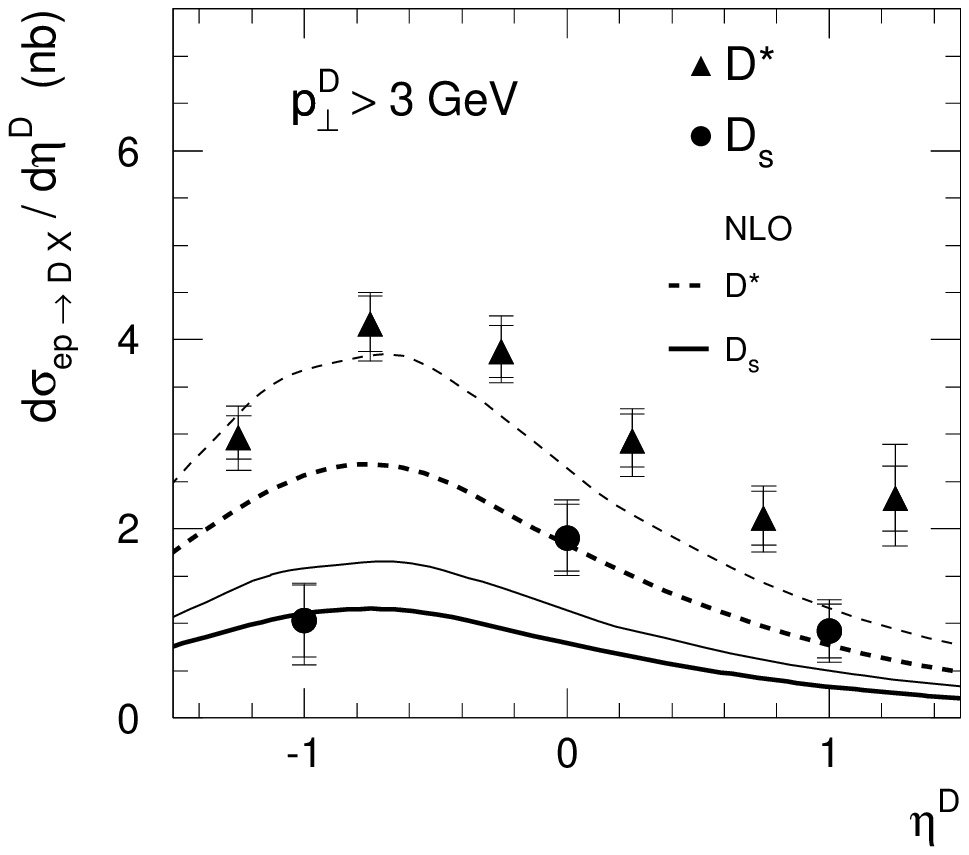,width=6.1cm%
%,bbllx=80pt,bblly=220pt,bburx=540pt,bbury=666pt,clip=%
}}
\raisebox{0mm}{\epsfig{file=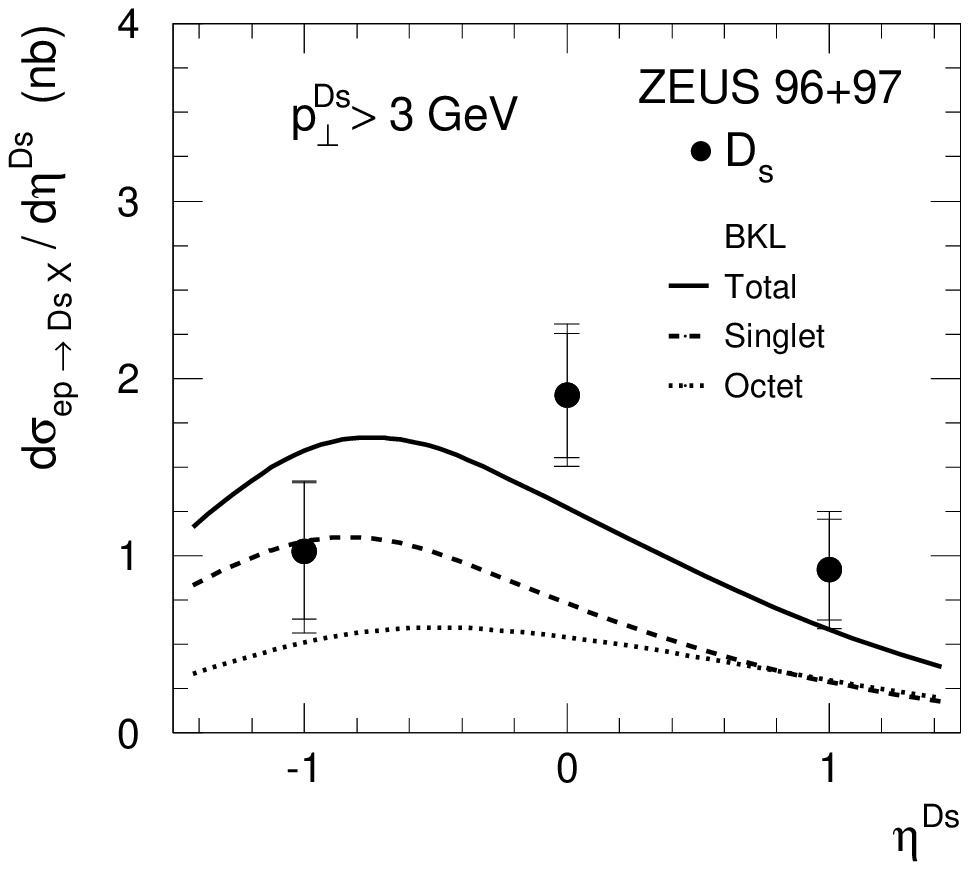,width=6.1cm%
%,bbllx=80pt,bblly=220pt,bburx=540pt,bbury=666pt,clip=%
}}}\vspace*{2mm}
\fcaption{Differential cross sections for inclusive $D_s$ production 
(ZEUS)  In the left-hand plot, the upper (thinner) curves of each type have
$m_c = 1.2$ GeV and the lower (thicker) have $m_c = 1.5$ GeV. 
The BKL calculation (right) uses  $m_c = 1.5$ GeV. 
}\label{p498}\end{figure}
%========================================================================
Measurements of $D_s$ production give further comparisons with theory.
In fig.\ \ref{p498}a, ZEUS first compare their data,\cite{p498}
together with the previous $D^*$ data,\cite{p499a} with the FMNR model
at NLO.  The fragmentation to the D states was performed with a
Peterson formula: the $f(c\to D^{*\pm})$ and $f(c\to D_s)$
distributions then differ by a simple scale factor.  As discussed
above, a standard calculation (heavy curves) fails to fit the $D^*$
data although a low $m_c$ value will almost suffice.  The $D_s$ data
are consistent with this, but the statistics are at present poor.
Another type of theoretical model has been given by Berezhnoy et al.\
(BKL),\cite{bkl} who relate the experimental value of $f(c\to
D^{*\pm})$ to $f(c\to D_s)$ by tuning the colour-singlet and
colour-octet contributions to a description of the hadronisation.  The
$D^{*\pm}$ data are used to obtain the relevant singlet/octet ratio,
yielding a prediction for the $D_s$ cross sections.  If both
components are used, there is a fair description of the ZEUS data.

\subsection{Charm in dijet photoproduction}
\noindent
More information can be obtained if the charm meson is associated with
the production of jets.  ZEUS have presented two analyses on this
topic; note that at present, the inclusive $D^*$-containing
events are measured with a defined experimental acceptance, but with
no specific association of the $D^*$ with a jet, or information about
a second charm particle in an event.  The two highest-\ET\ jets found
in an event are used.
%==============================================================
\begin{figure}[t!]
\vspace*{-5mm}
\centerline{\hspace*{6mm}
\raisebox{0mm}{\epsfig{file=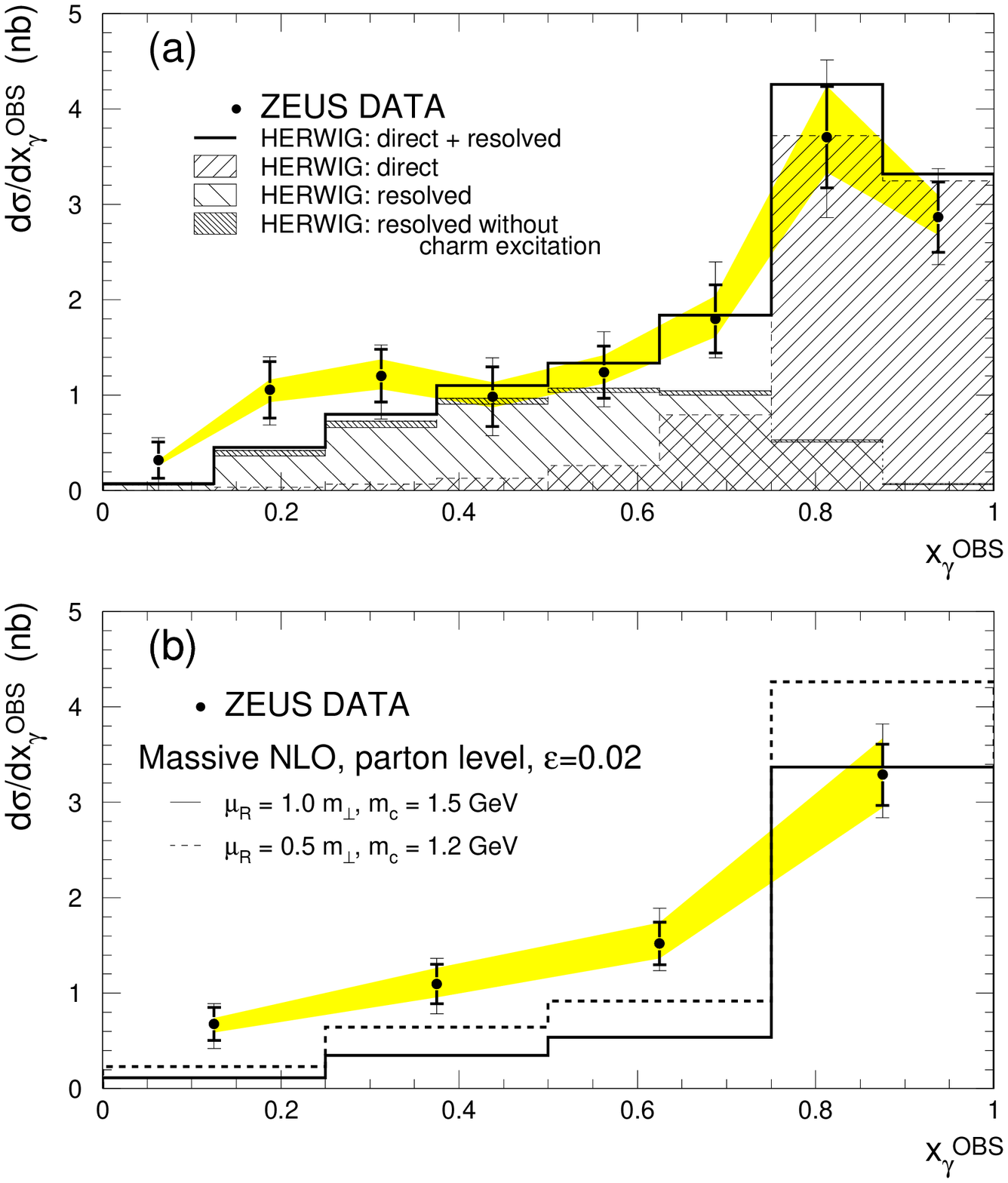,width=6.5cm%
%,bbllx=80pt,bblly=220pt,bburx=540pt,bbury=666pt,clip=%
}}
\raisebox{-20mm}{\hspace*{-3mm}\epsfig{file=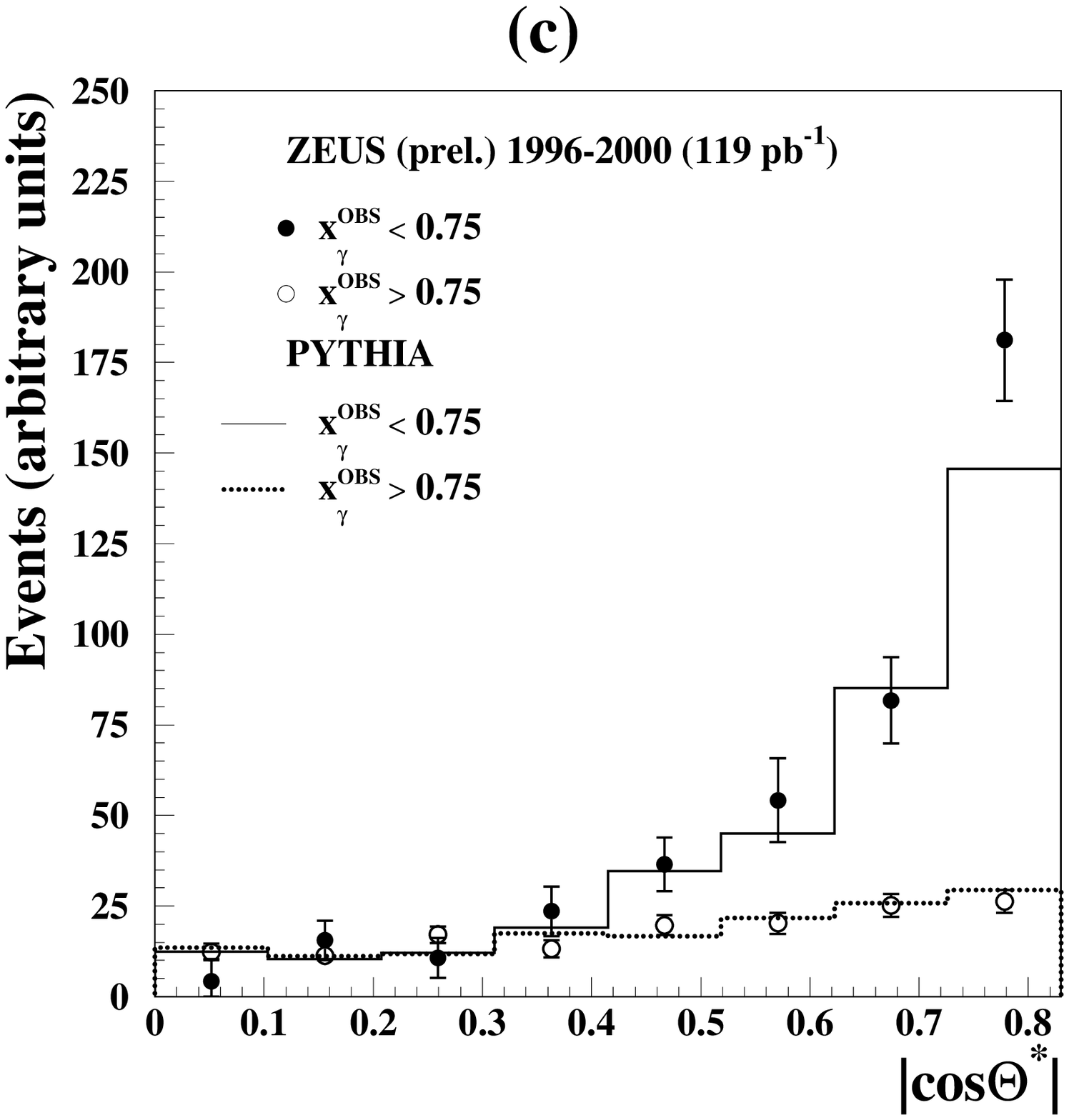,width=7.2cm%
%,bbllx=80pt,bblly=220pt,bburx=540pt,bbury=666pt,clip=%
}}}\vspace*{-15mm}
\fcaption{Differential cross section $d\sigma/d\xgO$ for dijets 
with an associated $D^*$ meson: (a) ZEUS
96-97 data\cite{p499a} compared with resolved and direct contributions
calculated at LO using HERWIG; (b) (rebinned) data compared with NLO
model of FMNR\protect\cite{fmnr}. Jets with $\ET > 6,$ 5 GeV and $D^*$
with $\pT > 3$ GeV are used. (c) Distribution in $|\cos\theta^*|$,
using increased data sample and jets with $\ET > 5$ GeV.\cite{p499} }
\label{p499aa}\end{figure}
%================================================================
% CONTINUE WITH THE TEXT.........
% 
In fig.~\ref{p499aa}a, the distribution in \xgO\ is compared with
predictions obtained using HERWIG, the total predictions of the model
being normalised to the data.  To a first approximation, HERWIG
describes the shape of the distribution well, despite the QCD
calculation being basically at LO.  Above $\xgO\approx 0.75$ a strong
direct peak is seen both in the data and the theory. A strong resolved
contribution is required.  The tail to lower \xgO\ of the HERWIG direct
contribution can be attributed to effects of hadronisation and
initial-state gluon radiation.  The small contribution of $D^*$'s
arising from $b$ quark production and decay is included.

To obtain this reasonable description of the resolved component, the
GRV-G HO photon pdf was used.\cite{grv} This has given good results in
many photoproduction analyses at HERA.  It is possible to separate
between the charm-excitation and the other charm contributions
(e.g. $gg\to c\bar c$) in the HERWIG resolved calculation; the latter
contributions are very small.  The charm-excitation here is believed to
come mainly from the photon: its nature is discussed further in
section 3.5.  The charm excited from the proton ends up mainly
at forward angles outside the present experiments' acceptance.\cite{urik}

A parton-level comparison with the massive NLO
calculation of FMNR is unsuccessful (fig.\ \ref{p499aa}b).  There is
no explicit charm excitation in this calculation, by definition, and
three-flavour pdf's are used.  The size of the direct peak is
accurately accounted for, bearing in mind that gluon radiation pushes
some of the events of direct origin into the low-\xgO\ tail.  The
effects of hadronisation would further increase this tail but are
estimated to be small.  So it is clear that the low-\xgO\ data are not
described by the FMNR model, even with the desperate recourse of
putting $m_c = 1.2$ GeV, which would anyway tend to spoil the
agreement in the direct peak.  This suggests as a conclusion that the
fixed flavour-number scheme is inadequate in the context of charm
photoproduction at moderate \pT.

This work has been extended by ZEUS, this time using PYTHIA as the
reference Monte Carlo, in a study of dijet angular distributions in
$D^*$-containing events.  The direct process proceeds by quark
exchange, while the resolved processes have a mixture of exchanges
dominated by that of a gluon.  These are respectively
spin-$\frac{1}{2}$ and spin-1 exchanges, which have very different
distributions in the scattering angle $\theta^*$ in the dijet rest
frame.  Experimentally they can be distinguished, to a reasonable
approximation, by comparing the direct-dominated and
resolved-dominated event samples with $\xgO > 0.75$ and $< 0.75$
respectively.\cite{p499}

Preliminary results in fig.\ \ref{p499aa}c show  a steep angular 
rise in the direct-dominated events, in marked contrast to the
gentler behaviour of the resolved-dominated events.  The agreement
with expectations is very satisfactory considering that the
calculations are at LO.  A next logical step will be to tag the jet
with which the charm is associated, and examine the forward and backward
distributions of the charm-tagged jets in the two classes of event, in
order to study charm excitation in the photon and in the proton in
more detail.

\subsection{Open charm in DIS}
\noindent
The production of charm in DIS processes is dominated by the LO
boson-gluon fusion diagram.  Both HERA collider experiments have
measured the inclusive production of the $D^{*\pm}$ in DIS and have
used this to measure $F_2^c$, the part of the proton structure
function $F_2$ which represents charm production.  The method employed
is to identify the $D^{*\pm}$ signal, correct it to an inclusive
$D^{*\pm}$ production cross section in bins of $x$ and $Q^2$, and then
use a theoretical model to relate the $D^{*\pm}$ cross section to
$F_2^c$, defined at the parton level.  One writes:
\begin{equation} 
F_2^{c\;\mathit{exp}}(\avval{x},\avval{Q^2}) =
\frac{\sigma^\mathit{exp}(x,Q^2)}{\sigma^\mathit{theor}(x,Q^2)}\;
F_2^{c\;\mathit{theor}}(\avval{x},\avval{Q^2}),
\end{equation}
to obtain an experimental $F_2^c$ result at bin-averaged values of
$x$, $Q^2$.  The theoretical terms take account of the heavy quark
fragmentation and the finite phase-space acceptance of the cross
section measurement.  The cross sections are taken to be sufficiently
smooth-varying so that bin migration effects can be neglected.

\begin{figure}
\vspace*{1mm}\centerline{\hspace*{2mm}
\epsfig{file=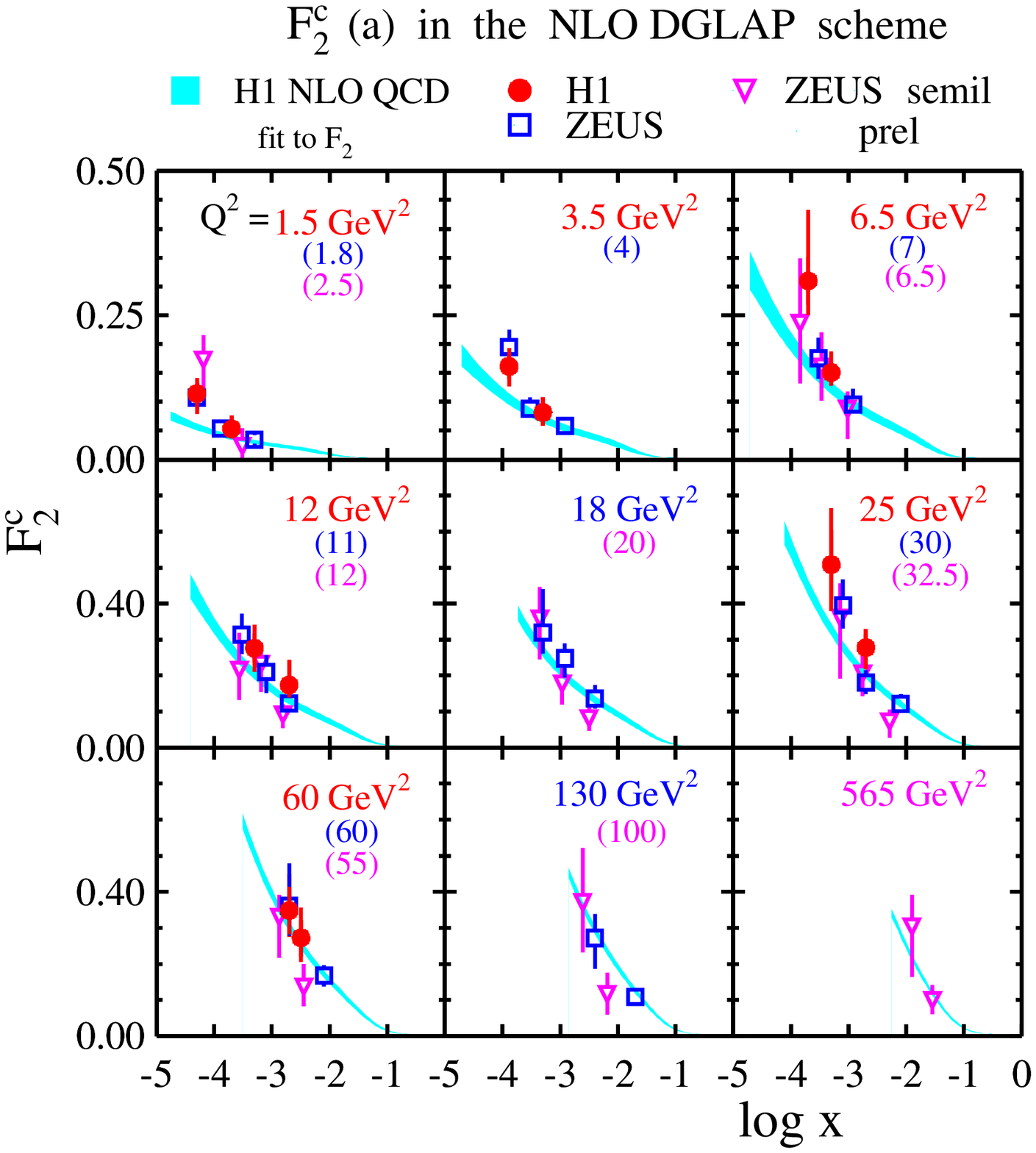,width=6.5cm%
,bbllx=45pt,bblly=138pt,bburx=540pt,bbury=683pt,clip=%
}
\raisebox{1.5mm}{\epsfig{file=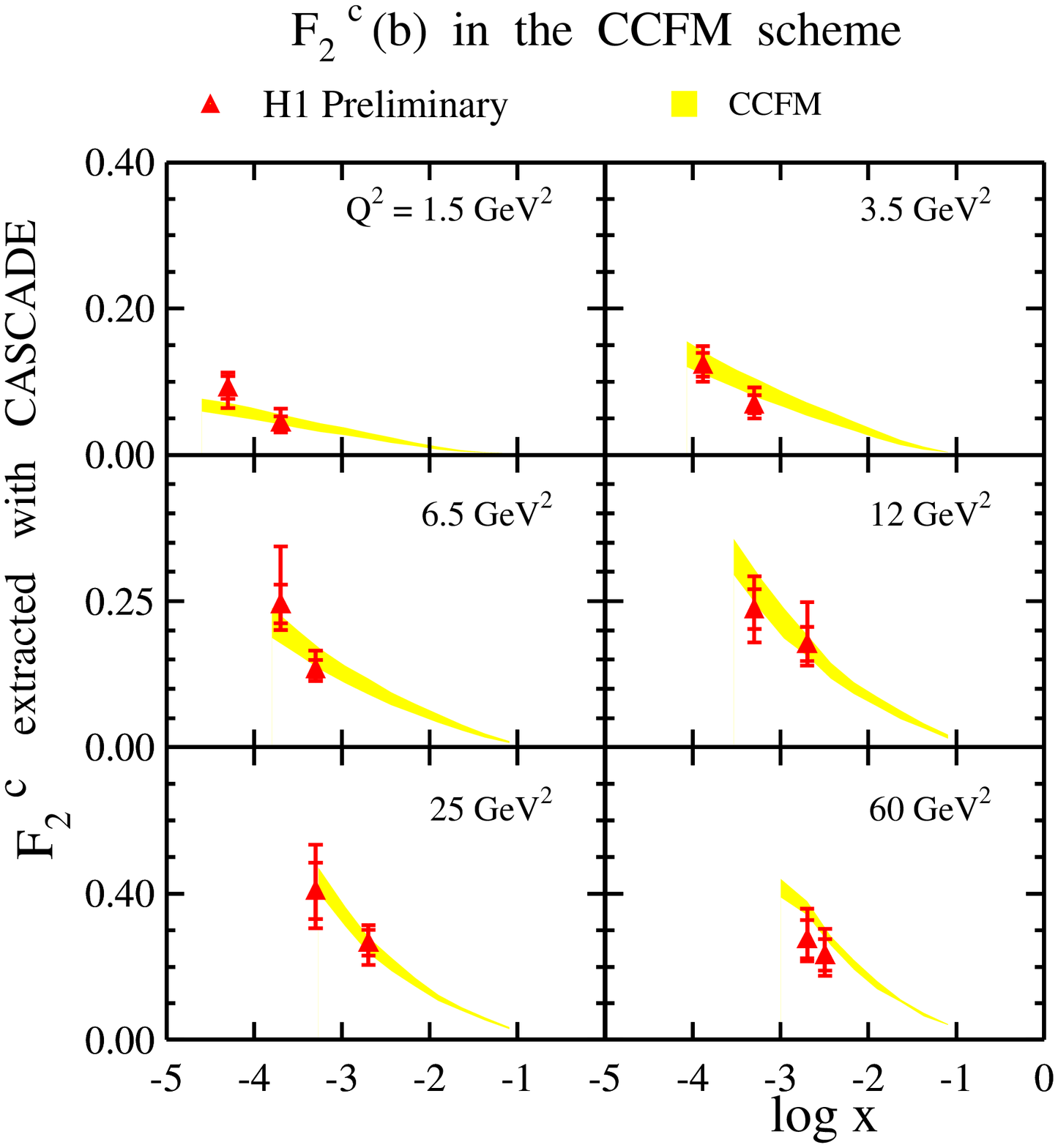,width=6.2cm%
,bbllx=0pt,bblly=0pt,bburx=530pt,bbury=560pt,clip=%
}} }\vspace*{1mm}
\fcaption{Structure function $F_2^c$ (a) measured by H1 and ZEUS using 
an NLO DGLAP scheme to correct to full phase space and (b) by H1 using
a CCFM-based scheme.  The preliminary results are compared with the respective
theories.  The shaded band represents the uncertainty on the $c$ quark
mass (1.3 - 1.5 GeV). Note that the ZEUS and H1 data are not at
identical \QQ\ values.}
\label{h1f2c}\end{figure}

\begin{figure}[t!]
\vspace*{1mm}
\centerline{\raisebox{-9mm}{\hspace*{55mm}{\sf(a)\hspace{55mm}(b)}}
\hspace*{8mm}}\vspace*{-9mm}
\centerline{\hspace*{-1mm}
\raisebox{0mm}{\epsfig{file=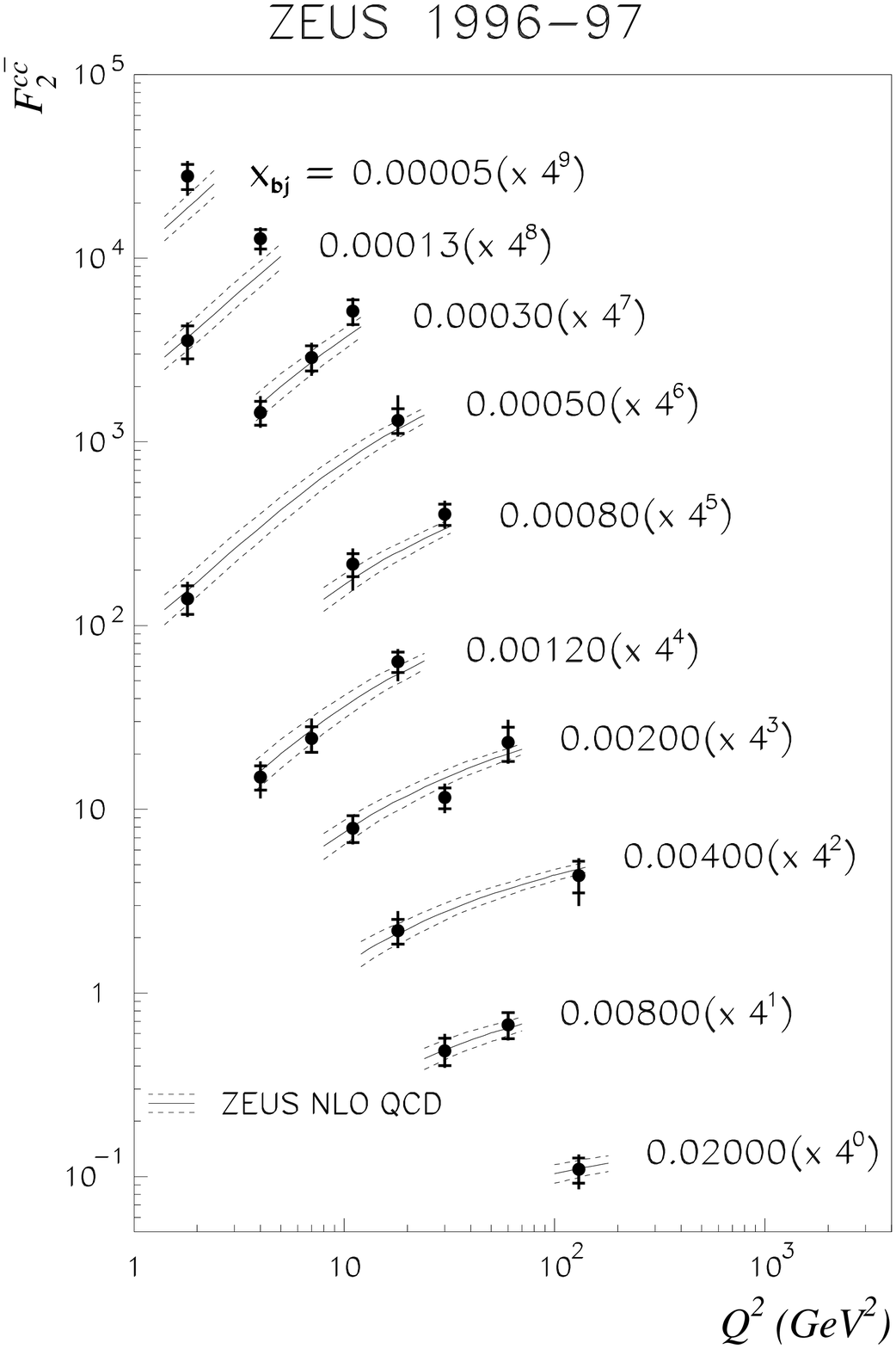,width=6.3cm%
%,bbllx=80pt,bblly=220pt,bburx=540pt,bbury=666pt,clip=%
}}
\raisebox{1mm}{\epsfig{file=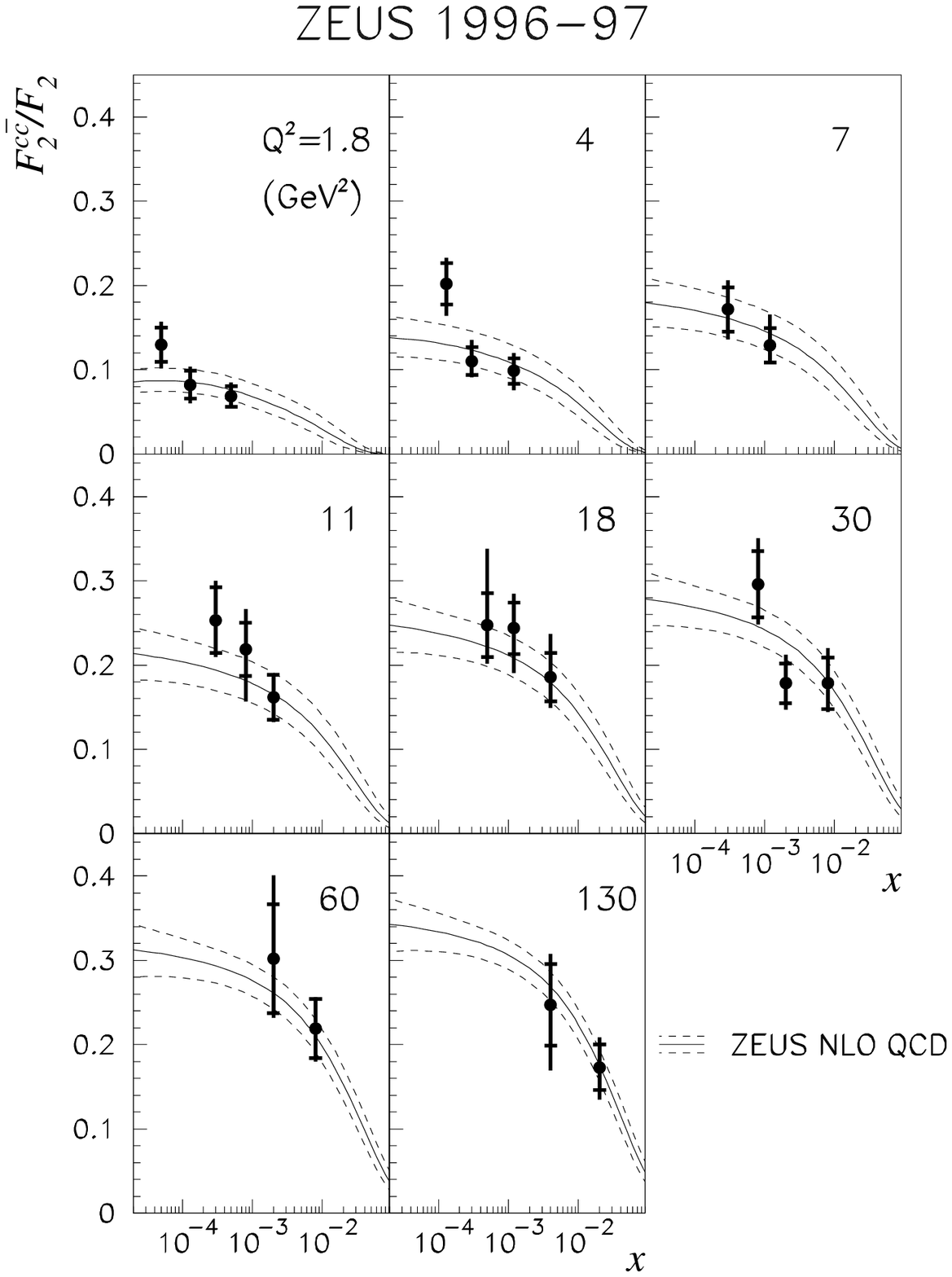,width=6.3cm%
,bbllx=0pt,bblly=0pt,bburx=402pt,bbury=565pt,clip=%
}}} \vspace*{2mm}
\fcaption{(a) Charm structure function measured by ZEUS using an 
NLO DGLAP scheme to correct to full phase space, and compared with
a fit using this scheme.\cite{p493a}  Uncertainties of around $\pm 10$\% 
in the normalisation due to 
the luminosity measurement, the charm meson branching ratios and the 
charm fragmentation function are not included.  (b) The ratio of the 
charm structure function to the total structure function $F_2$.
}
\label{zf2c}\end{figure}

H1 seek to compare their results with predictions from a DGLAP-based
evolution scheme and also from a CCFM-based scheme.\cite{p791} To do this,
they first make use of the hadronising LO Monte Carlo
AROMA\cite{aroma} to evaluate inclusive $D^{*\pm}$ cross sections from
the observed event numbers.  They then perform the subsequent
calculation of $F_2^c$ in two ways, by means of (a) HVQDIS, which uses
DGLAP, and (b) CASCADE, which uses CCFM.  The HVQDIS
program\cite{hvqdis} uses a fixed flavour-number model for DIS in
which the boson-gluon fusion process is calculated to NLO.  The
phase-space acceptance correction differs significantly in the two
approaches, and so H1 have presented two different sets of
experimental structure functions (fig.~\ref{h1f2c}), each compared
with the absolute predictions of a theory of a type similar to that
used to perform the correction.  The CCFM prediction includes proton
gluon densities measured by H1 in an overall $F_2$ determination using
this model.  The shaded band represents the effect of varying the $c$
mass between 1.3 and 1.5 GeV, which is the dominant uncertainty, and
in the CCFM case the effect also of using a proton gluon density
obtained with an NLO fit to $F_2$.

Overall, the CCFM-based calculation describes the data better than the
NLO-DGLAP based calculation,\cite{p945} the differences appearing at
lower values of $Q^2$ and $x$.  Similar conclusions were obtained when
comparing the inclusive $D^{*\pm}$ cross sections with these two
models.

In a like manner, ZEUS have obtained $D^{*\pm}$ cross sections using
RAPGAP to evaluate the hadronisation corrections and HVQDIS as the 
parton-level model.\cite{p493a}  The inclusive cross sections agree
well with H1 and with HVQDIS, but a JETSET-style fragmentation
function is required, rather than the Peterson formula, to perform the
hadronisation corrections.  There are some further technical
differences from the H1 approach.  ZEUS also find good agreement between
their $F_2^c$ measurements and results from an NLO QCD fit using ZEUS
proton pdf's.\cite{znlofit} This is illustrated in the $F_2^c$ results
plotted as a function of
\QQ\ (fig.\ \ref{zf2c}a) for different values of $x$.  The variation
of $F_2^c$ with \QQ\ --- namely, scaling violation --- is very much in
evidence, and is well described globally by the theoretical fit.  The
uncertainty on the theory (dashed lines) is dominated by that on the
$c$ quark mass.

%A preliminaryattempt to investigate some of these points has been presented by
%ZEUS.\cite{p493b} $D^{*\pm}$ data from the period 1995-2000 were
%compared with predictions from zero-mass variable-flavour calculations
%performed by several LO Monte Carlo generators.  While AROMA and
%RAPGAP, like HVQDIS, use a fixed flavour calculation which ignores
%charm excitation in the proton, DJANGOH and LEPTO allow an effective
%charm content in the proton. This is excited by means of $gamma*c\to
%gc$ scattering (QCD Compton, QCDC).  The distributions in $Q^2$ and
%$x$ were studied, the outcome being that the NLO calculation HVQDIS
%fitted the data well, as for the earlier data, while the similar LO
%calculations AROMA and RAPGAP overestimated the data, as did 
%DJANGOH and LEPTO. These programs do not treat the charm mass
%correctly, but provide no evidence that the excitation mechanisms are 
%important.

The ratio of $F_2^c$ to $F_2$ is presented by ZEUS in fig.\ \ref{zf2c}b.
The fraction of charm events is substantial, owing to the
flavour-insensitivity of the boson-gluon fusion process at high \pT.
The flattening of the theoretical ratio as $x$ becomes smaller is a
feature of the boson-gluon fusion model: the rising charm production
follows the rising gluon density in the proton, accompanying the
similarly rising quark density which steers $F_2$.  At high $x$ this mechanism
should fail, since the gluon density falls off and the proton has no charm
valence quarks.  The experimental data are consistent
with these theoretical expectations.  Overall, therefore, a clear physical
picture seems to emerge.

Using a larger event sample from the period 1998-2000, ZEUS have
confirmed that HVQDIS gives a good overall description of the
$D^{*\pm}$ production, up to \QQ\ values of 1000 GeV$^2$.\cite{p493}
An unexpected observation is apparent in the preliminary results,
namely that the cross sections measured using an incoming electron beam
are higher than those using a positron beam, although both
sets individually remain consistent with HVQDIS.  The effect appears
highest at high \QQ.  No explanation for the difference is yet on
offer, and further investigation is required. 

\begin{figure}
\vspace*{1mm}
\centerline{\hspace*{6mm}
\raisebox{0mm}{\epsfig{file=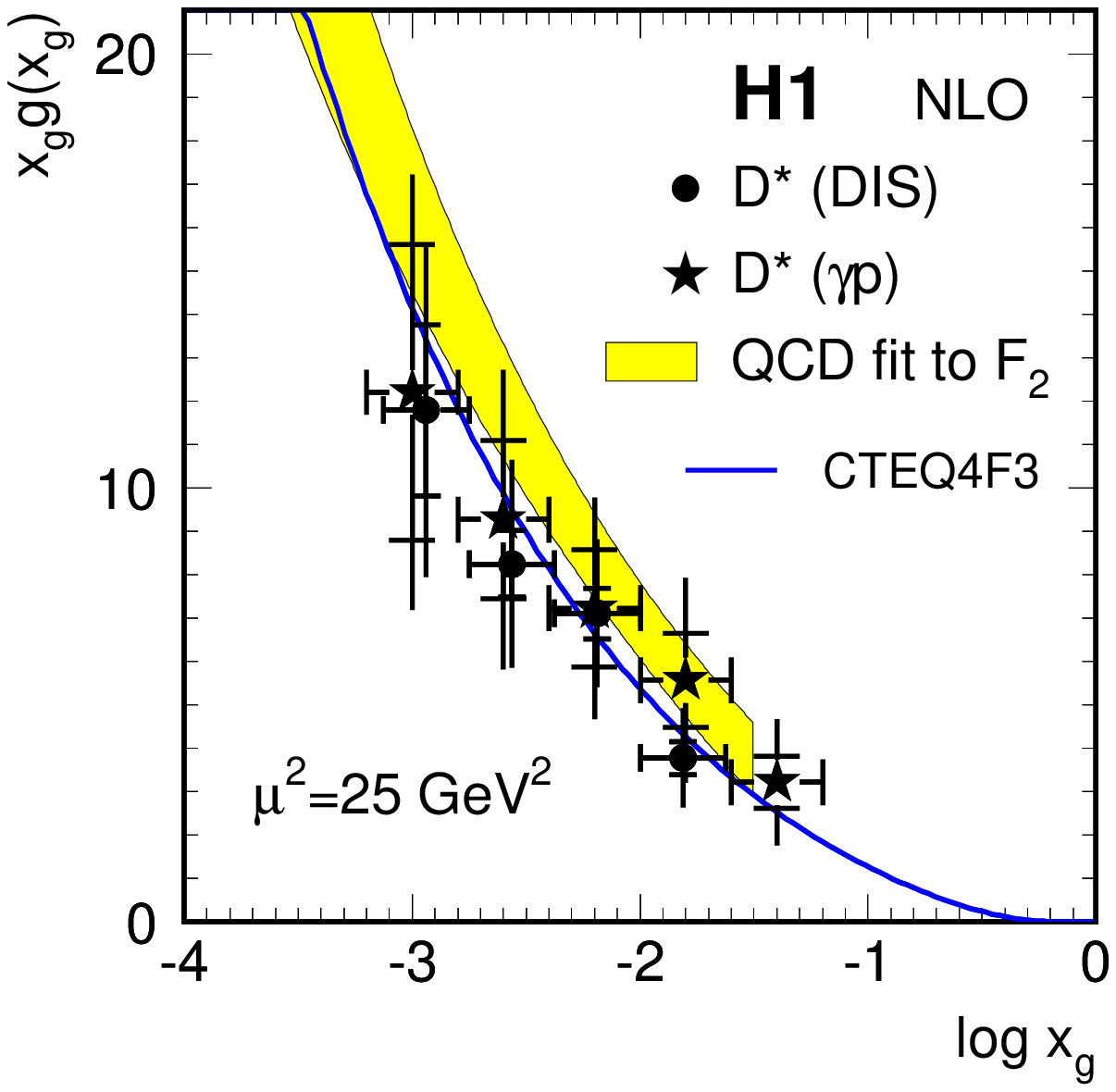,width=7.5cm%
,bbllx=55pt,bblly=26pt,bburx=404pt,bbury=372pt,clip=%
}}}
\fcaption{Proton gluon densities obtained from two $D^*$ analysis by H1,
namely from photoproduction and from DIS, 
compared with the gluon density extracted from a QCD fit to $F_2$ data.
}  \label{p502_9}\end{figure}

H1 have also extracted the gluon density in the proton on the basis
that the charm production cross sections are dominated by the
boson-gluon fusion process, and hence depend strongly on the proton
gluon content.  Results\cite{p502} using $D^*$ data from
photoproduction and DIS are plotted in fig.~\ref{p502_9} and are
compared with those from a QCD fit to $F_2$ data.  The $F_2$
measurements depend primarily on the coupling of the virtual photon to
quarks in the proton, with the DGLAP equations generating the gluon
distribution in the course of the fitting procedure.  The excellent
agreement seen between the two types of method represents a
significant triumph for the standard QCD description of the $ep$
system.

\subsection{Photoproduction-DIS relationship}
\begin{figure}[t!]
\vspace*{0.5mm}
\centerline{\hspace*{3mm}
\raisebox{0mm}{\epsfig{file=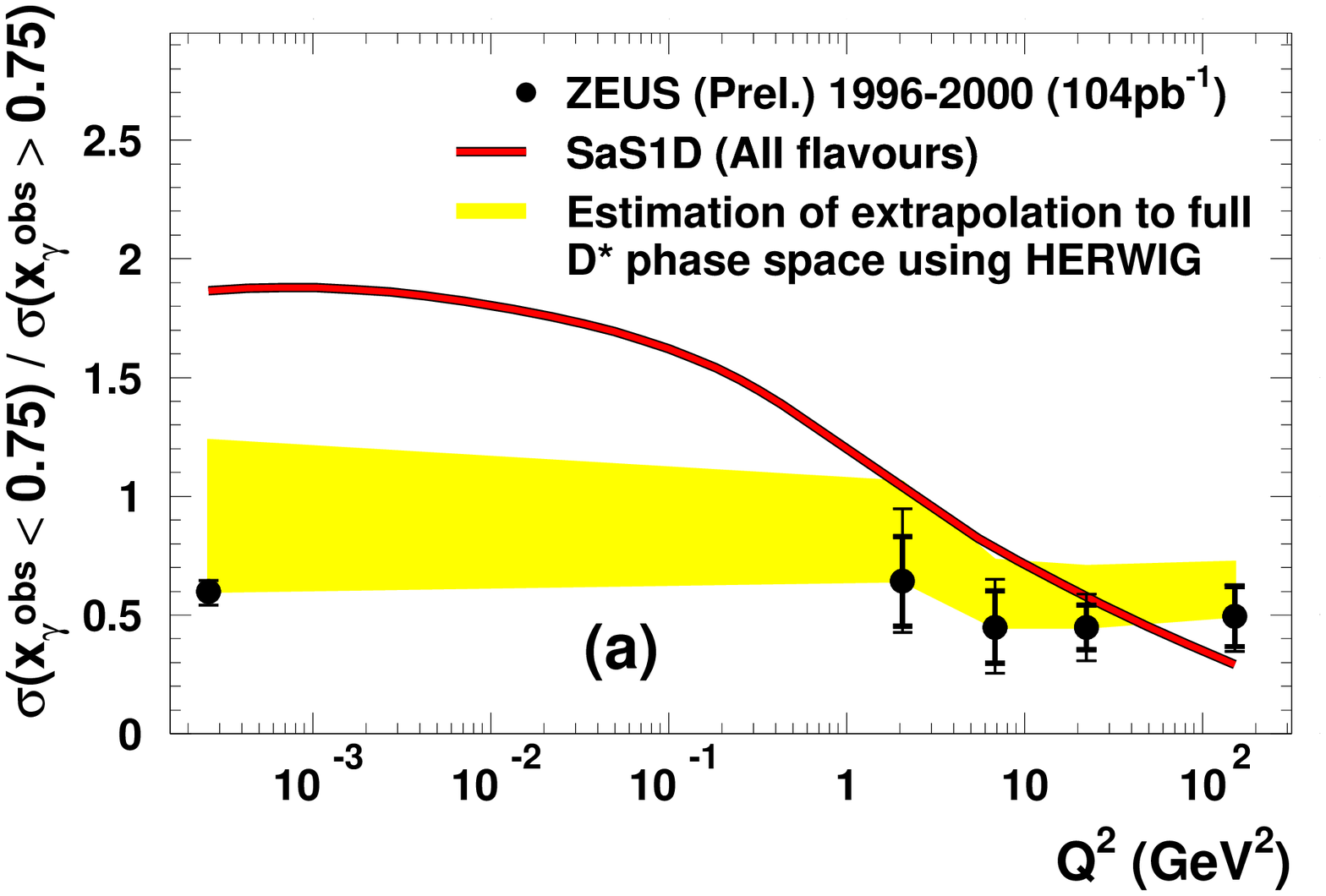,width=6.8cm%
%,bbllx=80pt,bblly=220pt,bburx=540pt,bbury=666pt,clip=%
}}
\epsfig{file=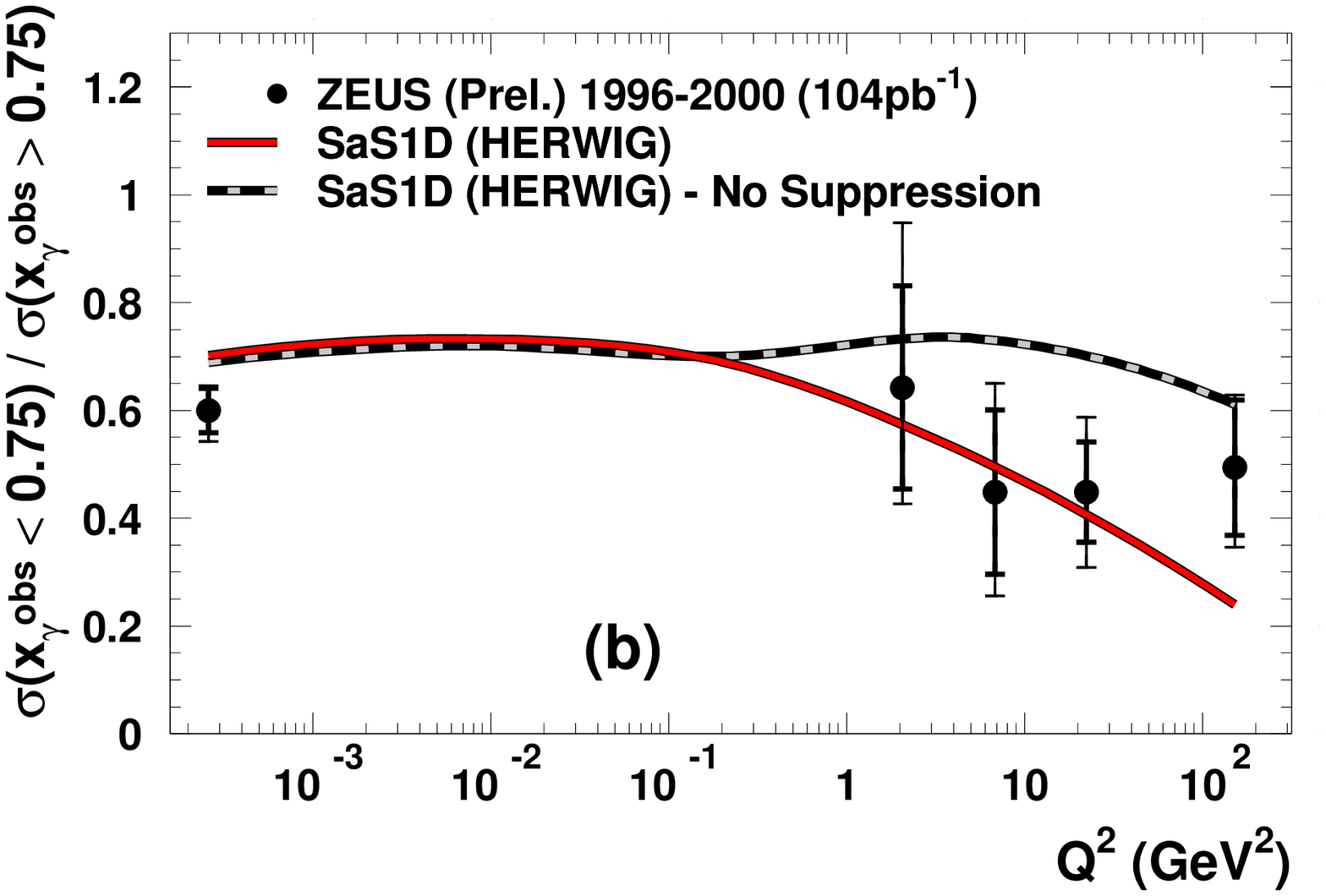,width=6.8cm%
%,bbllx=0pt,bblly=0pt,bburx=402pt,bbury=565pt,clip=%
}} \vspace*{1mm}
\centerline{\hspace*{1mm}
\raisebox{0mm}{\epsfig{file=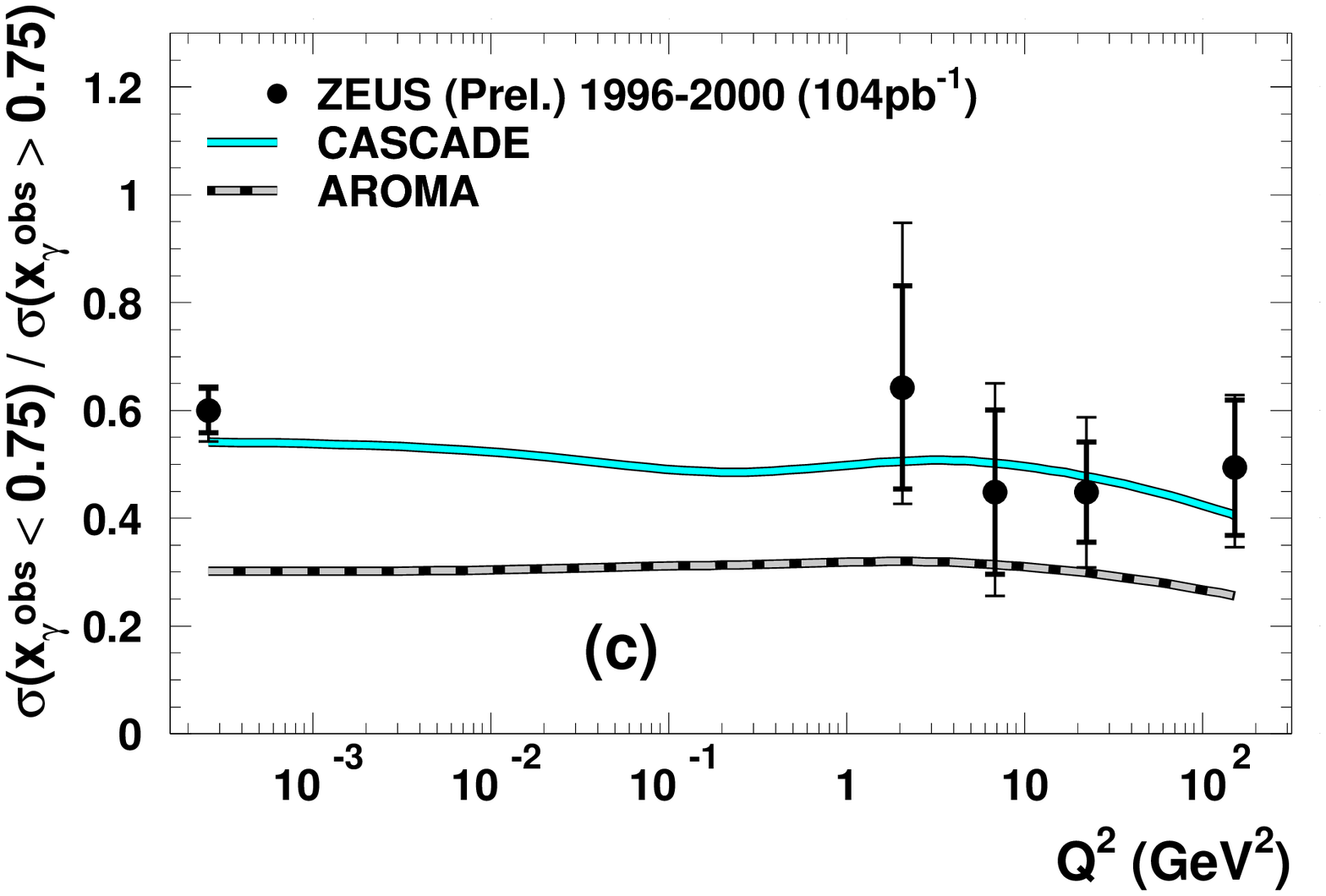,width=6.8cm%
%,bbllx=80pt,bblly=220pt,bburx=540pt,bbury=666pt,clip=%
}}}\vspace*{1mm} 
\fcaption{Ratio of low to high \xgO\ cross sections 
of dijet events containing a $D^*$ as measured in ZEUS (preliminary).
Jet pairs with $E_T > 7.5,\;6.5$ GeV are chosen, and $D^*$ with 
$p_T > 3.0$ GeV.
The results are compared with (a) HERWIG prediction calculated with no
charm requirement. The possible effect of acceptance corrections is
indicated.  (b) HERWIG (with charm requirement) using the SaS1D photon
structure with and without suppression of the photon parton structure
with \QQ, (b) CASCADE and AROMA, }
\label{p495}\end{figure}

\noindent
To probe the nature of the charm process further, ZEUS have compared
the production of charm in DIS with that in
photoproduction.\cite{p495} Dijet events containing a $D^{*\pm}$ meson
are selected, for $\QQ \approx 0$ and in the range 1.0 to 5000 GeV$^2$.
A value of \xgO\ is calculated for each such event, from which the
authors evaluate the cross section ratio between events in \xgO\
below and above 0.75, i.e.\ an approximate ``resolved/direct'' ratio.
This is plotted as a function of \QQ, and compared with theoretical
predictions in fig.~\ref{p495}.
%================================================================
%============================================================
The effects of uncertainties in the proton pdf largely cancel in
taking this ratio, so that we are mainly looking at the properties of
the photon.  The ratio is found to be approximately constant with \QQ,
a behaviour which contrasts strongly with that found in events lacking
a charm requirement, and which cannot be accounted for by acceptance
effects.  Such behaviour is predicted by the Monte Carlos HERWIG,
CASCADE and AROMA.  With HERWIG, the SaS-1D photon pdf's were used
with and without the option of suppressing the hadronic photon
contribution with increasing \QQ; however, the data are not yet able
to distinguish between these cases.  The AROMA prediction is flat, but
quantitatively too low, while CASCADE gives a good overall
description.

These observations are consistent with the expectation that, after
factorising out the proton pdf, the resolved photoproduction of charm
takes place mainly through perturbative QCD mechanisms as does the
direct process.  Thus the ``resolved/direct'' ratio should not vary
rapidly with \QQ.  This contrasts with the non-perturbative, hadronic
processes which dominate light-quark resolved photoproduction and
which show a strong fall-off with \QQ.  In other words, we are talking
about ``anomalous'' diagrams here, rather than about a
charm-containing hadronic photon structure.

Nevertheless, further statistics are clearly required to distinguish
between the two SaS models, and data at higher \QQ\ will also help.

\newpage
\subsection{Some theoretical comments}
\noindent
The HVQDIS model gives good results, but neglects the intrinsic charm
that can evolve in the proton, in other words charm excitation processes, and so could have
problems for $Q^2$ values above the charm mass squared.  However the
experiments do not yet seem to be sufficiently precise to be sensitive
to such an effect.  Also, HVQDIS does not attempt to describe $c\bar
c$ formation from the splitting of outgoing gluons, and the formation of final-state
$J/\psi$ mesons has been ignored in these measurements; this contribution is
not believed to be at an important level.  

Given the merits and drawbacks of the existing fixed and variable
flavour-number schemes, one might obviously wish for a more advanced
type of scheme that combines the advantages of both.  This would
require a model which allows the number of active flavours in the
proton and photon to vary with \QQ\ or \ET, while also treating the
charm quark mass satisfactorily.  In a composite approach, each of the
previous schemes will dominate in its own particular dynamic range, while
a suitable ``matching'' is achieved in the intermediate region.  A LO
model of this type was provided by Aivazis et al.\ (ACOT).\cite{acot}
Further developments have been presented by Collins,\cite{acotc} who
showed that the ideas can be made to work at all orders, and by
Amundson et al.\ (ASTW),\cite{astw} who report a program of
calculations at NLO.  The latter authors provide a Monte Carlo program
which can be used by experimentalists.  At present ASTW are able to
give a good account of the HERA $F_2^c$ measurements.  Their procedure
is less successful with the differential inclusive charm meson
distributions, owing to a strong sensitivity to an internal scale
parameter within the model; good qualitative agreement is
obtained with the data, but the theoretical uncertainties are very large.
ASTW state an intention to extend their calculations to higher order.

There are a number of further implementations of these ideas. This
approach offers a conceptually attractive way forward to a deeper 
understanding of charm production, provided that a sufficiently
high order calculation can be achieved. In principle it should be
applied also to beauty production.

It has been pointed out by Chuvakin et al.\cite{chuvakin} that the
fixed flavour-number schemes at NLO are very stable as regards scale
variations, and are therefore well suited for the description of DIS
processes over a wide $Q^2$ range.  On the other hand,  the cross sections
contain large logarithms of the form $\ln(Q^2/m^2)$, and  these can be
resummed in a variable flavour-number scheme, where the problematic
terms can be absorbed into the pdf's.  These authors demonstrate
that variable flavour-number schemes (BMSN\cite{bmsn} and
CSN\cite{csn}) are able to give good descriptions of $D^*$ production
at HERA.  However the results of the calculations are very similar to
the fixed flavour-number calculations of HVQDIS.  In view of the
latter's moderate scale dependence, there may be at present no
advantage in moving to the other schemes, at least insofar as the
$Q^2$ and $x$ description is concerned.

It should finally be remarked that considerable care is required in
defining $F_2^c$ at the theoretical level.\cite{cteq} For a further
discussion of some of the theoretical issues, the account by
Harris\cite{harrisringberg} may be consulted.

\section{Open Beauty}
\begin{figure}
\vspace*{-10mm}
\centerline{\hspace*{6mm}
\raisebox{2mm}{\epsfig{file=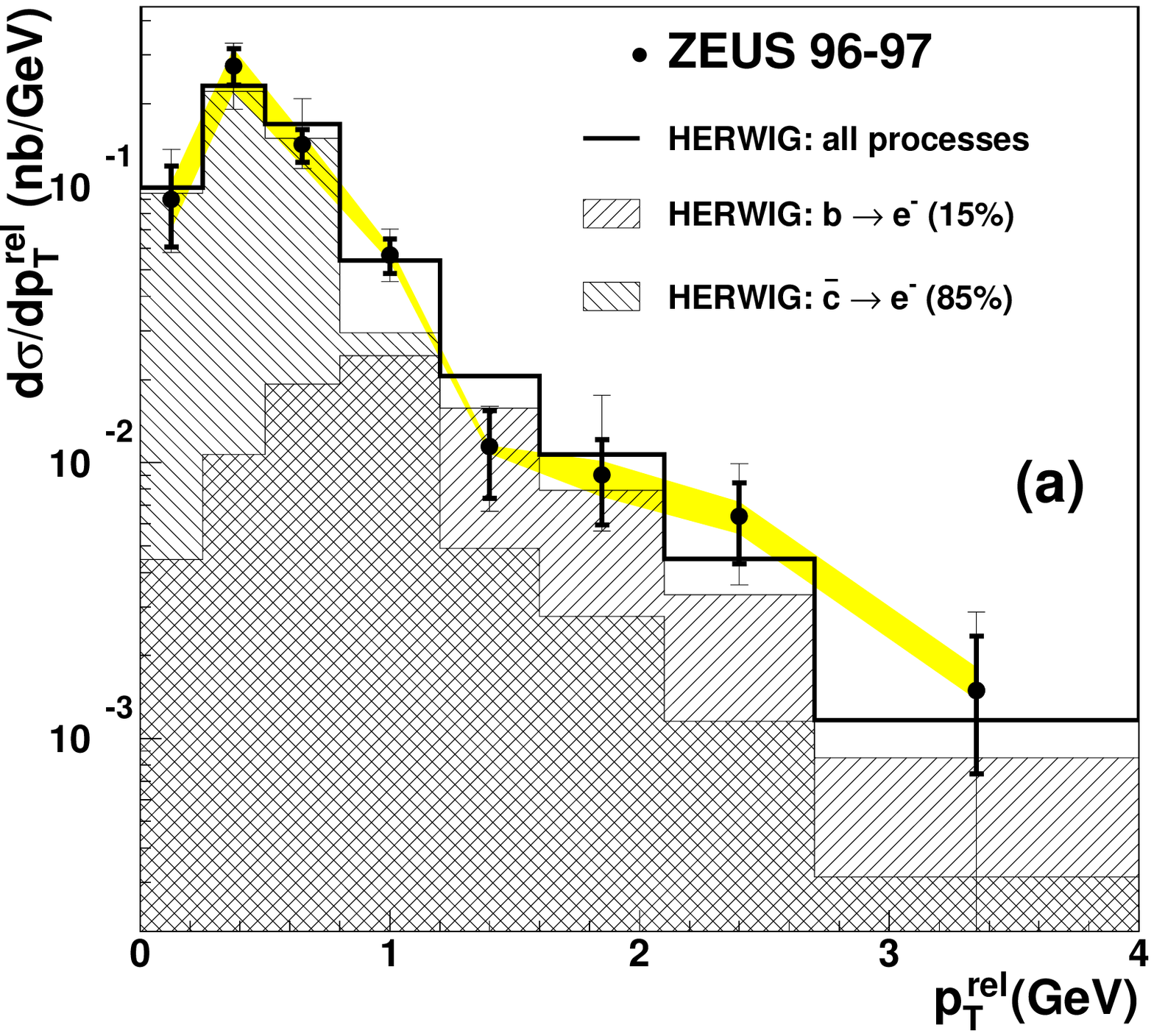,width=6.4cm%
,bbllx=0pt,bblly=0pt,bburx=470pt,bbury=430pt,clip=%
}}
\raisebox{0mm}{\epsfig{file=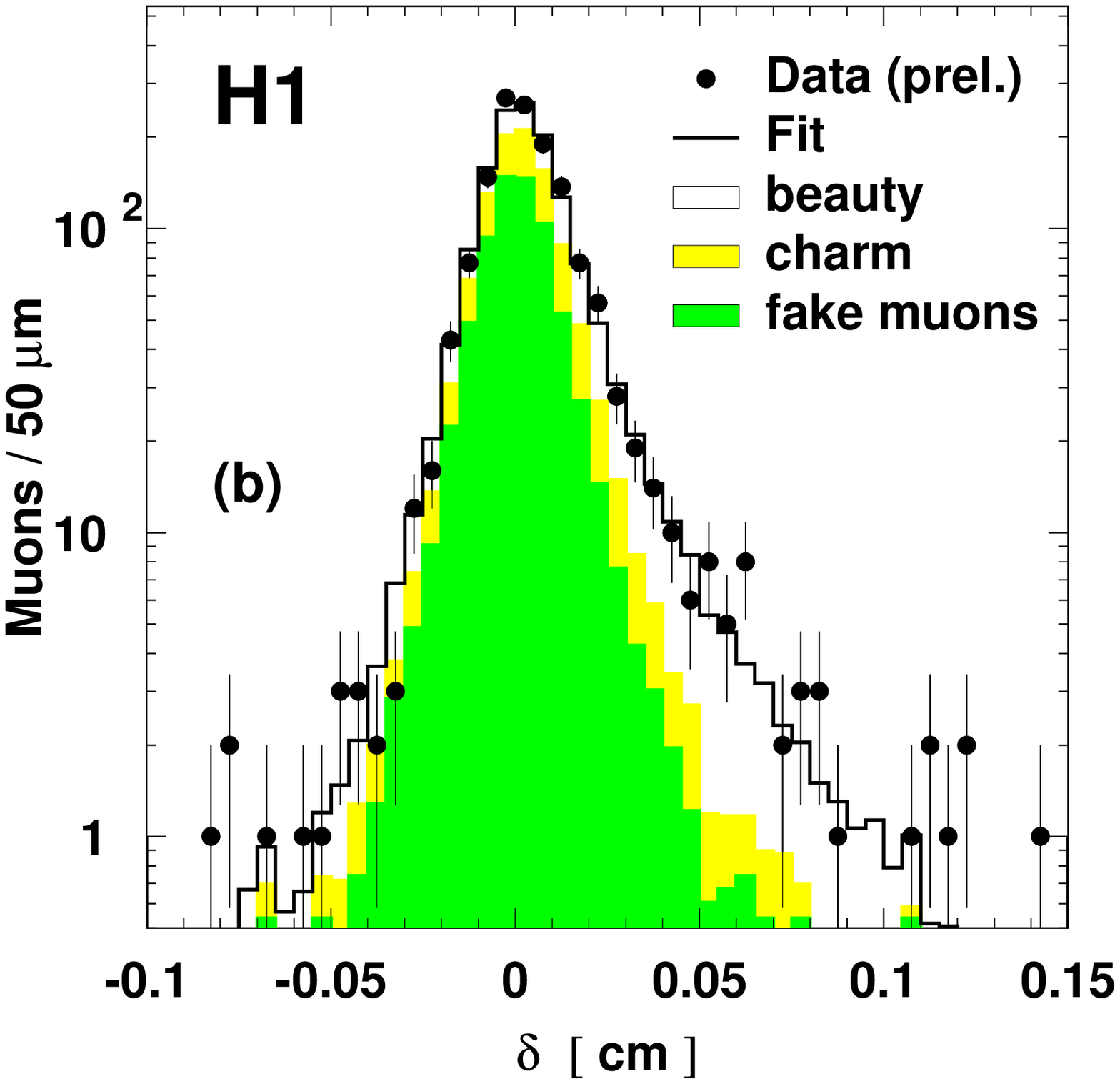,width=7.05cm%
%,bbllx=80pt,bblly=220pt,bburx=540pt,bbury=666pt,clip=%
}}}
\fcaption{
(a) ZEUS differential cross section for production of an electron with
transverse momentum $p_T^\mathit{rel}$ relative to the axis of a jet
($\ET > 6$ GeV) in photoproduction. The calorimeter energy uncertainty
is shown as a shaded band. The total HERWIG prediction for electron
production, and for electron production through $\bar c$ and $b$
decays, are indicated.  Possible backgrounds from DIS positrons
confined the analysis to electrons.  (b) H1 measurement of the impact
parameter $\delta$ of muons relative to the primary event vertex.  Two
jets of $\ET > 5$ GeV are required.  }\label{p487}
\end{figure}

\noindent
Both HERA collider experiments have reported measurements of open
beauty, i.e.\ the $B$ family of mesons.  The latest H1 results have
benefited from the use of a silicon tracking system, and ZEUS will use
such a system in future running.  In the absence of a precise vertex
measurement to help identify the $B$ mesons, with their relatively
long lifetime, the most common identification procedure is by means of
their leptonic decays.  The high mass of the $b$ quark compared with
the $c$ means that the $B$ meson decay products emerge at higher
angles relative to the direction of outgoing heavy-flavour meson or
its jet (fig.~\ref{p487}a).\cite{p487} Important backgrounds come from
the leptonic decays of $D$ mesons.

%=================================================
%=====================================================

ZEUS have used this method to obtain an inclusive $B$ photoproduction
cross section, where $B$ denotes all $b$-quark mesons, taking an
inclusive branching ratio to electrons.  H1 have useed
muons,\cite{p790} and have more recently demonstrated the
effectiveness of a selection on impact parameter (fig.~\ref{p487}b),
presenting preliminary measurements in both photoproduction\cite{p979982} and
DIS.\cite{p807}

\begin{figure} 
\vspace*{2mm}
\centerline{\sf \hspace*{0.6cm}(a)\hspace*{1.8cm}(b)}\vspace*{-7mm}
\centerline{\hspace*{3mm}
\raisebox{0mm}{\epsfig{file=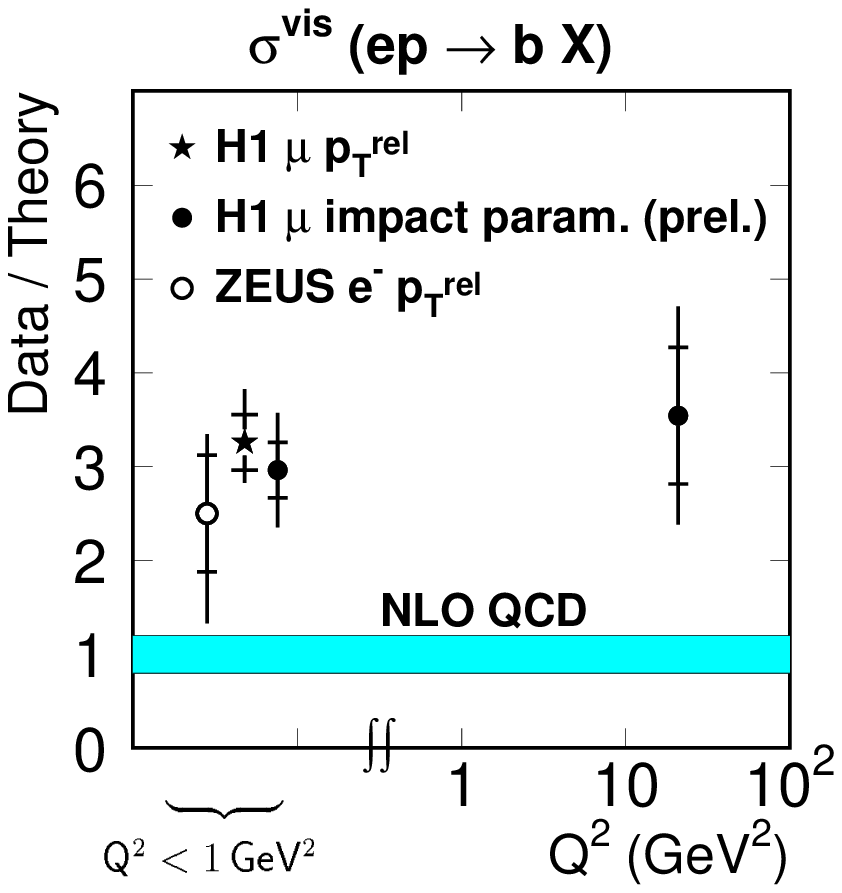,width=6.4cm%
%,bbllx=80pt,bblly=220pt,bburx=540pt,bbury=666pt,clip=%
}}
\hspace*{-1mm}
\raisebox{0mm}{\epsfig{file=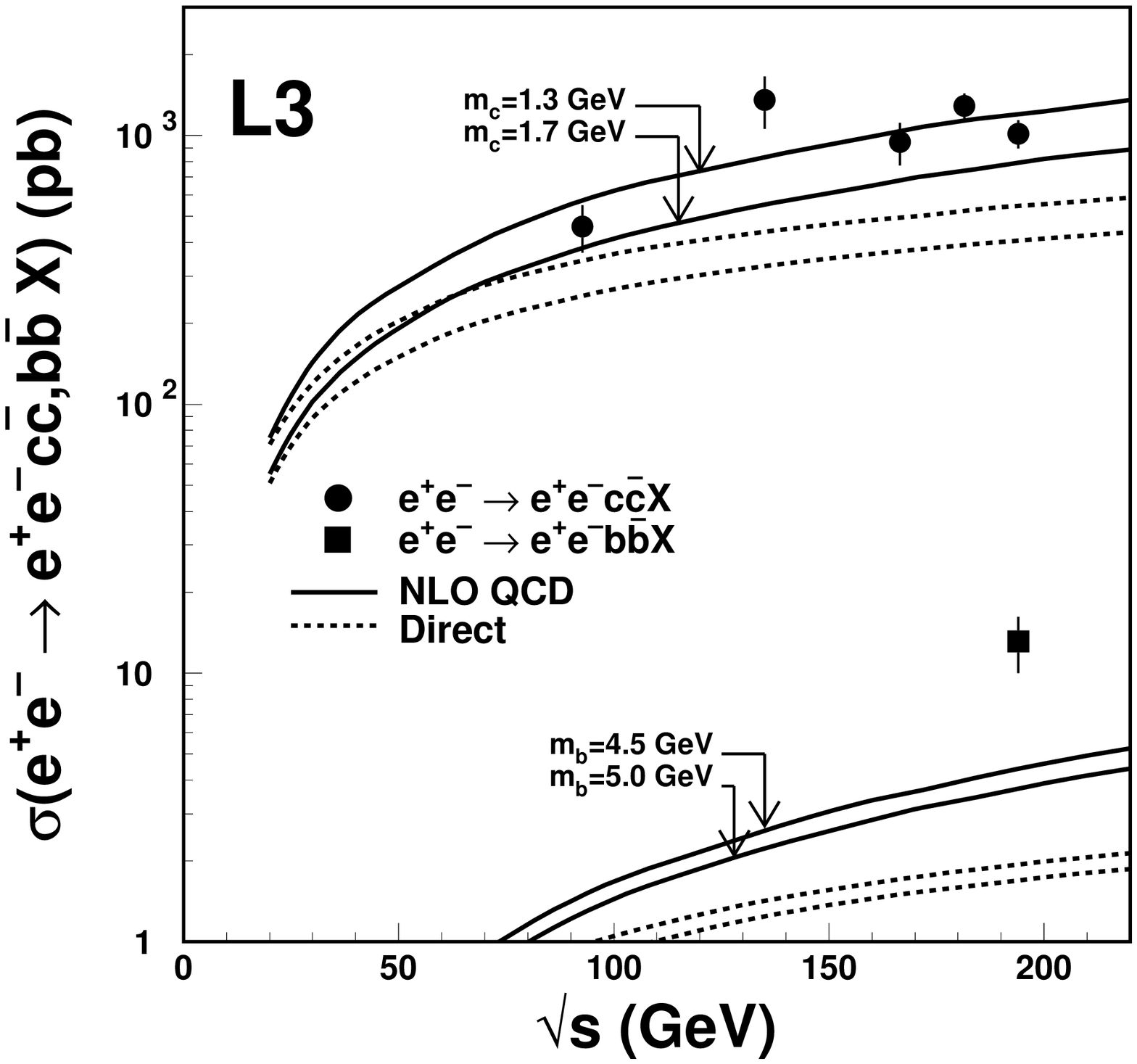,width=6.7cm%
%,bbllx=80pt,bblly=220pt,bburx=540pt,bbury=666pt,clip=%
}}}\vspace*{3mm}
% L3 is stored as p496a
\fcaption{Open beauty production in various hadronic processes.
(a) At HERA using H1 and ZEUS, as a function of \QQ.  (b) At LEP using
$\gamma\gamma$ collisions with the L3 detector, together with open
charm results (upper points and curves).  Statistical and systematic
uncertainties are added in quadrature.  }\label{l3}
\end{figure}

The results are summarised in fig.~\ref{l3}a.  Since the measurements
are defined using different experimental acceptances, they are plotted
as a ratio of the measured cross sections to the predictions at NLO
from the HVQDIS program.  A Peterson fragmentation function was used
to calculate $B$ production rates, from which lepton cross sections
were extracted using the AROMA generator --- in other words, a similar
method to that used by H1 for their $D$ measurements.  The HVQDIS
program yielded a predicted cross section of $11\pm2$ pb for the
production of $B$ hadrons at HERA, compared with the H1 DIS
measurement of $39\pm8\pm10$ pb.  AROMA itself gives a prediction of 9
pb and CASCADE, at LO, gave 15pb.  The photoproduction theoretical
result was calculated using the FMNR program.

%================= HERA and L3 =======================

In short, the integrated $B$ cross sections at HERA lie consistently
above the NLO theory whenever the experimental errors permit a
statement.  It is therefore worth making comparison with results from
other types of experiment.  At LEP, measurements of charm and beauty
production have been made in photon-photon collisions by L3\cite{l3}
and OPAL.\cite{opal} The acceptance is evaluated using PYTHIA and the
results compared to NLO predictions from Drees et al.\cite{Drees} The
charm situation is found to be not dissimilar to that at HERA.  L3
(\ref{l3}b, top) present charm cross sections in overall accord with
calculations, bearing in mind the $c$ mass uncertainty.  OPAL,
measuring $D^*$ production with \pT\ between 2 and 12 GeV,\cite{opalc}
find that the massless calculation by Binnewies, Kniehl and
Kramer\cite{kniehl2,bkk} fits the data well, while a massive
calculation by Frixione, Kr\"amer and Laenen\cite{fkl} underestimates
the data even with $m_c =1.2$ GeV. The direct process, here referring
to $\gamma\gamma\to q\bar q$, is insufficient to account for the
observed signal, and the authors argue that a resolved photon with
significant gluon content is required.

\begin{figure}
\vspace*{-6mm}
\centerline{
\raisebox{0mm}{\epsfig{file=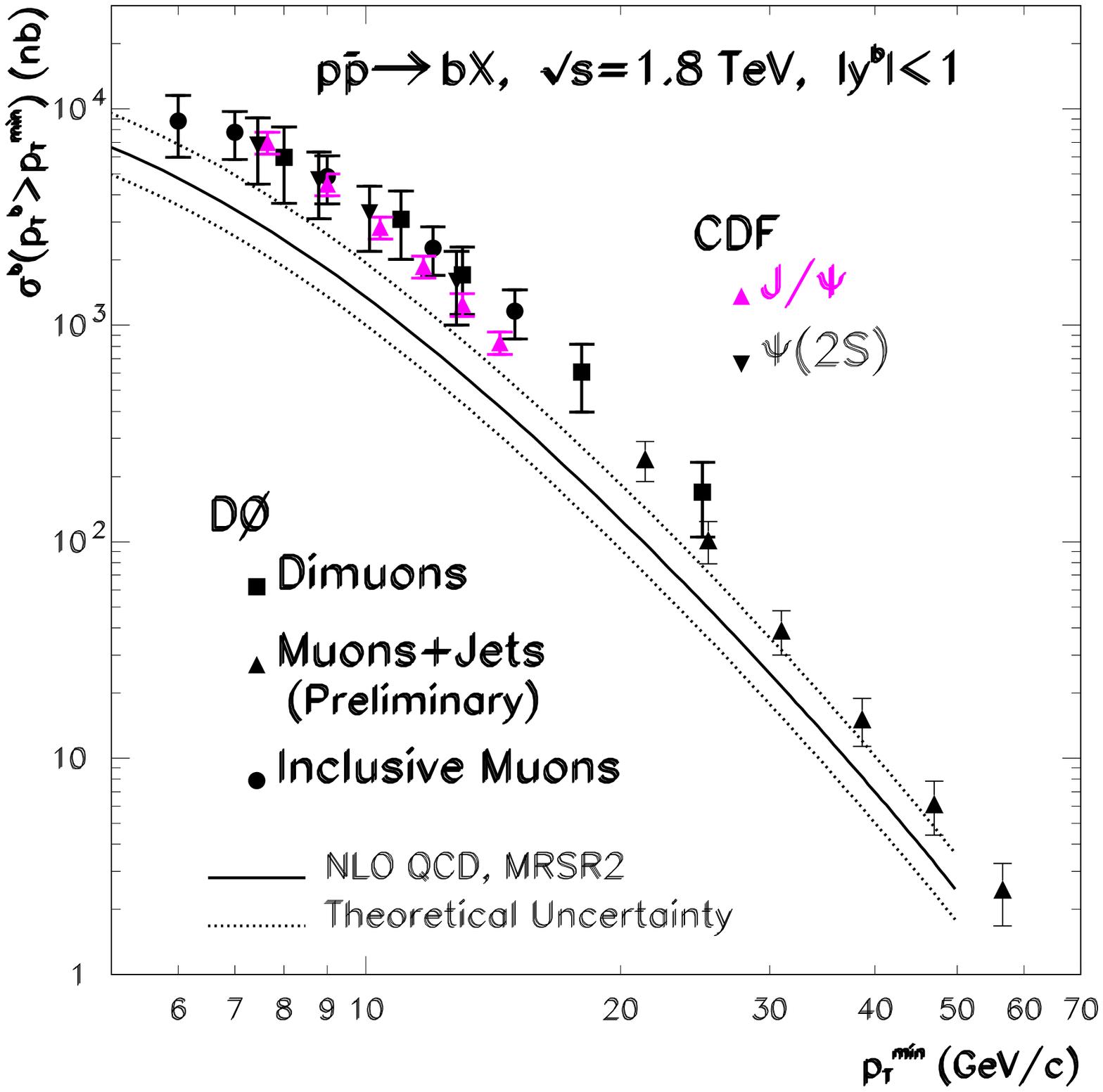,width=7.5cm%
%,bbllx=80pt,bblly=220pt,bburx=540pt,bbury=666pt,clip=%
}}}\vspace{2mm}
\fcaption{Beauty  production
at the Tevatron in CDF and D0, compared with an NLO
calculation\protect\cite{tevb}.  The different points denote different
ways of identifying $b$ events; the dashed curves indicate the
theoretical uncertainty.}\label{tevb}
\end{figure}

However the beauty rate in L3 is much higher than predicted. The OPAL
$B$ cross section, selecting on jets with $\ET > 3$ GeV, is in good
agreement with L3, whose kinematics are similar.  In $p \bar p$
collisions at the Tevatron, CDF and D0 have presented extensive data
on $B$ production\cite{cdfb,d0b} and, again, find that the NLO
calculation is an underestimate (fig.~\ref{tevb}).\cite{tevb} However 
it can be seen that at high \pT\ the disagreement is less serious,
 the largest discrepancies bein found at $\pT \sleq 20$ GeV,
roughly where the other experiments have obtained their
measurements.  This region, then, where the momentum scale of the hard
scatter is of the same magnitude as that of the $b$ quark, is where
the main trouble seems to lie.

\begin{figure}
\centerline{\hspace*{9mm}
\raisebox{0mm}{\epsfig{file=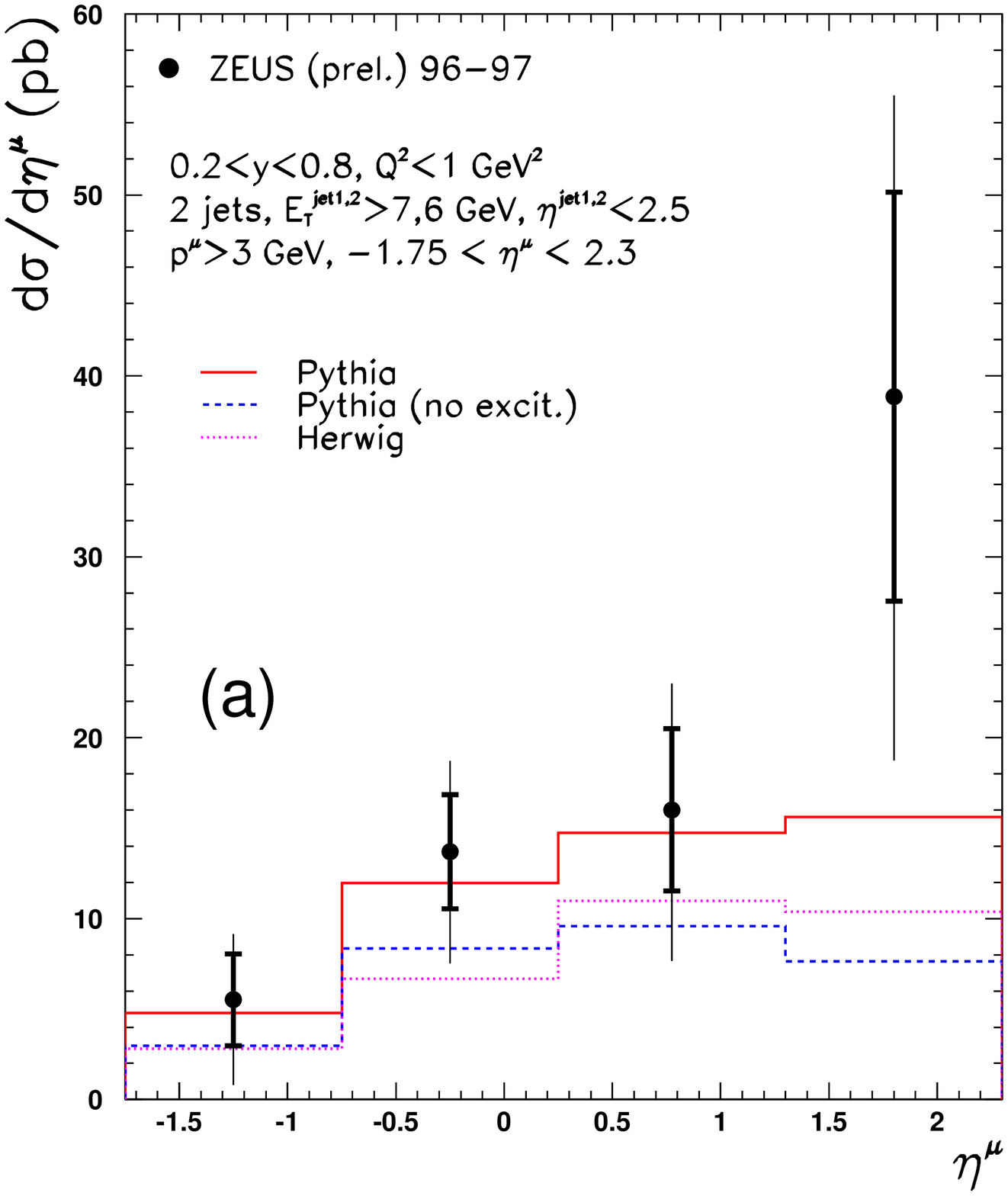,width=6.5cm%
%,bbllx=80pt,bblly=220pt,bburx=540pt,bbury=666pt,clip=%
}}
\raisebox{0mm}{\epsfig{file=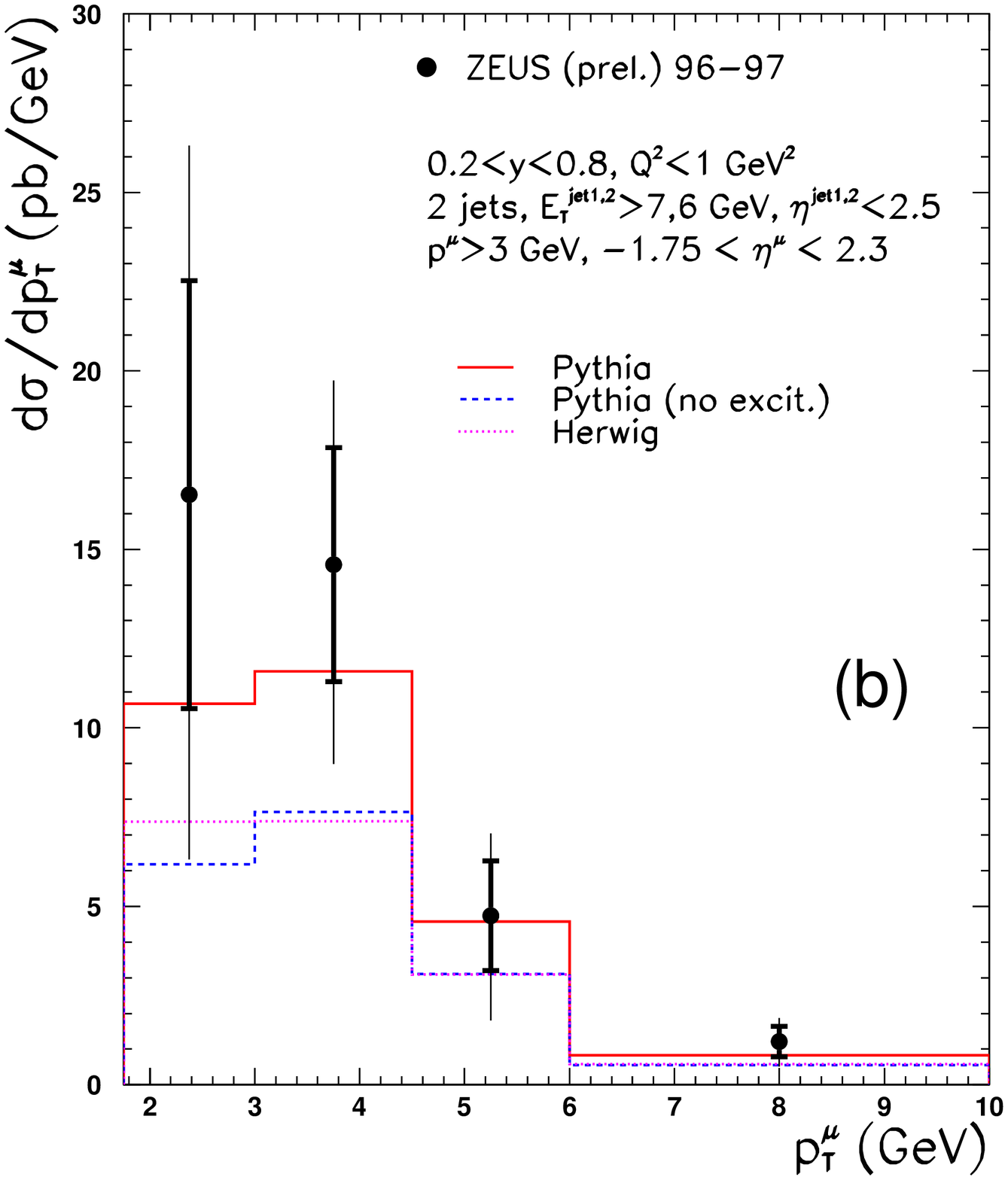,width=6.5cm%
%,bbllx=80pt,bblly=220pt,bburx=540pt,bbury=666pt,clip=%
}}}\vspace*{2mm}
\fcaption{ Results on open photoproduction of beauty (ZEUS, preliminary)
compared with predictions of PYTHIA with and without the excitation
contribution, and with HERWIG.  The $B$ signal is observed in the
muonic semi-leptonic decay mode.  The experimental cuts are indicated.
}\label{p496}
\end{figure}

There are suggestions,\cite{field} however, that a full treatment of
the excitation and fragmentation contributions within PYTHIA and
HERWIG may enable these models to fit the Tevatron $b$ data even at
lower transverse momenta.  It is important to generate all possible
final states where jet fragmentation may produce $B$ mesons.  The
excitation contributions can be substantial, as indicated in a recent
analysis from ZEUS (fig.~\ref{p496}).\cite{p496} Although the present
experimental errors prevent strong conclusions from being drawn, a
large disagreement with theory is not supported by the latter data,
although the agreement in the forward direction would be poor without
the excitation component.

The value of the $b$ quark mass requires comment.  The Particle Data
Group\cite{pdg} quote a value centred on 4.25 GeV at production
threshold.  In QCD, this mass runs with momentum scale, as
confirmed by experiments at LEP.\cite{brun} However, programs such as
FMNR are based on a different subtraction scheme which requires the
``pole'' mass, whose value is a little higher, e.g. 4.5 - 5.0 GeV, and
is fixed.  The Tevatron results are accompanied by an NLO curve
calculated with $m_B = 4.75$ GeV$^2$, and the dashed curves indicate
the uncertainties due to the $b$ mass, the renormalisation scale and
the Peterson parameter.

Lipatov et al.\cite{lipatov} have discussed BFKL-style models which
predict the photoproduction cross section for heavy quark-pair
production.  The essence of these schemes is to consider alternative
types of gluon distribution to those generated by DGLAP evolution.
Within the parameter space of this type of approach, these authors are
able to account for the H1 beauty cross section as well as for the
total charm cross sections measured by H1, ZEUS and lower energy
experiments.  A further account is given by Baranov, Lipatov and
Zotov,\cite{baranov} who demonstrate that such models are able to
describe the Tevatron $b$ production cross sections, again provided
that suitable variants of the theory are selected.

Although this sounds promising, it is not yet clear that any one
implementation of the theory is capable of describing the data from
all the experiments.  To get the H1 $B$ cross section right, the BFKL
model requires a $b$-quark mass value of 4.25 GeV,\cite{lipatov} which
is a little low, while the ``semihard'' LRSS model\cite{LRSS} is good
for both ZEUS and H1, but overshoots the CDF and D0 data if the same
parameter set is used.  The CASCADE Monte Carlo is able to give a good
description of the Tevatron data,\cite{jung,lonnblad} and, as seen, is
fairly satisfactory with ZEUS but not H1.  In this model,
initial-state gluon radiation gives rise to many low-\xgO\ events.
The authors tend to believe that the H1 and ZEUS data are in
disagreement.

At present a summary of this fluid situation is difficult: clearly,
$b$-quark production provides an testing-ground for a variety of
theoretical ideas.  As these develop, and as the experimental errors
from H1 and ZEUS decrease, we may expect to achieve serious
discrimination between the different approaches.

\section{Inclusive Quarkonium Production}
\noindent
The production of $c\bar c$ and $b\bar b$ mesons has been the subject
of extensive study at HERA.  Specifically, we are referring  to the
\Jpsi\ meson (i.e.\ the $\psi(1S)$ state), the radially
excited $\psi'$  (or $\psi(2S)$), and the set of analogous 
$\Upsilon$ states for the $b\bar b$ system.  The present section
discusses the inclusive production of the \Jpsi\ and $\psi(2S)$ states in
photoproduction and DIS. Like the production of open charm and 
beauty, this provides an interesting workshop for the study of QCD
mechanisms.  Experimentally, the \Jpsi\ is conveniently
studied at HERA only in its $\mu^+\mu^-$ and $e^+e^-$ decays, but
these are fully adequate to identify the particle
with reasonable statistics.

The other main area of quarkonium physics at HERA is in the
diffractive production of vector states $q\bar q$ $V$ through the
``elastic'' mechanism $\gamma^*p \to Vp.$ This will be discussed in
Section 6; at HERA, it is only in these processes that the $b\bar b$
states have so far been detected.

\begin{figure}[t!]
\vspace*{1mm}
\centerline{\hspace*{1mm}
\raisebox{0mm}{\epsfig{file=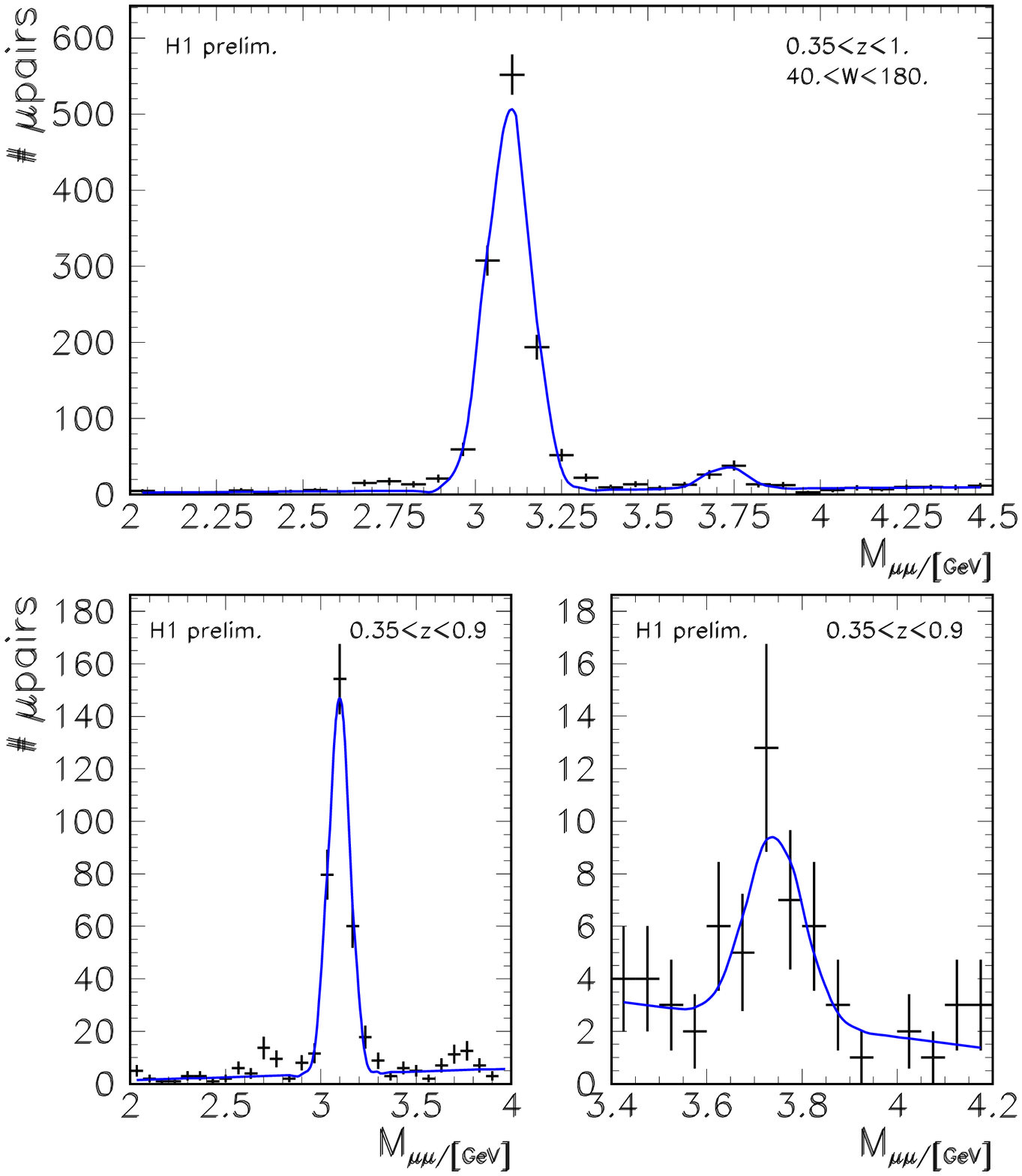,width=8.1cm%
%,bbllx=80pt,bblly=220pt,bburx=540pt,bbury=666pt,clip=%
}}} \vspace*{4mm}
\fcaption{
Dimuon mass spectra in H1 photoproduction.\protect\cite{p794} The full
1996-97 data sample is shown (top), followed by the selection used for
the inelastic \Jpsi\ measurement and the $\psi'$(2S) measurement.
The $\gamma p$ centre-of-mass energy range is 40 - 180 GeV.}
\label{p794}\end{figure}

\subsection{Production of $c\bar c$ mesons}  
\noindent 
In inclusive $c\bar c$ processes at HERA there is no necessity for the
bound quark pairs to appear in vector states; however only the vector
\Jpsi\ and $\psi(2S)$ states have been actually identified
experimentally.  Figure \ref{p794} illustrates the signals observed in
these channels by H1.\cite{p794}

As with the production of open heavy quark systems, these processes
are modelled theoretically in two stages: (a) a hard QCD subprocess,
which is calculable perturbatively and in which the production of the
$c\bar c$ pair is described, and (b) a soft hadronisation stage where
there is a certain probability for a given hadron to be formed.  This
may be represented as
\begin{equation}
d\sigma(\gamma^* p \to HX) = \sum_n\;d\sigma(\gamma^* p \to
\qh\qhbar(n) + x)\,\mathcal{O}^H(n) \label{qqbareqn} 
\end{equation} 
in which a series of intermediate \qh\qhbar\ states, labelled $n$, are
considered, and each has a certain matrix element $\mathcal{O}^H(n)$
for transition into the chosen hadronic state $H$.  The perturbatively
calculable subprocess acts at a momentum scale given by the mass of
the \qh\qhbar\ system, or the \pT\ of the scatter if higher; it is a
short-distance, fast process.  The intermediate \qh\qhbar\ system may
be in a number of angular momentum states, and is produced in one of
two colour states, namely singlet or octet.  If necessary, soft gluon
radiation ensures the colour-neutrality of the final-state particles.

A specific implementatiion of this scheme uses the so-called
non-relativistic QCD or NRQCD approach.\cite{nrqcd,nrqcd2} The initial QCD
calculation having first been performed, this effective field theory
model allows the $\mathcal{O}^H(n)$ factors to be expressed in terms
of known powers of $v$, the velocity of the heavy quark in the
quarkonium system, which should not be too large.  The matrix elements
are otherwise not calculable from first principles.  However, they
should be independent of the original QCD process that formed the
$q\bar q$ state, and the hope is that by measuring experimental cross
sections in one context we can obtain information on
$\mathcal{O}^H(n)$ terms which can be applied in another.

The chief experimental questions at HERA concern:
\begin{itemlist} 
\item The relative importance of the colour-octet state, or even whether it
is important at all; 
\item The effective value of the mass of the $c$ quark; 
\item The need for higher-order QCD calculations.
\end{itemlist}
There are expectations that the method will work better for
$b\bar b$ systems than $c\bar c$ systems since the non-relativistic
approximations will be more accurate.

Further details have been given by Frixione et al.\cite{frixetal} and
in an extensive review by Kr\"amer.\cite{KRAMER} This is a situation
where several QCD momentum scales apply in the same event, and in
particular it is assumed that $m_qv \gg
\LQCD$.  In general, the power of $v$ which multiplies the matrix
elements depends on the relative size of $m_qv$, $m_qv^2$ and
$\Lambda_\mathit{QCD}$.

The simplest version of these ideas is the {\it colour-singlet\/}
model, in which colour-octet (CO) terms are neglected, since they are
suppressed by a power of $v^4$ compared to the simplest colour-singlet
(CS) term.  However they are multiplied by a colour factor of 8 and
have a different \pT\ dependence, so that their unimportance cannot be
taken for granted.  Theoretically, the octet terms are required in
order to remove certain infra-red divergences due to soft gluon
emission.  For best accuracy, one would wish the sum in (\ref{qqbareqn})
to include as many terms as possible.  In practice, it is not possible
to determine more than a small number of the $\mathcal{O}^H(n)$
factors by experiment, and so the hope is that a model with just a few
terms included will be able to test the assumption of their
universality to sufficient accuracy to be useful.\cite{schuler} The
different behaviour of the octet from the singlet terms at large
transverse momentum should also provide a useful handle for our
understanding of how well these ideas describe the physics.

\begin{figure}
\centerline{\hspace*{1mm}
\raisebox{0mm}{\epsfig{file=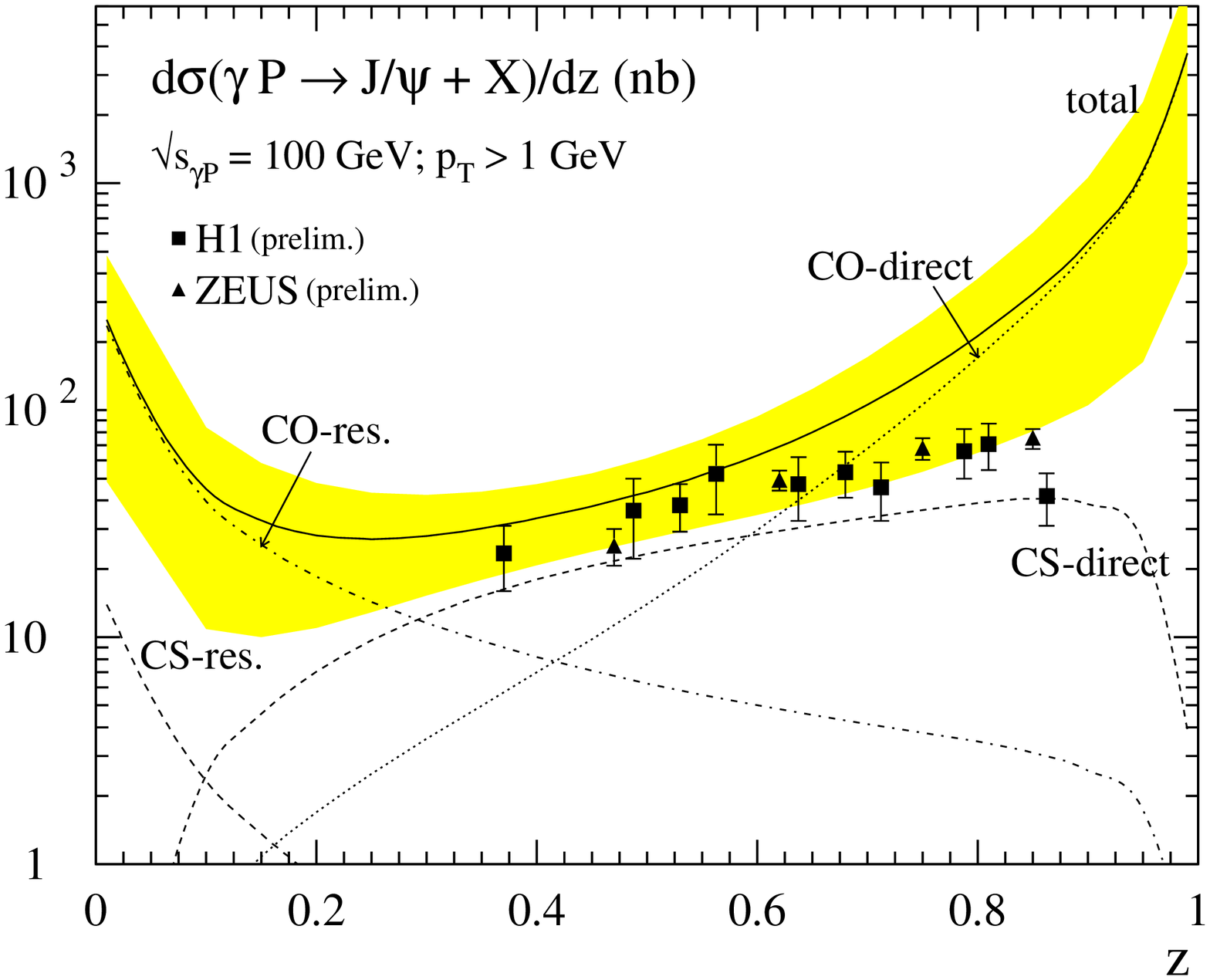,width=7.5cm%
,bbllx=40pt,bblly=30pt,bburx=600pt,bbury=485pt,clip=%
}}
\raisebox{1mm}{\epsfig{file=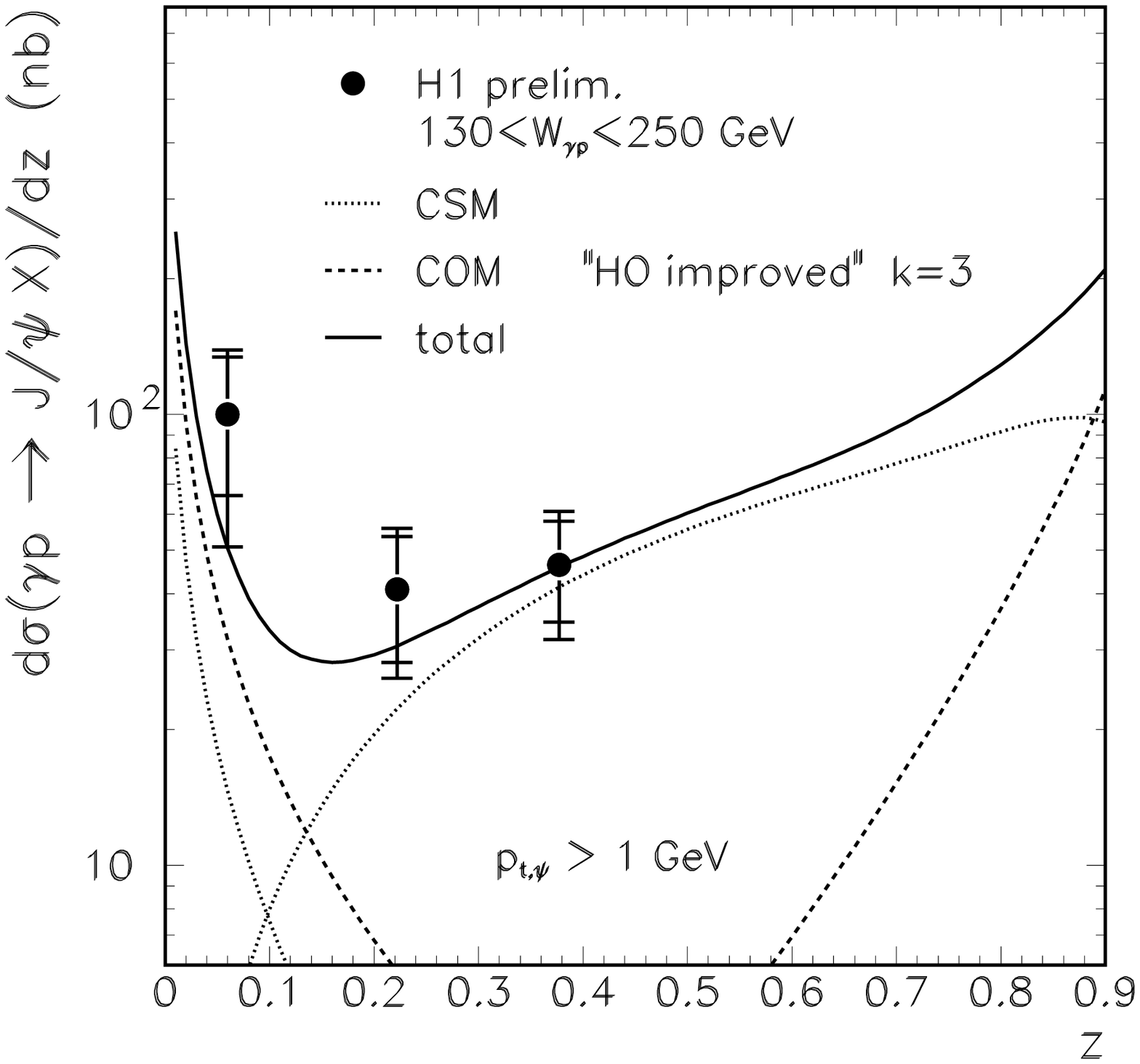,width=5.7cm%
%,bbllx=40pt,bblly=30pt,bburx=600pt,bbury=485pt,clip=%
}}}\vspace*{2mm} 
\fcaption{(left)
Experimental data from H1 and ZEUS on inelastic \Jpsi\
photoproduction, compared with theory (Kr\"amer).  (right) Data from
H1 (preliminary) at low $z$ values.  }
\label{k39}\end{figure}

Both H1\cite{p794} and ZEUS\cite{p851} have presented measurements of
inclusive \Jpsi\ photoproduction.  H1 have in addition made the first
observations of the $\psi'$ in this channel.  Cross sections are given
in terms of the \pT\ of the \Jpsi, its centre-of-mass rapidity $y^* =
\frac{1}{2}\ln(E+p_z)/(E-p_z)$ and its inelasticity $z = P\cdot P_\psi
/ P\cdot q$.  Here $P,\; P_\psi$ and $q$ are the four-momenta of the
beam proton, the \Jpsi\ and the incident virtual photon.  The $z$
parameter is the fraction of the photon energy taken by the \Jpsi\ in
the proton rest frame, and is unity for elastic events.

Figure \ref{k39}(left) shows the cross sections as a function of $z$,
with predictions at LO from the NRQCD model.  For $z>0.3$, the
theoretical cross sections are dominated by the direct process, with
the CO contribution becoming increasingly dominant as $z$ increases.
The CS resolved cross section is small compared to the CO.  The shaded
band represents the total theoretical uncertainty.  The dominant QCD
diagram is from photon-gluon fusion, $\gamma g \to c\bar c$.

H1 have made a special study\cite{p795} of the low-$z$ production of
\Jpsi's.  Figure \ref{k39}(right) shows results that are in excellent
agreement with the LO curves, although the theories quoted here and
previously are not identical.  The presence of the resolved photon
contribution seems clearly required.

\begin{figure}[p]
\begin{center}
\raisebox{0mm}{\epsfig{file=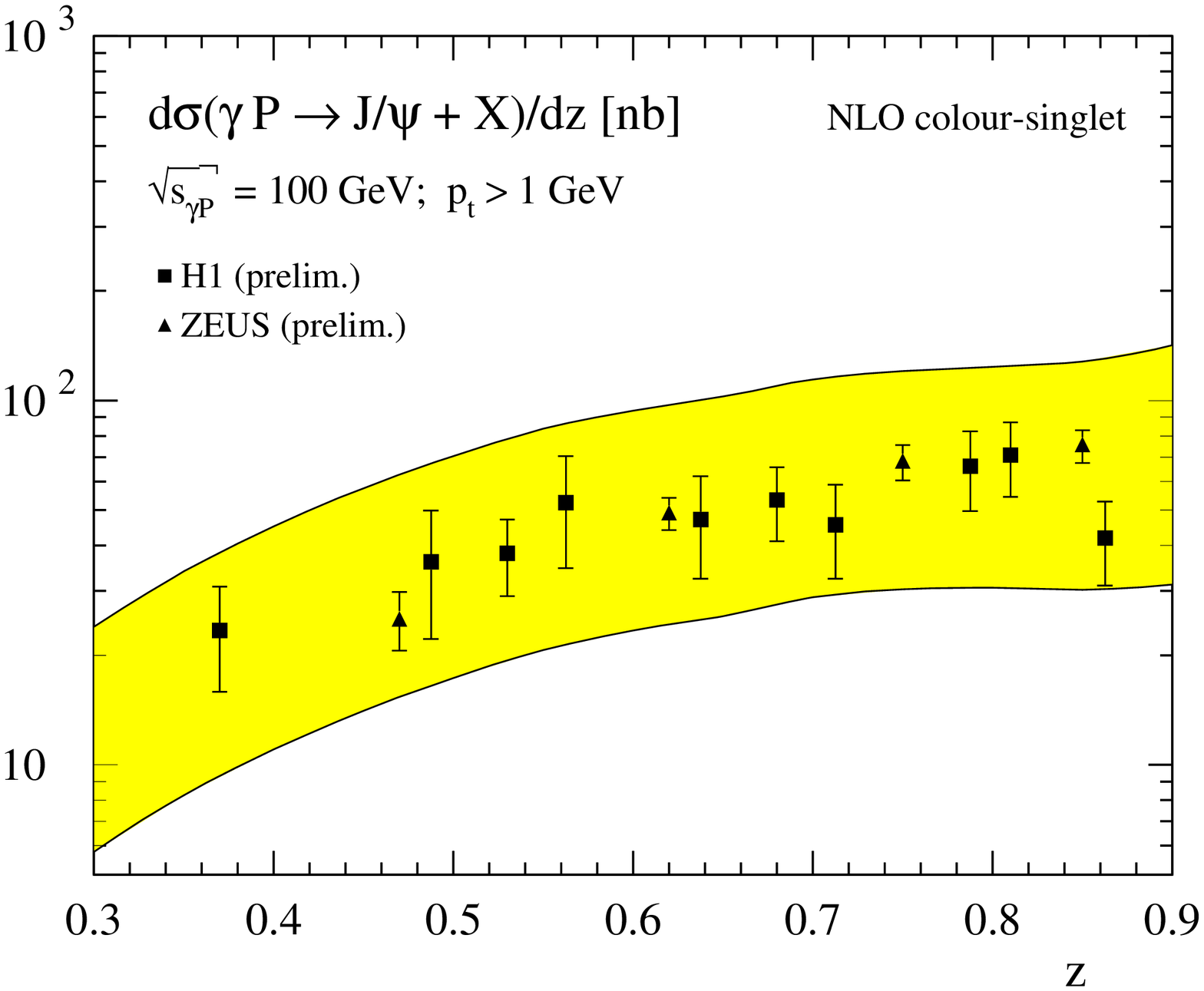,width=7.3cm%
,bbllx=40pt,bblly=30pt,bburx=610pt,bbury=500pt,clip=%
}}
\raisebox{0mm}{\epsfig{file=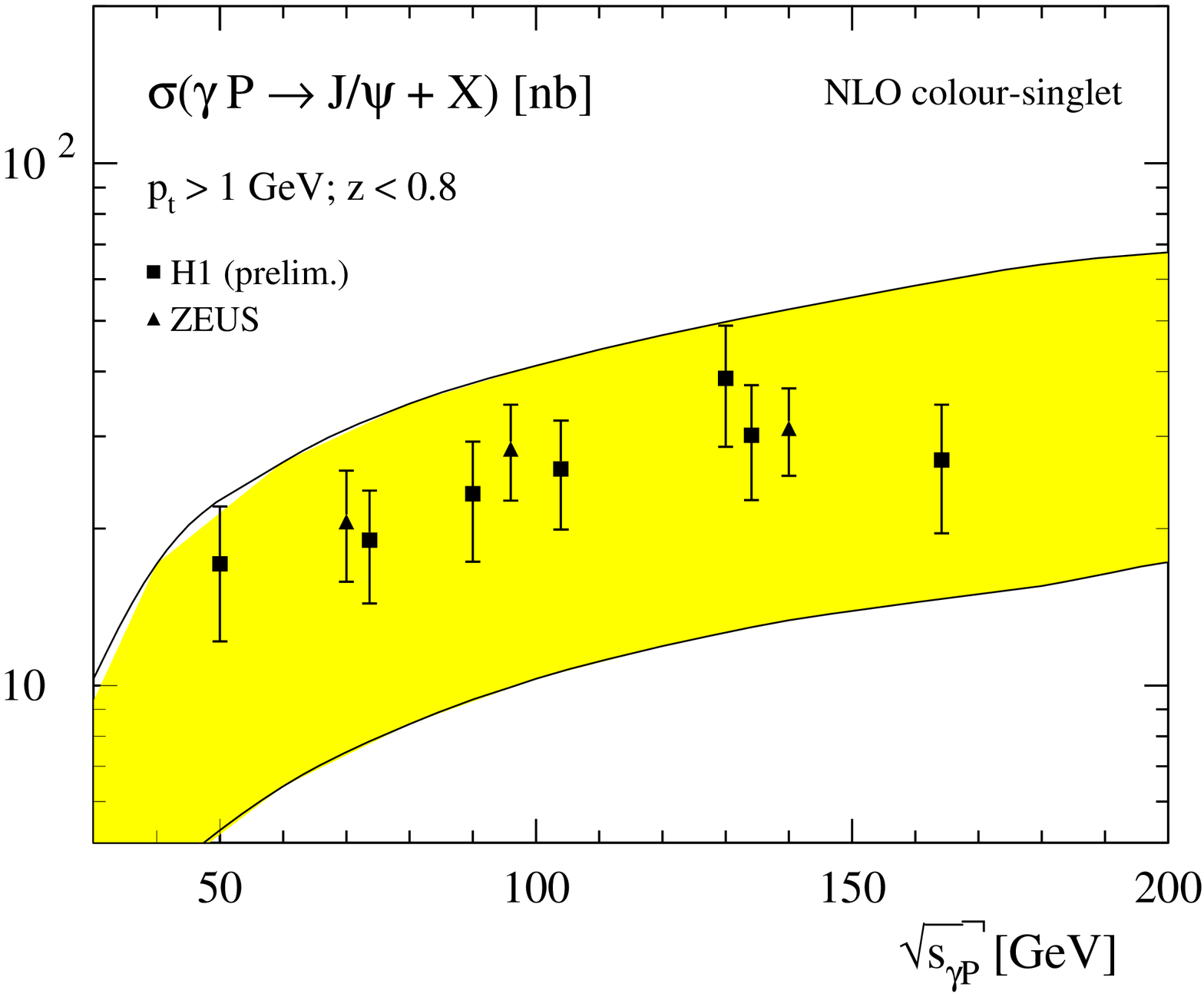,width=7.3cm%
,bbllx=40pt,bblly=30pt,bburx=610pt,bbury=500pt,clip=%
}}
\raisebox{0mm}{\epsfig{file=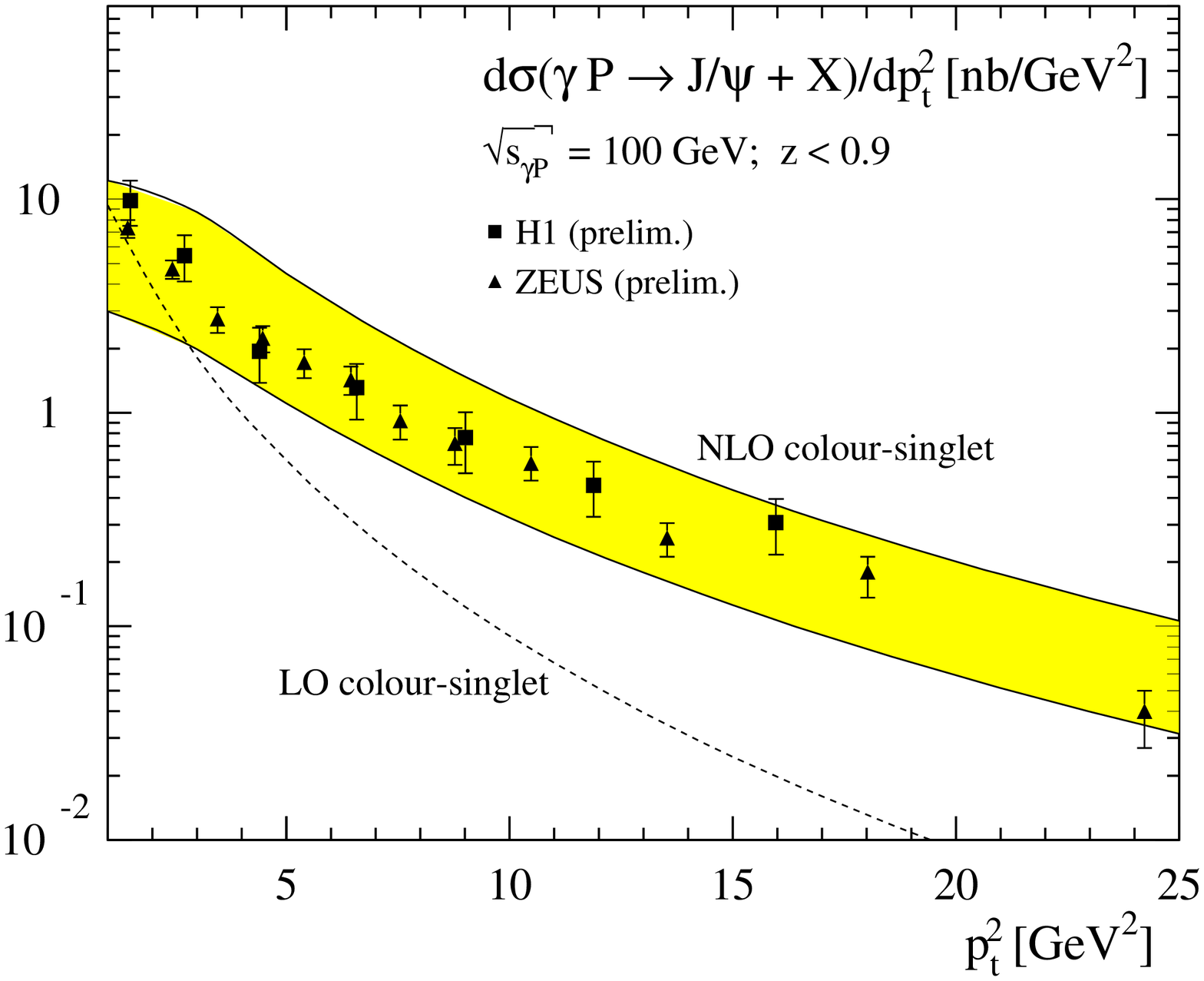,width=7.8cm%
%,bbllx=80pt,bblly=220pt,bburx=540pt,bbury=666pt,clip=%
}\hspace*{6mm}}
\end{center}
\fcaption{
Experimental data from H1 and ZEUS on inelastic \Jpsi\ photoproduction,
compared with NRQCD theory (Kr\"amer) 
}
\label{k41}\end{figure}

It is evident that, while not in disagreement with the LO theory, the
data lie at the low edge of the predicted range for CS + CO, and at
the high edge (if a similar error band is allowed) for CS alone.  The same
data are compared to a NLO calculation in fig.~\ref{k41}(a,b), where
good agreement is found with a CS prediction alone, although the
normalisation uncertainties are again large.  The theoretical
uncertainties include those on the $c$ quark mass and the QCD coupling
constant $\alpha_s$.  ZEUS\cite{p851} have also compared their results
with a NLO CS calculation\cite{kramer} and with a LO CS calculation
supplemented by a LO CO model using fits from CDF and from CLEO data.
The conclusions are similar: NLO CS succeeds, the others sometimes
succeed and sometimes fail, but the theoretical uncertainties are
large.  Clearly, this does not yet amount to a powerful test of
these ideas, especially for the presence of the CO term.  The
\pT\ distribution (fig.~\ref{k41}c) demonstrates the importance of 
an NLO calculation.  It includes terms which are dominated by
$t$-channel gluon exchange and scale as \pT$^6$ instead of \pT$^8$,
but similar $t$-channel exchanges occur in the LO CO contribution.
%==============================================================
\begin{figure}[t!]
\vspace*{0.5mm}
\centerline{\hspace*{1mm}
\raisebox{0mm}{\epsfig{file=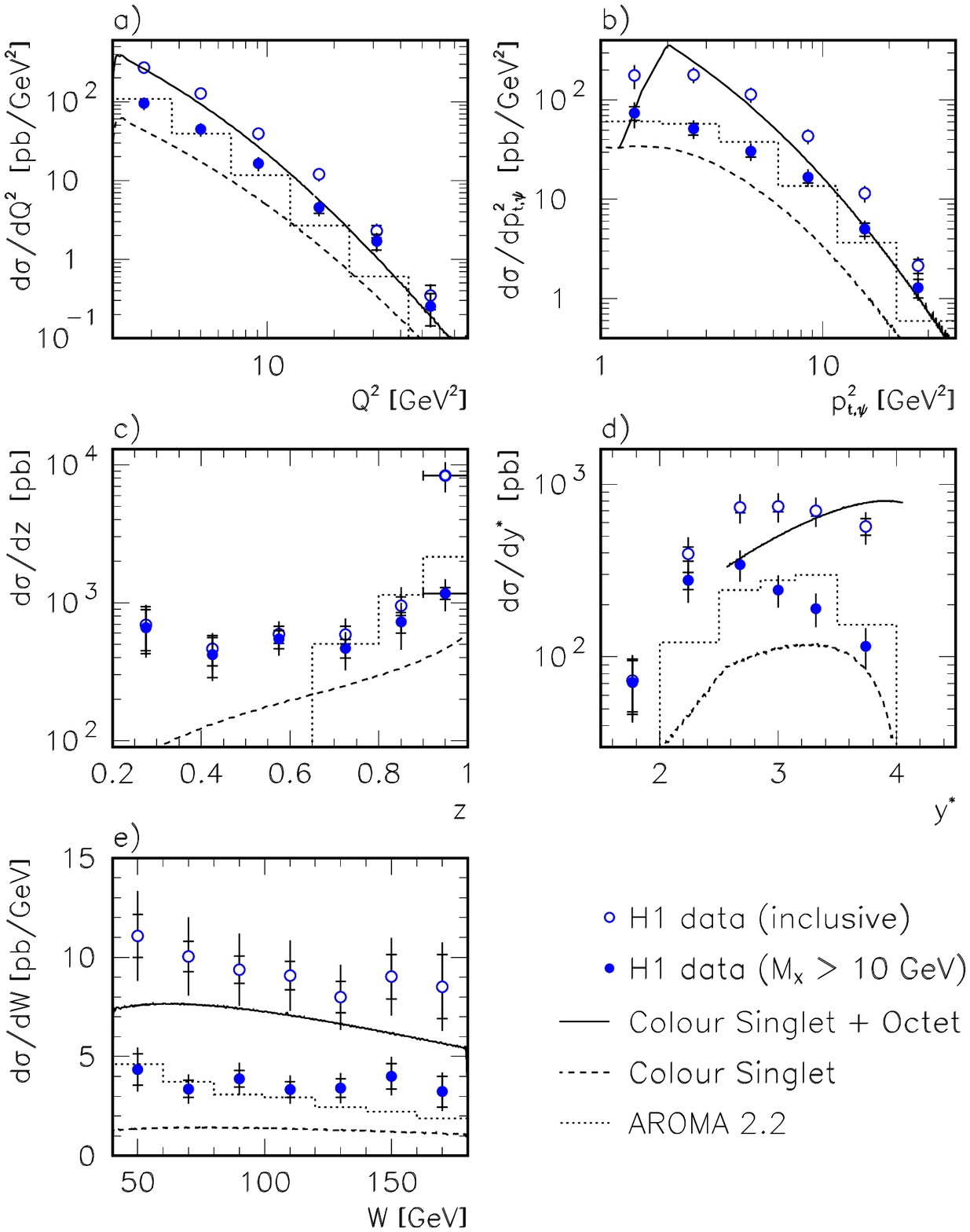,width=9cm%
,bbllx=20pt,bblly=0pt,bburx=520pt,bbury=660pt,clip=%
}}}
\vspace*{-2mm}
\fcaption{
H1 DIS $ep$ cross sections for 
various distributions in inclusive \Jpsi\ production (open circles),
to be compared with the AROMA histogram,
and for inelastic ($M_X > M10$ GeV) production (solid circles), 
to be compared with NRQCD curves from Fleming and Mehen.\cite{fleming} 
}
\label{k42}\end{figure}
%============================================== TEXT CONTINUES ===>

In an analysis of DIS \Jpsi\ production,\cite{p560} H1 have presented
results with and without a mass cut which effectively removes most
events at $z > 0.9$, where the elastic and CO effects are largest
(fig.~\ref{k42}).  They find that the CO contribution, as calculated
by Fleming and Mehen,\cite{fleming} still dominates the remaining
phase space.  Fragmentation is not included in this LO calculation.
The CS term is inadequate to describe the data, but including the CO
term makes the total too high.  H1 attribute these discrepancies to
the matrix elements $\mathcal{O}^H$ used in the calculation.
Meanwhile the Soft Colour Interaction model,\cite{SCI} calculated in AROMA, is
inadequate at the quantitative level, although the shapes of most of the
distributions are described reasonably well.

One path to a better test of these ideas is to measure the
polarisation of the \Jpsi, where the effects of the CO terms should be
more unambiguously visible.\cite{kramer} It would also be interesting
to make an approximate separation between the resolved and direct
contributions, by means of $x_\gamma$, using the events with a \Jpsi\
jet and a gluon jet.

The production of $\psi'$(2S) has been measured by H1 relative to that
for \Jpsi\ in the same kinematic range of inelastic
photoproduction.\cite{p794} The cross section ratio was found to be
$0.210\pm0.048\pm0.032$ (preliminary).  A substantial fraction of
observed \Jpsi\ mesons can be direct or indirect decay products of
$\psi'$(2S) mesons.  Allowing for this, H1 calculate the ratio of
$\psi'$(2S) to directly produced \Jpsi\ as $0.24\pm0.08$, in
agreement with an expectation of 0.15.\cite{kramer}

The NRQCD model is itself being developed and studied, with a variety
of ideas on the market.\cite{morenrqcd} These will in due
course invite further experimental test.

\section{Diffractive Processes}    
\noindent
The final topic of this survey concerns the diffractive production of
heavy quark systems at HERA.  As illustrated in fig.~\ref{resdir}d,
such processes occur through the photon first fluctuating into a
hadronic state $V$, where $V$ must preserve the quantum numbers of the
photon.  The assumption of {\it vector meson dominance\/} asserts that
the relevant hadronic states are the members of the vector meson
family, namely the $L=1$ $q\bar q$ mesons $\rho$, $\omega$, $\phi$,
\Jpsi, $\Upsilon$ etc.  The state $V$ may interact inelastically with the
proton, but the Optical Theorem requires the existence of an
elastic channel, in which the fluctuation $\gamma^*\to V$ is followed
by the elastic process $Vp \to Vp$ and the $V$ emerges on mass shell.
A full theory must also take into account excitation processes whereby
the $V$ or proton may emerge in an excited state.

From a QCD standpoint, the essential feature of diffractive processes 
is that in the primary scatter, the exchanged object is  
colour-neutral (colour-singlet).  At HERA, this condition permits the
proton to remain intact, to end up in an excited nucleon state, or to
become dissociated.  However there must be no transfer of colour that requires
subsequent neutralisation by the generation of a string of hadrons along
the rapidity range between the forward nucleon system and the
remainder of the event.  This rapidity range usually ends up
underpopulated or unpopulated, therefore; in fact a
so-called ``large rapidity gap'' is commonly found in the forward
region of  a diffractive event.

\begin{figure}
\centerline{
\raisebox{0mm}{\epsfig{file=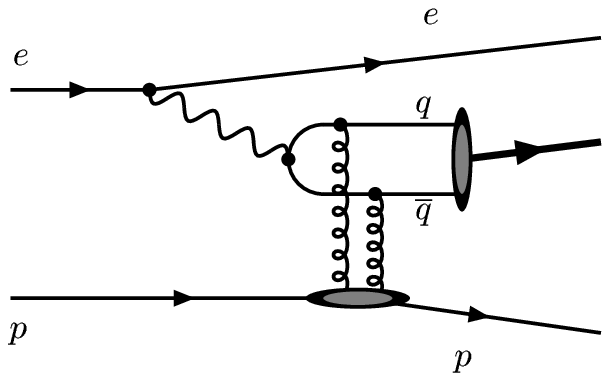,width=5cm%
,bbllx=150pt,bblly=414pt,bburx=324pt,bbury=525pt,clip=%
}}} 
\vspace*{2mm}
\fcaption{
QCD-modelled production of a vector meson in elastic diffraction
(from ZEUS\cite{p559a}).}
\label{qcddiag}\end{figure}
%==========================================================================
Diffractive processes may be interpreted either in terms of Regge
theory, where the exchanged object is a pseudoparticle with the
quantum numbers of a vacuum, known as a pomeron, or else in terms
of QCD models with the typical exchange of two or more gluons
(fig.~\ref{qcddiag}).  We shall here, for convenience, refer to the
exchanged entity generically as a ``pomeron''.

An account of heavy quark production in diffractive $ep$ scattering
must place it against the backdrop of what is now a broad area of
``pomeron'' physics.  Indeed, the manifestations of the pomeron
radiated from the proton parallel in many ways the behaviour of the
photon radiated from the positron.  The pomeron is hadronic,
and it may display parton structure and give rise to ``resolved''
interactions.  In the present context, though, we regard it
simply as an exchanged object with a possibly complex nature,
capable of being modelled in various ways.  In the elastic
photoproduction of vector mesons the pomeron exchange is a soft
process so long as the vector mesons are made of light quarks.
However, when $c\bar c$ or $b\bar b$ mesons are produced, the process
cannot be entirely soft from a QCD viewpoint, and so its
characteristics change.  The situation gives rise to the possibility
of perturbative QCD (pQCD) calculations.

In the  elastic photoproduction of $\rho$, $\omega$ and $\phi$
mesons, the fluctuation into a vector meson is taken to occur {\it
before\/} the scatter off the proton.  For the vector meson to remain
intact, the scattering process must then be essentially soft.  In the
QCD-modelled process 
(fig.~\ref{qcddiag}), the photon couples to a virtual $q\bar q$ pair,
which exchanges gluons with the proton, and then $\it finally\/$ the
emerging $q\bar q$ pair may hadronise into a vector meson.  In elastic
vector-meson production, the coupling to the proton selects the
latter's gluon content at low mean values of Bjorken $x$.  Since $W^2
\propto 1/x$, higher $W$ means lower $x$, corresponding to a rapidly
increasing gluon density in the proton pdf (fig.~\ref{p502_9}). One
expects to observe in general, therefore, that elastic cross sections
will increase strongly with $W$ if perturbative QCD is operating.  In the
Regge approach, a weak rise with $W$ is expected.\cite{DLbasic}

Thus the important difference between the elastic photoproduction of
light and heavy vector mesons is that the production mechanism of the
former is fundamentally ``soft'' at all stages, whereas the
latter require a ``hard'' phase initially.  The situation changes in
DIS.  The elastic formation of even a light vector meson from an incoming
highly virtual photon, with $\QQ \gg m_V$ requires a hard vertex
somewhere, which renders the process in part perturbatively
calculable while depressing the cross section.  

\begin{figure}[p]
\centerline{
\hspace*{2mm}
\raisebox{-3mm}{\epsfig{file=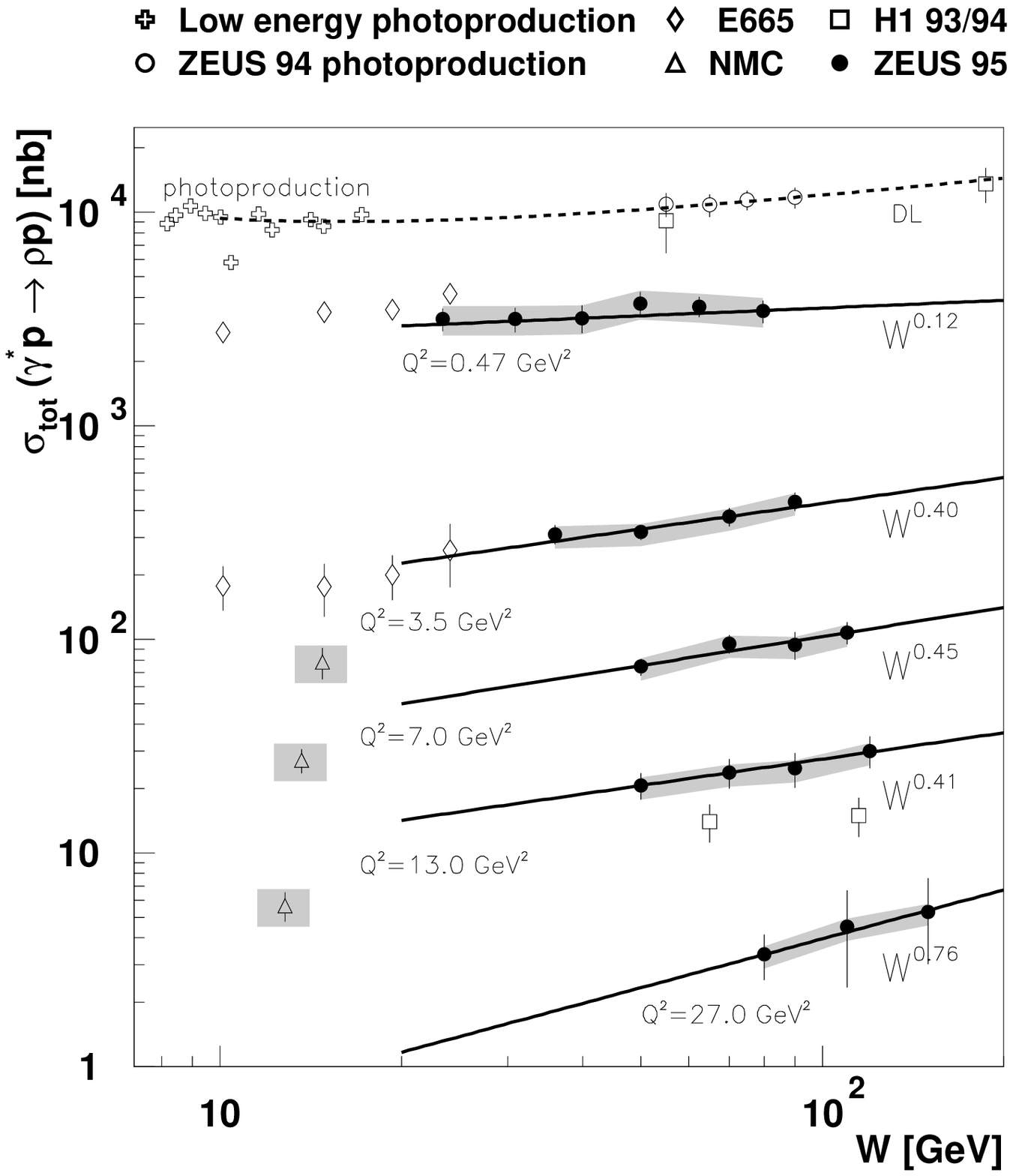,width=7.5cm%
%,bbllx=80pt,bblly=220pt,bburx=540pt,bbury=666pt,clip=%
}}
\hspace*{-7mm}
\raisebox{0mm}{\epsfig{file=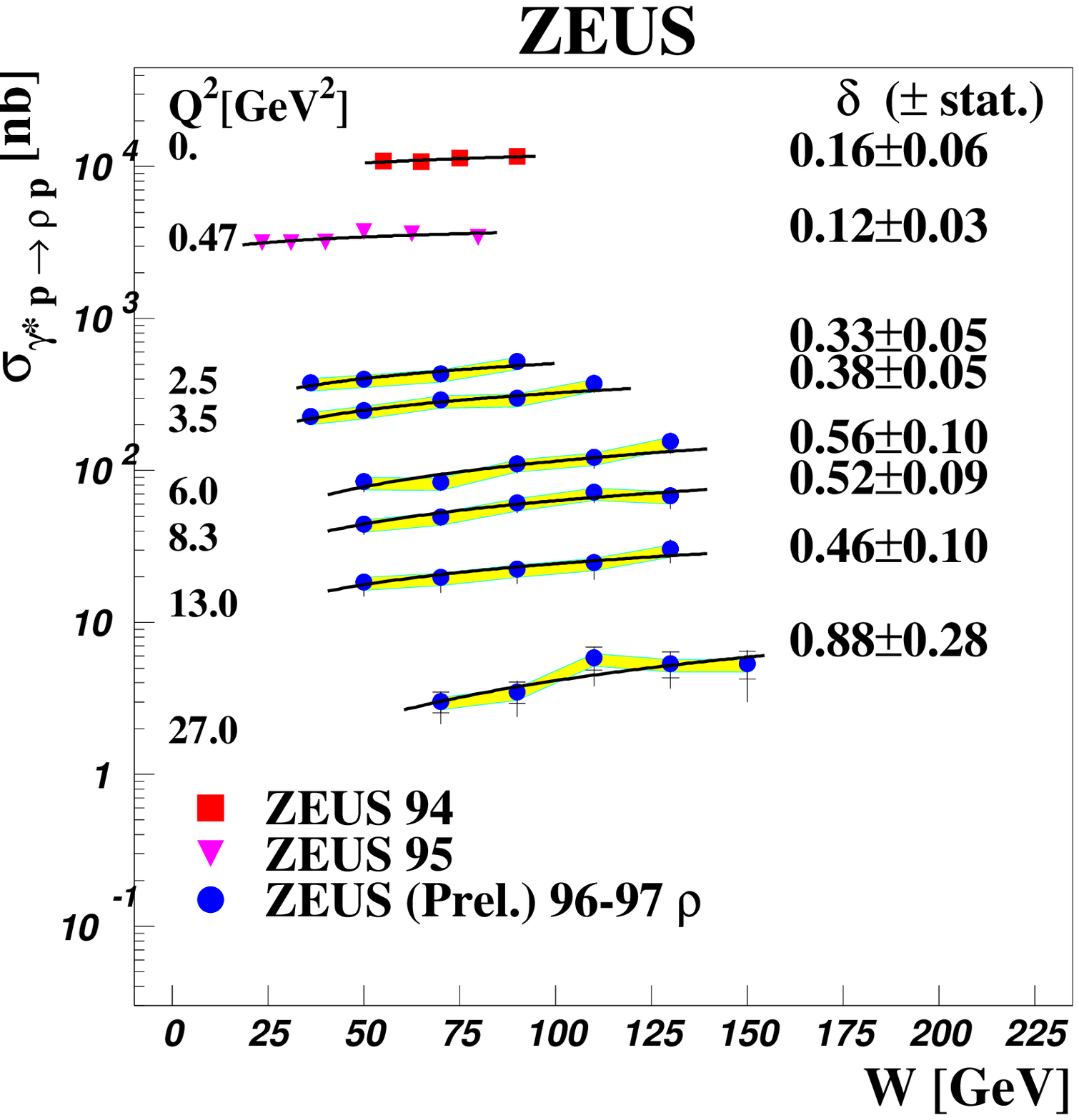,width=6.8cm%
%,bbllx=80pt,bblly=220pt,bburx=540pt,bbury=666pt,clip=%
}}
\vspace*{3mm}
}\centerline{
\raisebox{0mm}{\epsfig{file=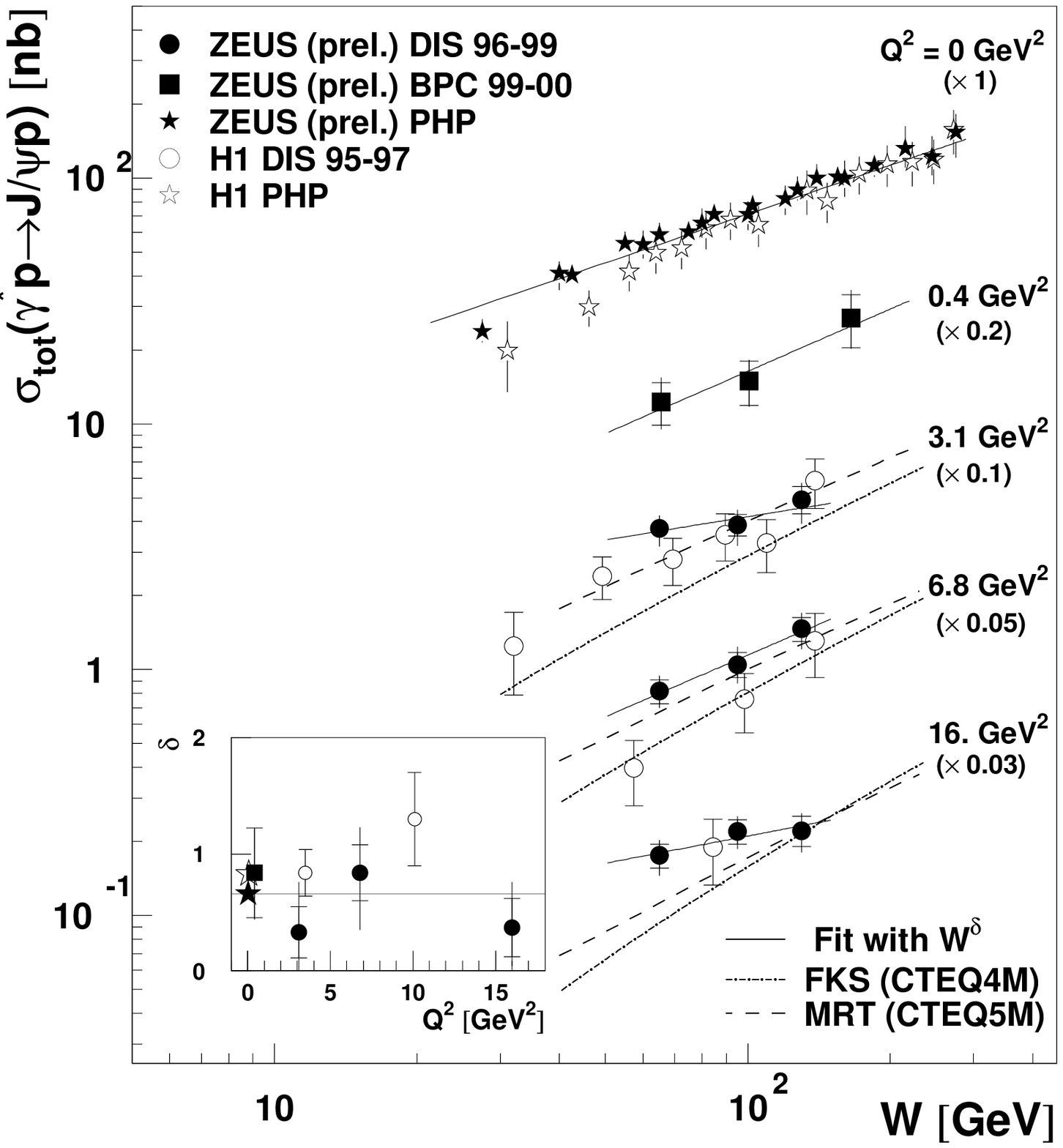,width=6.5cm%
%,bbllx=80pt,bblly=220pt,bburx=540pt,bbury=666pt,clip=%
}}
\raisebox{2mm}{\epsfig{file=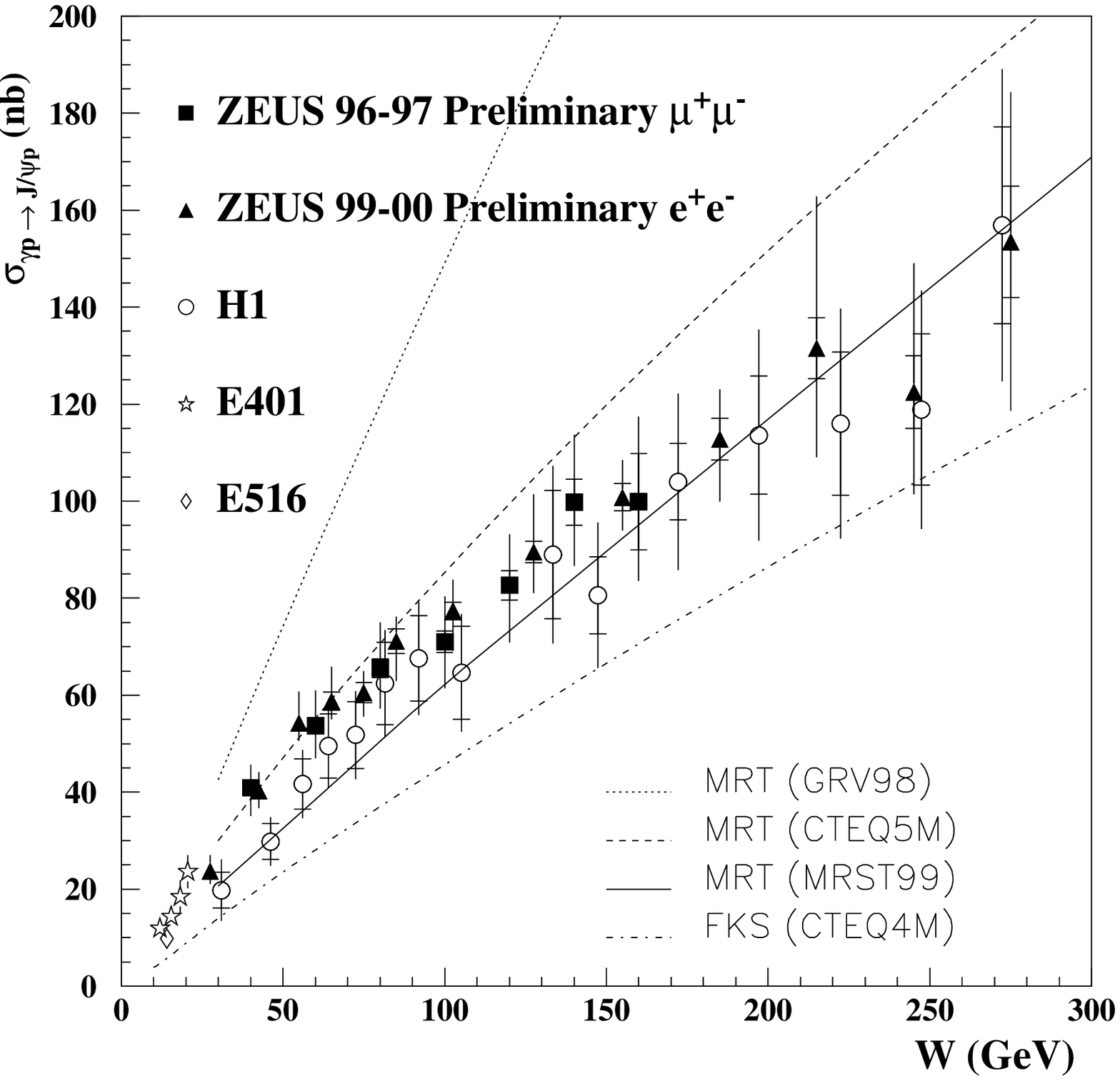,width=6.2cm%
,bbllx=0pt,bblly=10pt,bburx=525pt,bbury=520pt,clip=%
}} }
\fcaption{
Total cross sections for $\gamma^*p\to\rho p$ (top)\cite{p559a,p594}
 and $\gamma^*p\to J/\psi p$ (bottom),\cite{p559,p878} as measured by
 ZEUS and H1 as a function of $\gamma p$ centre-of-mass energy and
 for different values of photon virtuality $Q^2$.  In the first plot,
 earlier ZEUS data is compared with earlier experiments.  The inset in
 the third plot shows the variation of the parameter $\delta$ with
 \QQ.  In the final plot the photoproduction data are compared with
 pQCD predictions using different proton pdf's in two theoretical
 models.  }
\label{p559}\end{figure}

\subsection{Measurements of vector charmonium}
\noindent
The interplay of the different hard scales in diffractive vector meson
production is illustrated in the sets of total elastic cross sections
shown in fig.~\ref{p559}.  In the first two plots, earlier\cite{p559a}
and more recent\cite{p594} ZEUS results on $\rho$ production as a
function of $W$ are compared with H1.  In the third plot, \Jpsi\ data
are presented.\cite{p559} Recent H1 $\rho$ results\cite{p594H} are in
agreement with ZEUS; the agreement seen in the \Jpsi\ case is likewise
reasonable.  The cross sections are presented for the cases of
photoproduction (\QQ\ = 0) and for DIS at various \QQ\ values.  At a
given value of \QQ\ the $W$-dependence of the total cross sections can
be parameterised as $W^\delta$, where $\delta$ is a power to be
measured experimentally.

In the case of $\rho$ photoproduction, the data over a wide range of
$W$ are well fitted by the Regge-theory formalism of Donnachie and
Landshoff,\cite{DLbasic} and indicate $\delta$ values in the range
0.12 to 0.16 at low \QQ.  As \QQ\ rises, $\delta$ increases to values
in the range 0.4 to 0.9, but with a large error for the highest \QQ\
data set. The \Jpsi\ data sets, in contrast,  show a large value of
$\delta$ already at \QQ\ = 0.  A value of $\delta$ in the range 0.6 -
0.8 is indicated, with no evidence from H1 and ZEUS for variation with
\QQ.  The suggestion is that by the time \QQ\ has reached a suitably
high value, a common type of hard behaviour may be found in this $W$ range.

In the fourth plot,\cite{p878} \Jpsi\ data at \QQ\ = 0 are combined
and compared with pQCD predictions using different proton pdf
sets,\cite{pdfs} all of which give good fits to the proton structure
function $F_2$.  The data show a clear sensitivity to the pdf set ---
that is, to its gluon component --- and suggest a preference for one of
them.  However this should be treated with caution since the two
models chosen, FKS and MRT,\cite{fks,mrt} have large uncertainties on
their normalisation.  They differ in their treatment of the gluon
distributions in the photon, and in the description of the charmonium
state.  Overall, these models may be said to confirm the shape of the
variation with $W$, and the general validity of the pQCD approach to
this topic.  The FKS model describes the data well when renormalised
by a factor of $\approx 1.6$.

The coupling of the photon to the vector mesons $V$ is through their
quarks; the quark content of the $\rho$, $\omega$, $\phi$,
\Jpsi\ and $\Upsilon$ mesons predicts total elastic cross sections for 
$\gamma p\to Vp$ in the respective ratios 1 : $\frac{1}{9}$ :
$\frac{2}{9}$ : $\frac{8}{9}$ : $\frac{2}{9}$, respectively, if
flavour independence otherwise applies.  For five quark types, this is
referred to as SU(5) symmetry.  The $\omega$ : $\rho$ cross section
ratio is found to be in good agreement with this prediction, both in
photoproduction and DIS.\cite{zomega}

\begin{figure}[t!]
\centerline{\hspace*{1mm}
\raisebox{0mm}{\epsfig{file=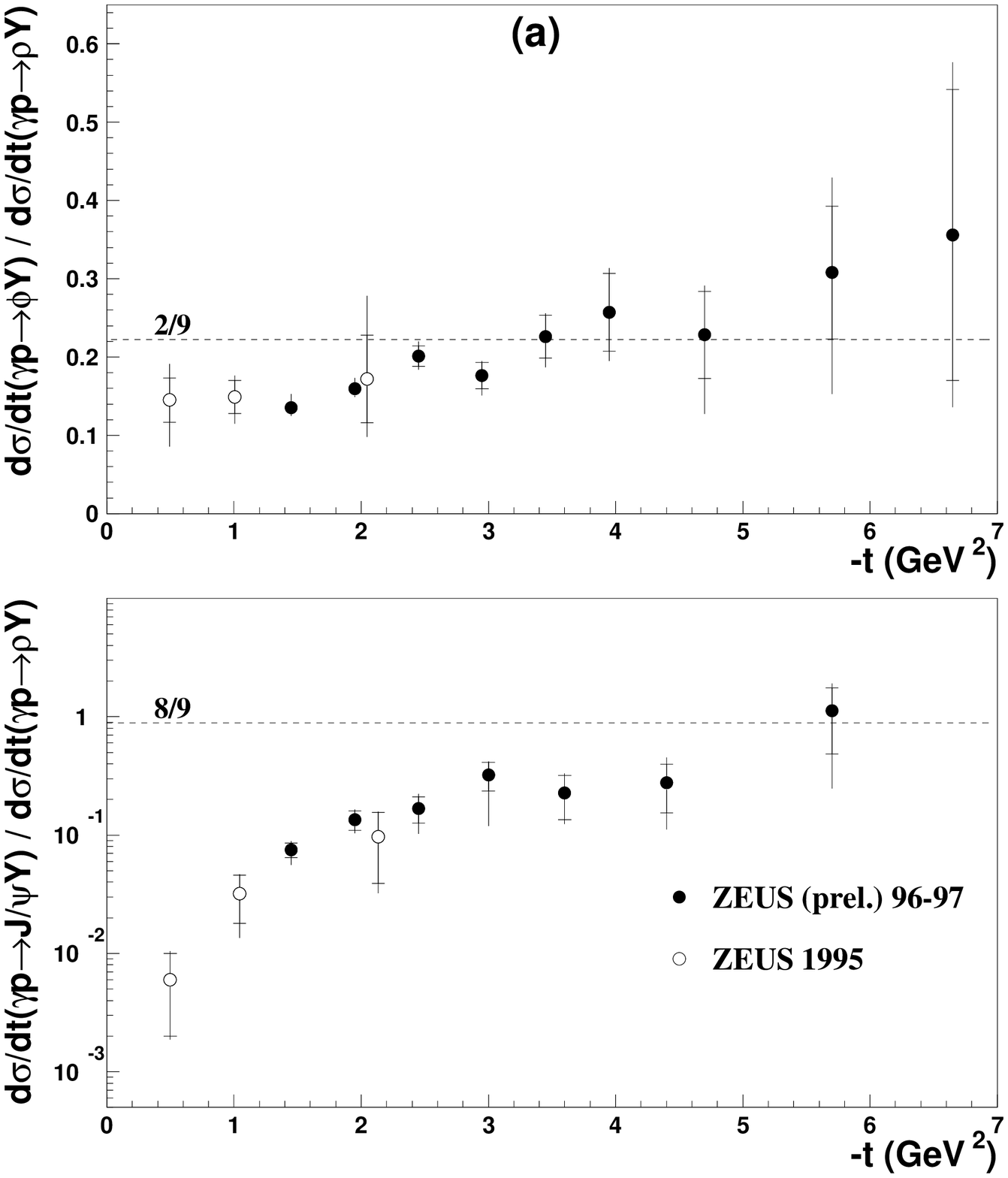,width=6.5cm%
,bbllx=0pt,bblly=100pt,bburx=540pt,bbury=710pt,clip=%
}}\hspace*{2mm}
\raisebox{7.5mm}{\epsfig{file=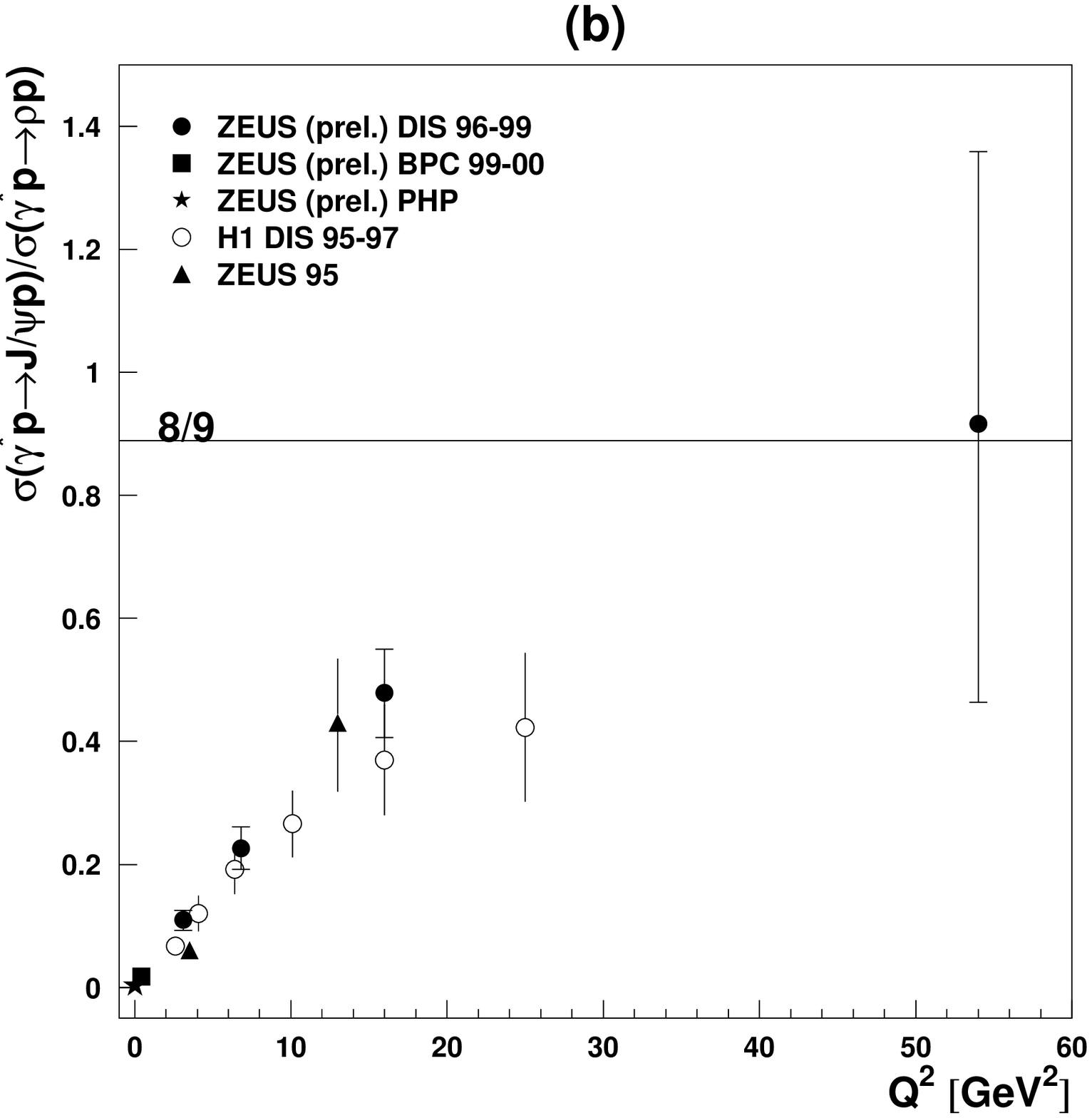,width=6.3cm%
%,bbllx=80pt,bblly=220pt,bburx=540pt,bbury=666pt,clip=%
}}\vspace*{3mm}} 
\fcaption{
(a) Ratio of cross sections $ d\sigma/dt$ for $\phi$ and $\rho$
(upper) and for \Jpsi\ and $\rho$ (lower) measured by ZEUS in
proton-dissociative elastic scattering $\gamma p\to V Y$.  (b) Ratio
of elastic cross sections for \Jpsi\ and $\rho$,
%Elastic scattering ratio  $\sigma_\mathit{tot}^{\gamma^*p\to J/\psi p}$/
% $\sigma_\mathit{tot}^{\gamma^*p\to\rho p}$
as a function of photon virtuality $Q^2$ as measured by ZEUS and H1.
}
\label{p556}\end{figure}

However at \QQ\ = 0 both the $\phi$ and \Jpsi\ cross sections are
lower than predicted.  This must be attributed to different production
mechanisms, even the $\phi$ being sufficiently heavier than the $\rho$
and $\omega$ to follow significantly different dynamics.  To expect
SU(5) symmetry to hold, we must operate in a kinematic regime where
uniform production mechanisms operate, and it would appear necessary
that this be either one where all the mechanisms are soft, or one
where pQCD applies but quark mass effects are unimportant.  In the
latter case this means there must be a hard momentum scale
considerably greater than \LQCD\ and the vector meson masses.

A first area to check is scattering at high momentum transfer $t$,
where ZEUS have measured proton-dissociative diffractive production of
vector mesons.\cite{p556} The predictions appear to work (fig.\
\ref{p556}a): for $|t|$ above approximately 3 GeV or 6 GeV,
respectively, the differential cross section ratios
$d\sigma/dt(\gamma^*p\to\phi Y)\,/\,d\sigma/dt(\gamma^*p\to\rho Y)$
and $d\sigma/dt(\gamma^*p\to J/\psi\:
Y)\,/\,d\sigma/dt(\gamma^*p\to\rho Y)$ become consistent with
asymptotic values of 2/9 and 8/9.  More high-$|t|$ data are desirable
to confirm this to greater precision.  The corresponding ratios for
the purely elastic $\phi$ and \Jpsi\ production have also been
measured by ZEUS:\cite{zel95} they are consistent with expectations
but the experimental errors are substantial.

Likewise, the total cross section ratio for $\phi$ relative to $\rho$
plateaus to a ratio of 2/9 when \QQ\ exceeds about 6 GeV (which seems
to be a higher value than in the case of $t$, indicating that $-t$ and
\QQ\ are not equivalent scales).\cite{h1mq,zphi} For \Jpsi, the data
are too sparse at present to make a definite statement; however there
is a steady rising trend with \QQ, and an asymptotic value of 8/9 is
possible (fig.~\ref{p556}b).

%The variation of the \Jpsi\ cross section with \QQ\ at a fixed value
%of $W$ = 90 GeV is shown in fig.\ \ref{539-5}.  Also plotted are the
%predictions of the theories considered in fig.\ \ref{YYY}, the FKS
%curve having been renormalised by a factor $\approx 1.6$ to pass
%through the ZEUS photoproduction point; thus treated it is a
%reasonable fit to the rest of the data.  The DIS data can also be
%fitted as a power of the quantity $(\QQ + m_{J/\psi}^2)$.

\begin{figure}
\centerline{
\raisebox{0mm}{\epsfig{file=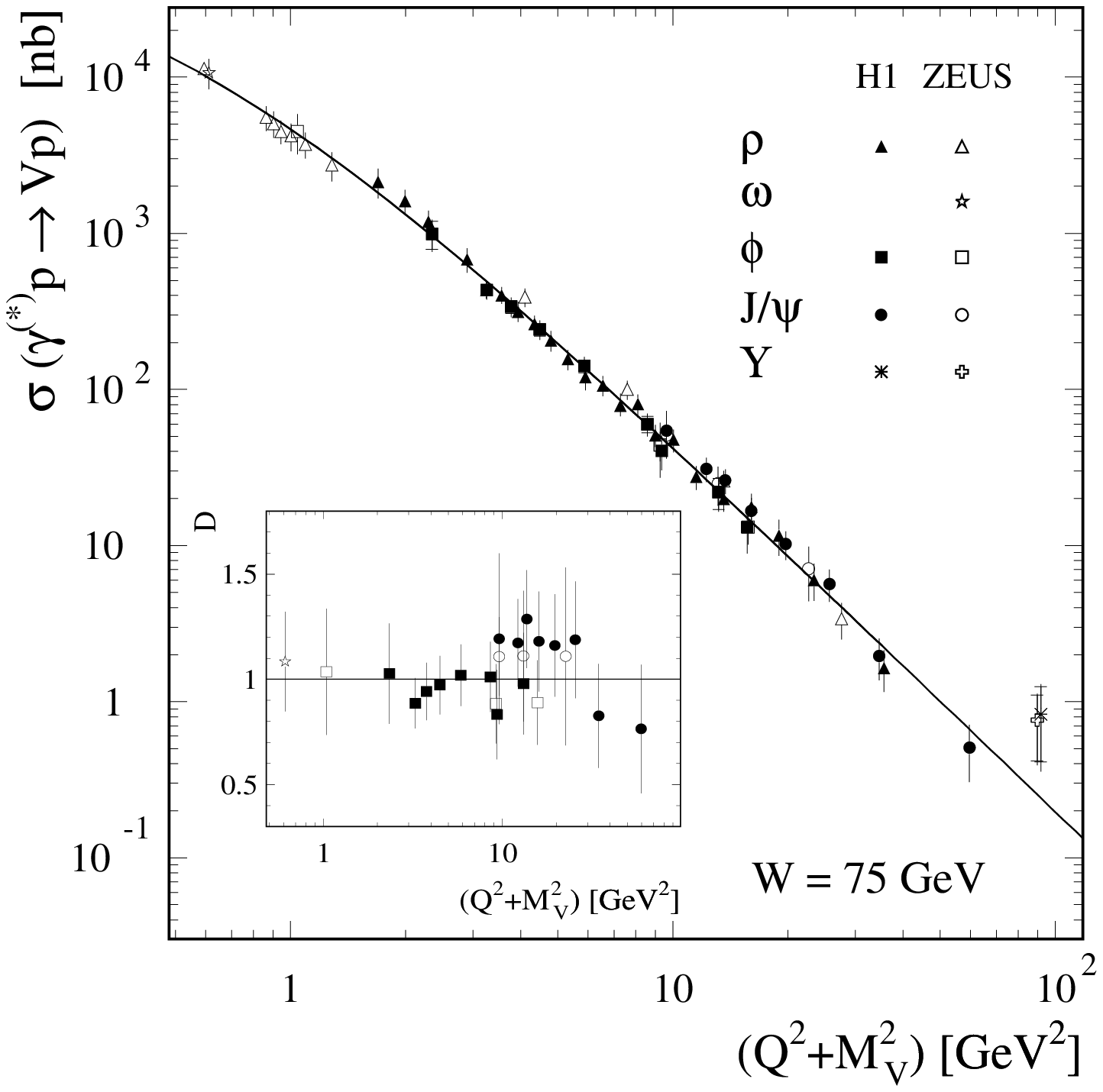,width=8cm%
,bbllx=73pt,bblly=220pt,bburx=485pt,bbury=630pt,clip=%
}}
} 
\fcaption{
Total elastic cross sections for vector meson production on a common
scale (from H1).  }
\label{h1mq}\end{figure}

H1 have argued\cite{h1mq} (fig.\ \ref{h1mq}) that a universal curve
can be plotted through all the total cross sections for $\gamma^* p
\to Vp$.  The plotted results were taken at a fixed $W$ value of 75 GeV and
the different cross section values were scaled by the reciprocal of
their respective SU(5) factors, as listed above.  (The $\Upsilon$
cross section needed to be extrapolated in $W$ in a possibly
questionable way.) After this, a function given by
$\sigma_\mathit{fit} = a_1(\QQ + m_V^2 +a_2)^{a_3}$, with suitable
values for the constants $a_i$, passes impressively though the data
points.

No explanation was offered by H1 for this observation.  ZEUS have
studied the matter further\cite{p548A} with the conclusion that the
postulate of a universal scale variable $(\QQ + m_V^2)$ is at best an
oversimplification.  It works for $\rho$, $\omega$ and $\phi$
production, but fails for \Jpsi\ production when the matter is studied
in detail.  Taken over a broad $W$ range, and at a series of $(\QQ +
m_V^2)$ values, the scaled \Jpsi\ cross sections are consistently
higher by around 50\% than the light meson data.  At least SU(5) is
broken.

%\begin{figure}
%\centerline{\hspace*{1mm}
%\raisebox{0mm}{\epsfig{file=p562-3.eps,width=7cm%
%,bbllx=80pt,bblly=220pt,bburx=540pt,bbury=666pt,clip=%
%}}
%} 
%\fcaption{
%Elastic scattering ratio  $\sigma_\mathit{tot}^{\gamma^*p\to\psi(2s) p}
% /\sigma_\mathit{tot}^{\gamma^*p\to\Jpsi p}$
% as a function of $\gamma p$ centre-of-mass energy as measured by ZEUS and H1.
%}
%\label{p562}\end{figure}

As well as scattering elastically in diffractive processes, the
virtual photon may dissociate into an unbound $q\bar q$ system.  ZEUS
have presented measurements of $D^*$ production in such processes,
both for photoproduction and DIS.\cite{korzh,p489} The kinematical
distributions, at present with large errors, are approximately
consistent with boson-gluon fusion in resolved-pomeron predictions
using RAPGAP.\cite{rapgap} H1 have measured this process in DIS\cite{p489b}
and have obtained satisfactory agreement with QCD model predictions.
In both experiments, a suitable parton model of the pomeron is needed, 
with a dominant gluon component.

\subsection{Regge phenomenology}
\noindent
The previous sections have treated diffraction principally from a pQCD
viewpoint, which has an obvious relevance to heavy quark production.
Elastic light vector meson photoproduction is described well within
the older Regge framework, however, and this must now be brought into
the discussion.  A recent account of this area of HERA physics has
been given by Abramowicz.\cite{halina}

\begin{figure}
\centerline{\hspace*{1mm}
\raisebox{0mm}{\epsfig{file=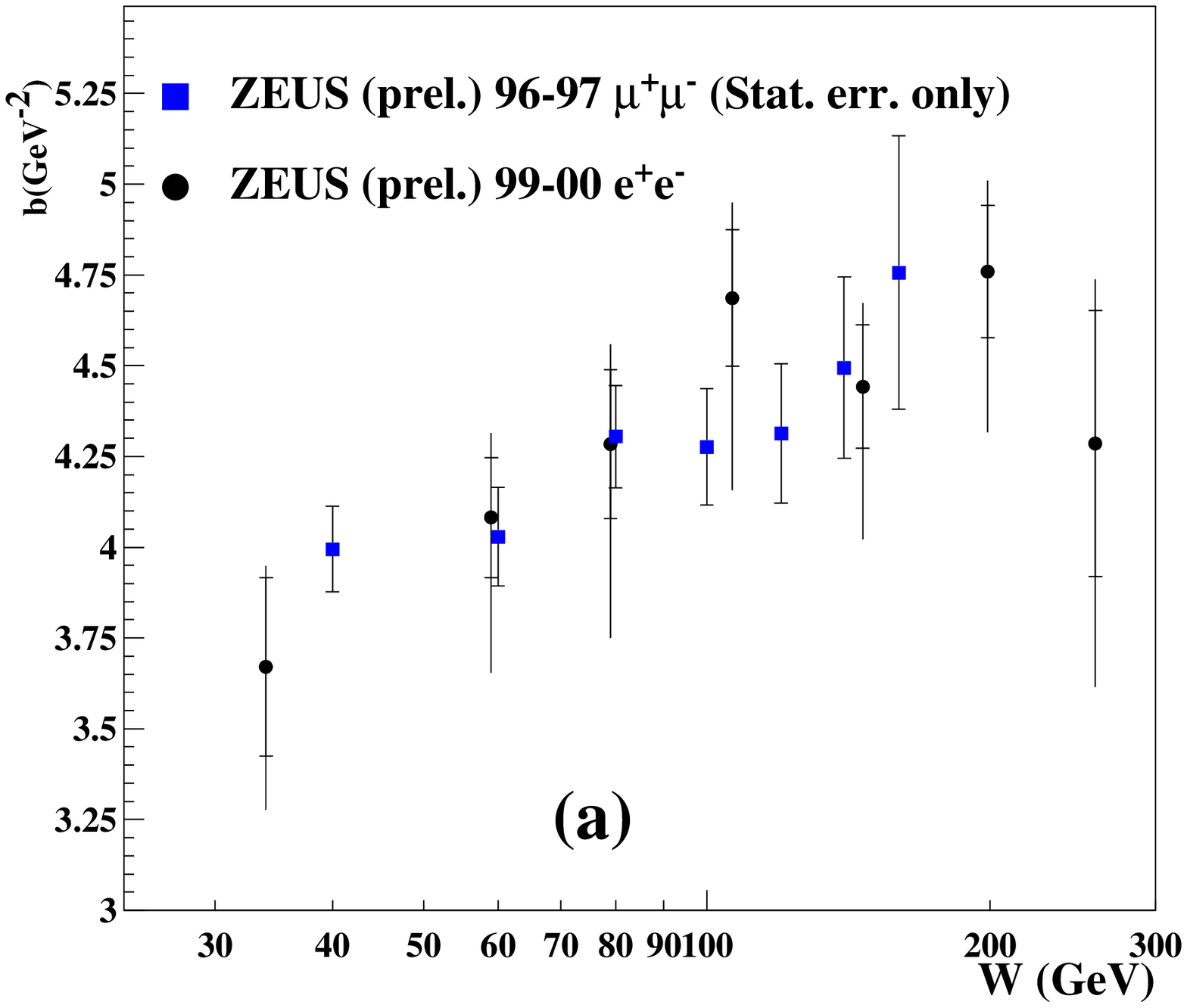,width=6.5cm%
%,bbllx=80pt,bblly=220pt,bburx=540pt,bbury=666pt,clip=%
}}
\raisebox{0mm}{\epsfig{file=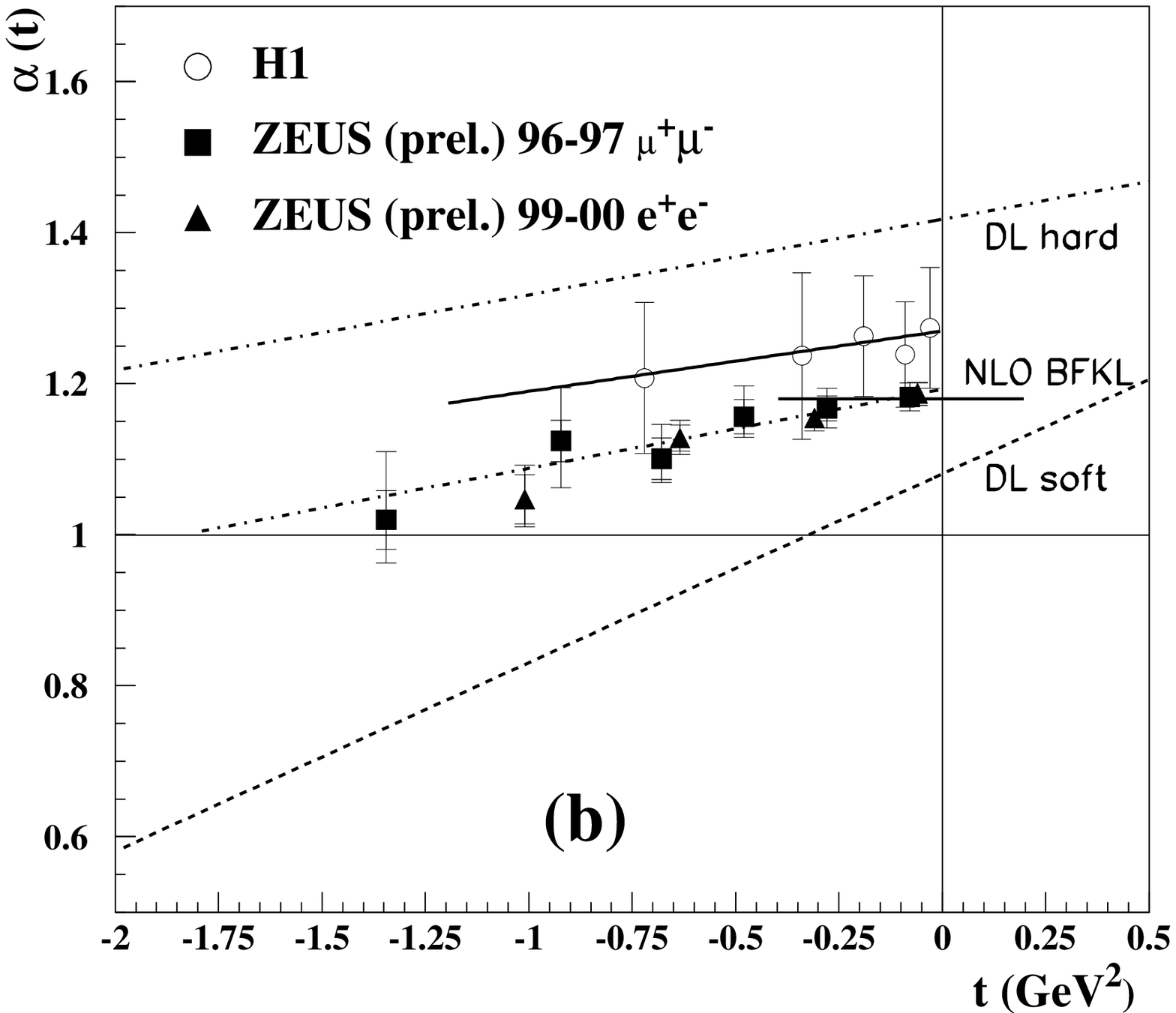,width=6.5cm%
%,bbllx=80pt,bblly=220pt,bburx=540pt,bbury=666pt,clip=%
}}
}\vspace*{2mm} 
\fcaption{
Elastic scattering parameters $b$ and $\alpha$ in the 
process $\gamma p \to\Jpsi p$ measured by ZEUS (preliminary) and H1. 
}
\label{p548}\end{figure}

The differential cross section for Regge pomeron exchange
is given by
\begin{equation}
\frac{d\sigma(\gamma p)}{d|t|} \;\propto\; 
\exp^{b_0t}\left(\frac{W^2}{W_0^2}\right)^{2(\alpha(t) -1)}
\label{eqt}\end{equation}
where $\alpha(t) = \alpha(0) + \alpha't$.  The constants $\alpha,
\alpha'$ are believed to be universal, and characteristic of the
pomeron, while $b_0$ and $W_0$ depend on the particular
process.  The $t$ distribution depends on
\begin{equation}
b \; = \; b_0 + 4\alpha'\ln(W/W_0).
\end{equation}
A positive value of $\alpha'$ means that the cross section falls more
rapidly with $|t|$ as $W$ increases, a characteristic known as
``shrinkage''.  The total elastic cross section, on integrating
(\ref{eqt}), varies as $W^\delta$ where $\delta=4(\alpha(0) -1
-\alpha'/b)$.  Experimentally one finds $\alpha(0)\approx1.08$ and
$\alpha'\approx0.25$ GeV$^{-2}$.  Light vector mesons have $b\approx
10$ GeV$^{-2}$.

There is already a difficulty here in \Jpsi\ photoproduction, because a
value $b\approx 4$ GeV$^{-2}$ is measured (fig.~\ref{p548}a), which
would give $\delta\approx 0.1$, while the measured value of $\delta$
is around 0.8; in other words the actual $W$ dependence is
steeper than the Regge formula requires.  On the other hand, if pQCD
were the whole story, we should not expect to see shrinkage in the
\Jpsi\ differential cross section, since in pQCD the $W$ and $t$
dependences are not coupled.\cite{fks2} From fig.~\ref{p548}b it is
evident that shrinkage is indeed observed.\cite{p878}  This implies
non-perturbative physics, which is good news for Regge theory.

Nevertheless, the $W$ dependence represents a problem.  In order to
rescue Regge theory, Donnachie and Landshoff have proposed a ``hard''
pomeron\cite{DLhard} as well as the familiar ``soft'' one which
describes light vector meson production well.\cite{DLsoft} The hard
pomeron has $\alpha(0)\approx1.4$ while the soft one still has
$\alpha(0) \approx 1.08$.  However, the experimental $t$ variation
gives an intercept of $\alpha(0)=1.193\pm0.011\;^{+0.015}_{-0.010}.$
This agrees with a BFKL calculation,\cite{brodsky} but with neither
the soft nor the hard pomeron on its own.

In a more complex approach one may attempt to combine the two types of
pomeron.  The above authors and H1 have shown that a fit
involving both pomerons is able to describe the observed
variation of the cross section with $W$.\cite{DLsoft,h1ups} If this is
becoming a little complicated, perhaps the pomeron {\it is\/}
complicated.

One aspect of simple Regge theory which is found to hold in \Jpsi\
production is $s$-channel helicity conservation (SCHC). This has been
reported by ZEUS from an analysis of the relevant spin density-matrix
elements using the angular distributions of the \Jpsi\ decay
products.\cite{p559}

\subsection{Proton dissociation}
\noindent

%===================================================================
\begin{figure}
\centerline{\hspace*{5mm}
\raisebox{-0.5mm}{\epsfig{file=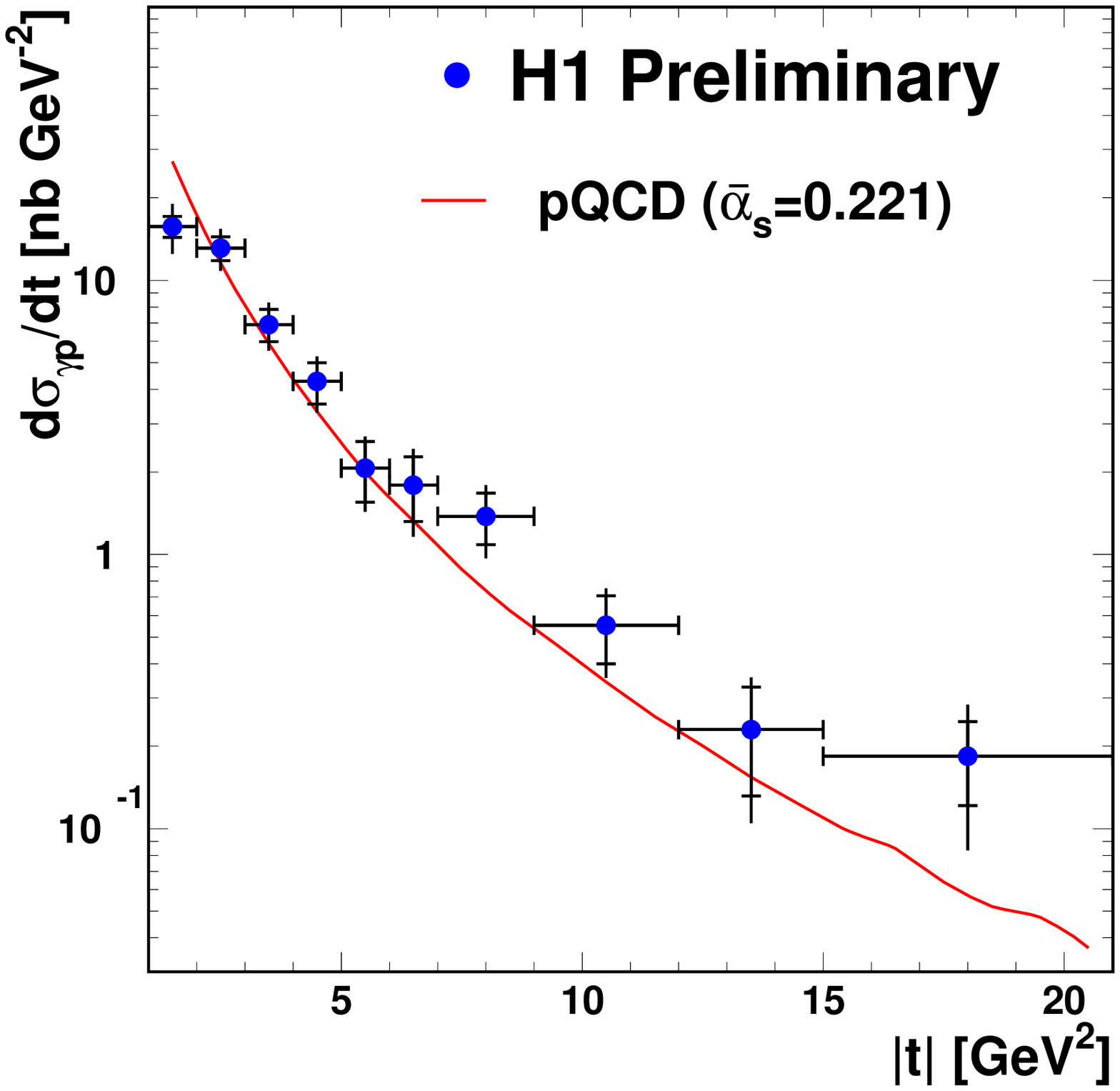,width=6.2cm%
%,bbllx=125pt,bblly=500pt,bburx=565pt,bbury=721pt,clip=%
}}
}\vspace*{3.5mm}
\fcaption{Differential cross section in $|t|$ 
for \Jpsi\ production in DIS with proton dissociation (H1).
A QCD-based theoretical prediction is shown.}
\label{p806}\end{figure}
%==================================================================
In diffractive events at high momentum transfers $t$, the proton can
dissociate while the vector meson still emerges elastically, i.e.\
with no additional associated final-state products.  Recent ZEUS
measurements of this process have already been discussed.\cite{p556}
In \Jpsi\ photoproduction, H1 have made a study\cite{p806}
selecting events in which the proton dissociates into a final state
with a mass in the approximate range 1.6 - 30 GeV, as
measured in their forward calormeter system.  The value of $t$ is
obtained from the transverse momentum of the measured
\Jpsi.  The pQCD-based event generator HITVM\cite{hitvm} describes
the $W$ and the $t$ distributions well (fig.~\ref{p806}).  The shape of
the variation with $t$ appears to be independent of $W$; at given $t$ the
cross sections rise as $W^\delta$ with $\delta\approx 1$.

\subsection{$\psi(2S)$ production}
\noindent
%===================================================================
\begin{figure}
\centerline{{\sf (a) \hspace{5cm} (b)}}
\centerline{\hspace*{1mm}
\raisebox{0mm}{\epsfig{file=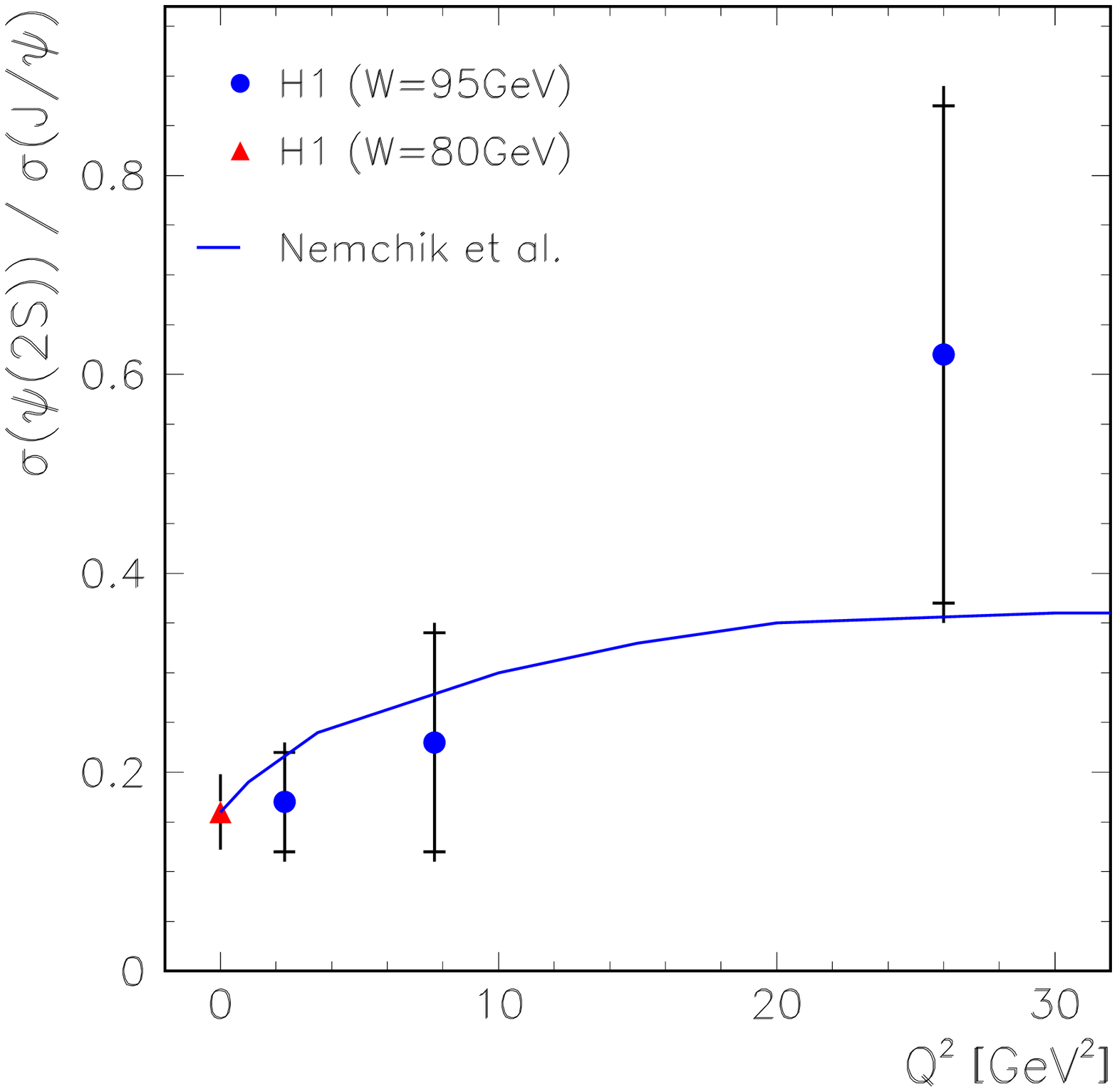,width=6.2cm%
,bbllx=0pt,bblly=0pt,bburx=560pt,bbury=520pt,clip=%
}}
\raisebox{3mm}{\epsfig{file=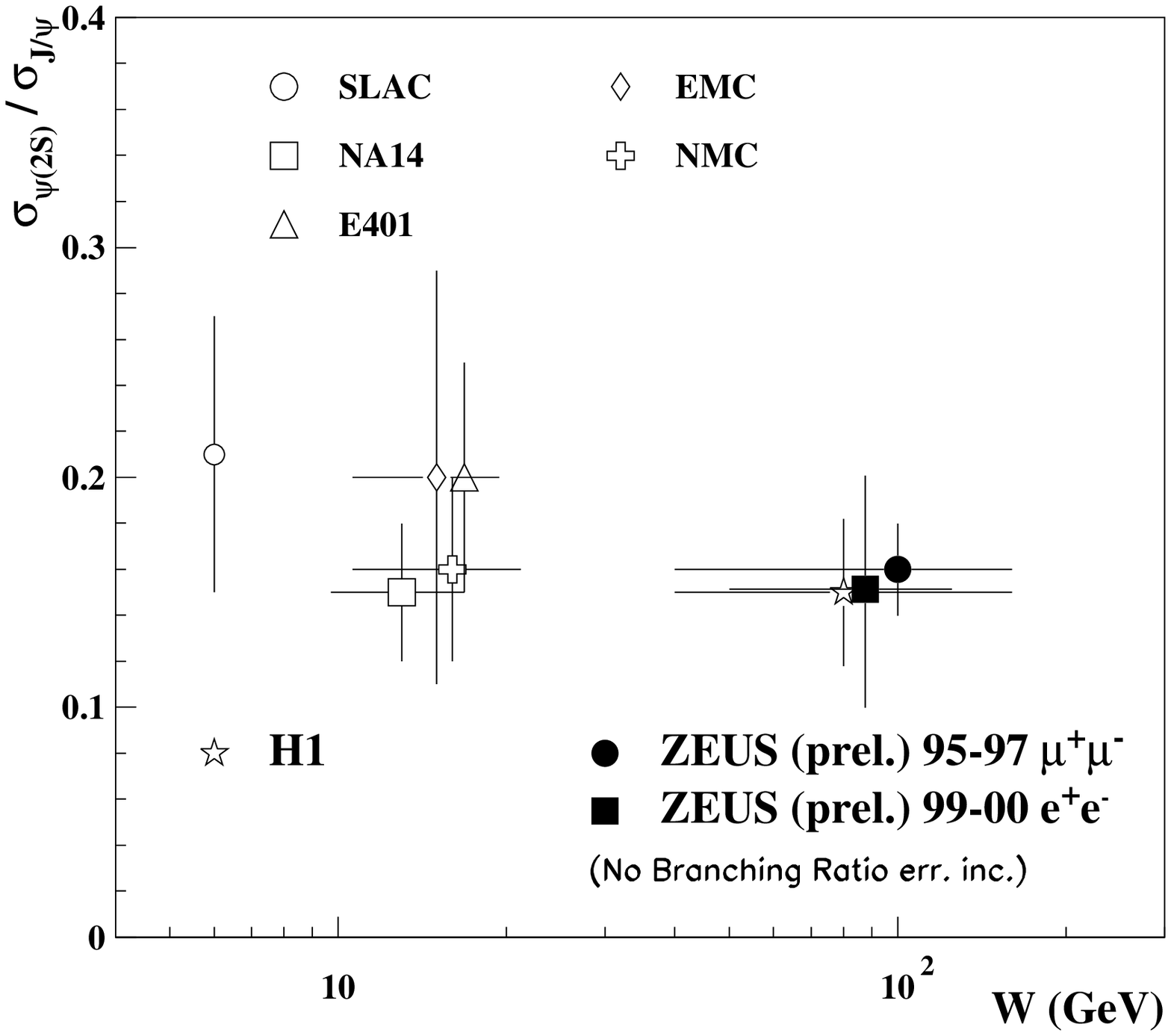,width=6.2cm%
%,bbllx=125pt,bblly=500pt,bburx=565pt,bbury=721pt,clip=%
}} }\vspace*{1mm}
\fcaption{Differential cross section 
for $\psi(2S)$ production in DIS with proton dissociation: (a)~results
from H1 with theoretical comparison, (b)~results from ZEUS
(preliminary) compared to other experiments.}
\label{p562}\end{figure}
%==================================================================
H1 have measured the elastic production of the $\psi(2S)$ state both
in photoproduction and in DIS.\cite{h1psip,p560} The decay channel
$\psi(2S) \to \Jpsi\pi^+\pi^-$ was used, and the results include the
contribution from proton dissociation.  A colour-dipole based QCD
prediction\cite{nemchik} describes the data well
(fig.~\ref{p562}). There is also a prediction\cite{fks2} that for
$\QQ\gg m_{J/\psi}^2$, the ratio of the elastic $\psi(2S)$ and \Jpsi\
cross sections should tend to 0.5.  ZEUS have measured $\psi(2S)$
production through the $e^+e^-$ decay channel and obtain a result that
is consistent with H1 and with lower energy experiments.\cite{p562}

\subsection{Measurements of $b\bar b$}
\noindent
%===================================================================
\begin{figure}
\vspace*{1mm}
\centerline{\hspace*{1mm}
\raisebox{0mm}{\epsfig{file=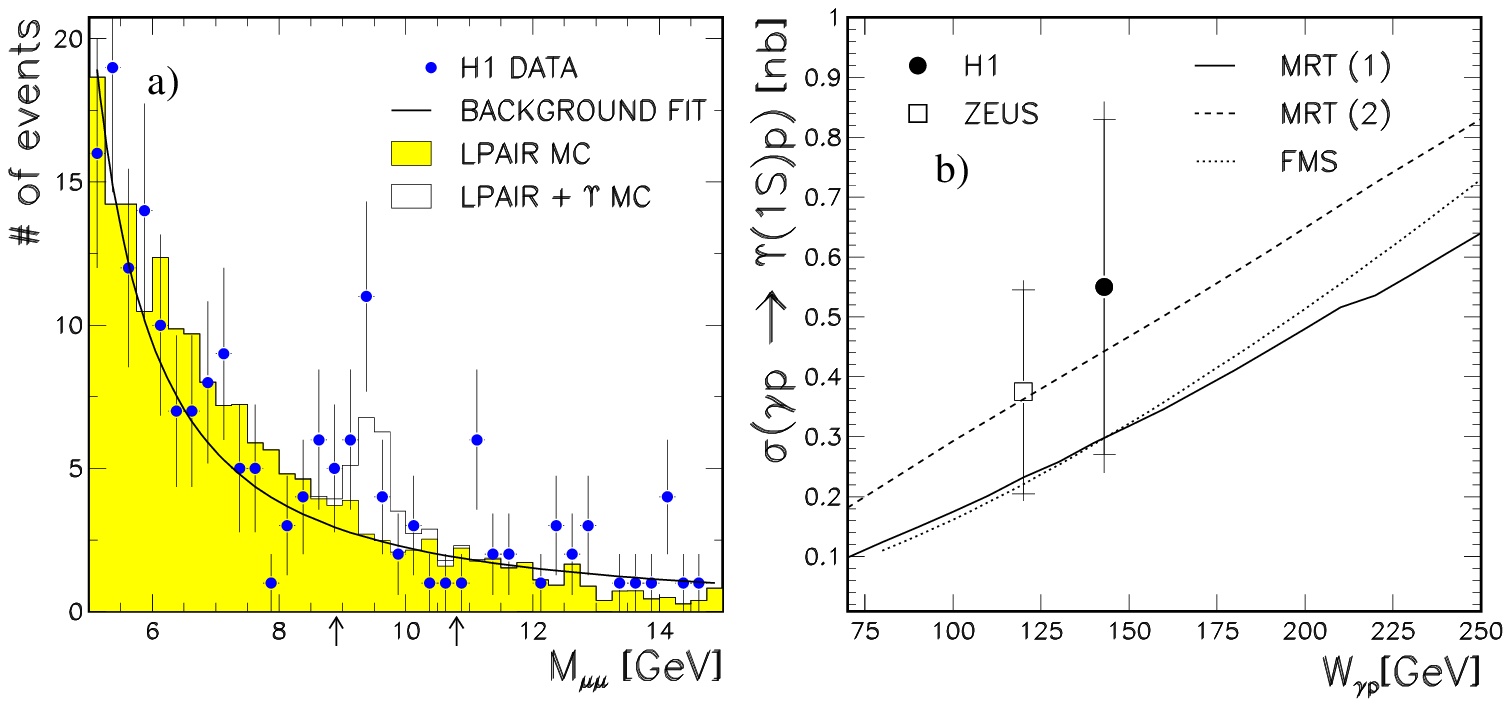,width=12cm%
,bbllx=125pt,bblly=500pt,bburx=565pt,bbury=721pt,clip=%
}} }\vspace*{-2mm}
\fcaption{(a) Dimuon mass distribution in photoproduction from H1.
The histogram represents the simulated Bethe-Heitler background.
(b)~$\Upsilon$ total cross section in photoproduction: HERA results 
compared to theory.}
\label{p805}\end{figure}
%==================================================================
The $\Upsilon$ family are the only $b\bar b$ states measured at HERA,
and have been observed only in the  elastic channel in photoproduction and
DIS, through their $\mu^+\mu^-$ decays.  The dimuon spectrum from
H1 is illustrated in fig.~\ref{p805}a.\cite{h1ups} A $\Upsilon$
enhancement is seen with a little  difficulty above a background 
due to the Bethe-Heitler process $\gamma\gamma\to\mu^+\mu^-$, where one
of the photons is radiated from the proton.  ZEUS have observed a similar
signal.\cite{zups} The mass resolution is not sufficient to separate the
$\Upsilon(1S),$ $(2S)$ and $(3S$) states, which cover the mass range
9.46 - 10.36 GeV, and so the signal-background subtraction is
performed over a range that covers all three resonances together.
 
The cross section is evaluated assuming that the standard branching
rate to muons applies to all three channels and taking 70\% of the
signal as $\Upsilon(1S)$.  The results from the two experiments are
shown in fig.~\ref{p805}b, compared with QCD-based
calculations by Martin, Ryskin and Teubner (MRT)\cite{mrt} and by
Frankfurt, McDermott and Strikman (FMS).\cite{fms} Within present
errors, these models all describe the data.  A calculation by
FKS\cite{fks} was found by ZEUS to be too low.

\section{Conclusions}
\noindent
The HERA collider offers a rich breadth of perspectives through which
many areas of particle physics can be studied, and in so doing stands
as an experimental facility unmatched in the world.  This is
particularly the case in the study of QCD processes, where the
production of heavy quark systems provides a laboratory for the
probing of many important details of the theory.   Mesons
containing charm quarks have been extensively measured, and a start
has been made on the study of beauty meson production.  These
measurements have made a useful start in distinguishing between
different perturbative and semi-perturbative QCD models.  Nevertheless
it is clear that better experimental statistics at HERA will be
exremely valuable in the study of charm physics, and indispensible for
the study of beauty.  The ability to study the production of $B$ and
$\Upsilon$ mesons at transverse momenta of the same order as the heavy
quark mass -- yet still ``hard'' compared with \LQCD\ -- can give
penetrating tests of theoretical ideas.

Although the available energies are not as high as at the Tevatron,
and top quark pairs lie outside HERA's reach, the recent improvements
to the HERA collider, together with a variety of enhancements in the
experiments, make for exciting prospects in this area.  Following the
machine upgrades, a factor of up to ten in integrated luminosity may
be achieved by 2005.  No serious evidence against QCD has been found
at HERA, but the phenomenological implementations of the theoretical
ideas do not yet seem to be perfect.  An important onus lies on the
theoreticians to provide an understanding of these areas of QCD that
will match the quantity and quality of the promised experimental data.

\nonumsection{Acknowledgements}
\noindent
I should like to express appreciation of the many fruitful contacts
with colleagues in ZEUS and H1 over the years, which have greatly
deepened my understanding of QCD physics at HERA, and with particular
thanks to D. H. Saxon for a critical reading of the manuscript.  I am
also grateful to S. Frixione for informative help.

\newpage
\nonumsection{References}

\end{document}